\documentclass[useAMS,usenatbib]{mn2e}

\usepackage{graphicx} 

\title[The Mg/Fe abundance ratio in the MILES library]{Element abundances in the stars of the MILES spectral library: the Mg/Fe ratio} 

\author[A. de C. Milone, A. E. Sansom and P. S\'anchez-Bl\'azquez] 
{A. de C. Milone$^{1,2}$\thanks{E-mail: acmilone@das.inpe.br, andre.milone@inpe.br}, 
A. E. Sansom$^{2}$ and P. S\'anchez-Bl\'azquez$^{3}$\\ 
$^1$Divis\~ao de Astrof\'\i sica, Instituto Nacional de Pesquisas Espaciais, \\ 
Av. dos Astronautas 1758, S\~ao Jos\'e dos Campos, SP, 12227-010, Brazil \\ 
$^2$Jeremiah Horrocks Institute, University of Central Lancashire, \\ 
Preston, PR1 2HE, United Kingdom \\ 
$^3$Departamento de Fisica Teorica, Universidad Aut\'onoma de Madrid, \\
Cantoblanco, 28049, Madrid, Spain} 
 
\begin{document} 
 
\date{Accepted 2011 January 31. Received 2011 January 28; in original form 2010 October 15} 
 
\pagerange{\pageref{firstpage}--\pageref{lastpage}} \pubyear{2009} 
 
\maketitle 
 
\label{firstpage} 
 
\begin{abstract} 
 
We have obtained [Mg/Fe] measurements for 76.3\% of the stars in the MILES spectral library used
for understanding stellar atmospheres and stellar populations in galaxies and star clusters. 
These abundance ratios were obtained through
(1) a compilation of values from the literature using abundances from high-resolution spectroscopic studies and
(2) a robust spectroscopic analysis using the MILES mid-resolution optical spectra.
All the [Mg/Fe] values were carefully calibrated to a single uniform scale,
by using an extensive control sample with results from high-resolution spectra.
The small average uncertainties in the calibrated [Mg/Fe] values
(respectively 0.09 and 0.12 dex with methods (1) and (2))
and the good coverage of the stars with [Mg/Fe] over stellar atmospheric parameter space of the library
will permit the building of new simple stellar populations (SSPs) with empirical $\alpha$-enhancements.
These will be available for a range of [Mg/Fe], including both sub-solar and super-solar values, 
and for several metallicities and ages.
These models will open up new prospects for testing and applications of evolutionary stellar population synthesis.

\end{abstract}


\section[1]{Introduction} 
 
Evolutionary stellar population synthesis, 
i.e. modelling spectral energy distributions emitted by evolving stellar populations,
is a natural approach to studying the stellar content of different galaxies.
One of the main ingredients of these models are the stellar libraries, which can be empirical or theoretical.
Stellar population models usually consider only the total metal content of stars
and, therefore, ignore the different chemical abundance patterns that are present in individual stars.
However, different chemical abundance patterns have a strong influence on the shape of the spectra.
In particular, it is well known 
that stellar population models based on empirical libraries
(which are mostly composed of solar neighbourhood stars)
cannot reproduce the high values of Mg abundances found in giant elliptical galaxies.
This is commonly interpreted as a consequence of high [Mg/Fe] in these systems,
most likely due to a rapid star formation history compared to the more quiet one of the solar neighbourhood
(Tinsley 1980).

An obvious solution, explored recently by several authors
(e.g. Coelho {\it et al.} 2007;
Walcher {\it et al.} 2009;
Lee, Worthey \& Dotter 2009),
would be to use theoretical libraries with the desired coverage
in chemical abundances. However, while theoretical libraries have improved dramatically in the last few years  
(Chavez, Malagnini \& Morossi 1997;  
Murphy \& Meiksin 2004; 
Rodriguez-Merino {\it et al.} 2005; 
Munari {\it et al.} 2005; 
Martins {\it et al.} 2005; 
Coelho {\it et al.} 2005; 
Fr\'emaux {\it et al.} 2006;
Bertone {\it et al.} 2008),
they still do not reproduce real stars of all spectral types, with especial problems depending on the wavelength range
(i.e. Martins \& Coelho 2007;
Bertone {\it et al.} 2008).
Some of the remaining problems are
the incompleteness of the atomic and molecular line opacity lists in the blue region of the spectrum 
and for cool stars (T$_{\rm eff}$ $<$ 4500 K) as well.
 
Another approach is to compute, with the help of model atmospheres, response functions 
to characterise the variation of specific spectral characteristic (usually Lick indices) to variation of different elements
(see Trippico \& Bell 1995;
Korn, Maraston \& Thomas 2005).
Models using these response functions are those of
Tantalo, Chiosi \& Bressan (1998),
Trager {\it et al.} (2000a), and
Thomas, Maraston \& Bender (2003).
However, the accuracy of these theoretical predictions have not been tested empirically yet.

It is the main objective of this work to provide [Mg/Fe] abundance ratios
for one of the most complete empirical stellar libraries currently available 
(MILES) ({\bf M}id-resolution {\bf I}saac Newton Telescope 
{\bf L}ibrary of {\bf E}mpirical {\bf S}pectra, 
S\'anchez-Bl\'azquez  {\it et al.} 2006).
The MILES database, which was especially designed for stellar population modelling,  
contains flux calibrated optical spectra of high signal-to-noise ratio (S/N) for 985 stars
covering $\lambda\lambda$3525-7500 {\AA} with a homogeneous resolution $\Delta\lambda$ = FWHM = 2.3 {\AA}. 
The parametric coverage of sample stars
in the three-dimensional H-R diagram is quite wide:
2800 $\leq$ T$_{\rm eff}$ $\leq$ 50400 K,
0.0 $\leq$ $\log$ g $\leq$ $+$5.0,
and $-$2.7 $\leq$ [Fe/H] $\leq$ $+$1.0 dex,
where [Fe/H] = log(Fe/H)$_{\star}$ $-$ log(Fe/H)$_{\odot}$
such that formally log(Fe) = log($n$(Fe)/$n$(H)) + 12, log(H) = 12 and
$n$(Fe) and $n$(H) are the numerical densities (cm$^{-3}$) of iron and hydrogen atoms respectively.
The scales for these photospheric parameters were carefully defined by
Cenarro {\it et al.} (2007).
Their precisions, respectively $\pm$100 K, $\pm$0.2 and $\pm$0.1 dex,
makes MILES good for SSP modelling.
For the present work, we excluded those stars with uncertain or wrong atmospheric parameters
(see Vazdekis {\it et al.} 2010
for details about how these stars were identified).

The MILES [Mg/Fe] catalogue presented here consists of two measurement sets.
The first one is composed of measurements obtained from the literature
from high spectral resolution analyses properly calibrated to a common system.
The second set assembles abundances measured by us directly
from the MILES mid-resolution (hereafter MR) spectra and calibrated using the high-resolution (hereafter HR) sample.
The catalogue is represented in two separate tables for field and cluster stars.
Both tables are only available in electronic form.
The paper layout is as follows:
Section 2 describes the compilation of HR abundance measurements from the literature and their calibration;
Section 3 shows the Mg abundance measurements from the MILES spectra;
Section 4 compiles the MILES [Mg/Fe] catalogue and analyses its coverage over the library parameter space;
Section 5 compares our stellar data with predictions of theoretical models focusing on the behaviour of some Lick indices with [Mg/Fe];
and, finally,
Section 6 plans for applications to building new simple stellar population models with variable $\alpha$-enhancement.
Section 7 summarizes the whole paper and final conclusions.
There are also three appendices:
Appendix A confronts the compiled HR data with a well-known stellar spectrum library, 
Appendix B presents comparisons of the MILES photospheric parameter scales with those from the compiled HR studies, 
and Appendix C compares the results for cluster stars with HR studies.

\section[2]{Compilation of magnesium abundances from high-resolution studies} 

The first step of this work consisted of performing a bibliographic compilation of magnesium abundances
from high-resolution spectroscopic analyses for the MILES library stars.
To guarantee
homogeneity between the measurements provided by several studies
we performed a calibration and correction of systematic differences among sources and a chosen standard reference system,
following a similar procedure as in 
Cenarro {\it et al.} (2001, 2007).
For instance,
Feltzing \& Gustafsson (1998)
give a detailed error analysis of elemental abundances for G and K metal-rich dwarfs
also including comparisons with other studies.
In this section, we describe the chosen reference sample and the procedures
we followed to homogenize the measurements to a single uniform scale of [Mg/Fe].

\subsection[2.1]{A reference scale for [Mg/Fe]} 

Our reference sample to define a scale for the Mg/Fe abundance ratio is from
Borkova \& Marsakov (2005, hereafter BM05).
The catalogue of Borkova \& Marsakov is a robust compilation
of the atmospheric parameters T$_{\rm eff}$, log $g$ and [Fe/H] plus [Mg/Fe]
from high-S/N high-resolution analyses of field stars published between 1989 and 2003
(covering 36 studies with Mg abundance determinations for around 900 stars).
BM05 computed weighted average values and their errors
through an iterative procedure in order to correct for the systematic deviations
of each data set relative to reduced mean homogeneous scales.
The uncertainty of [Mg/Fe] in BM05 is 0.05 and 0.07 dex, respectively,  
for metal-rich ([Fe/H] $>$ $-$1.0 dex) and metal-poor stars ([Fe/H] $\leq$ $-$1.0 dex). 
The BM05 catalogue contains 218 stars in common with the MILES library
(all with log $g$ $\geq$ 3.0).

BM05 was also the reference work adopted in
Cenarro {\it et al.} (2009), 
where Mg and Ca abundances were compiled for 192 stars of their calcium triplet stellar library (hereafter CaT) of 706 objects.
As there are many MILES stars in common with the CaT sample (132 stars),
this work provides values that we can compare with
(see Appendix A).

\subsection[2.2]{Calibration of the high-resolution [Mg/Fe]} 

We first checked the possible presence of systematic differences in the scales of MILES and BM05 atmospheric parameters
([Fe/H], log $g$ and T$_{\rm eff}$), however we did not find any (see Appendix B for details).
Apart from the BM05 compilation, we obtained [Mg/Fe]
for 97 more stars from 15 other HR studies, as listed in Table 1.
Their abundance ratios were then carefully transformed onto the adopted scale as described next.

\begin{table*} 
\caption{
Parameters of the linear calibrations of different sets of [Mg/Fe] to the BM05 scale 
and other information about the consulted high-resolution spectroscopic works. 
References are shown in the first column as:  
CGS00 (Carretta, Gratton \& Sneden 2000), 
F00 (Fulbright 2000), 
Ge03 (Gratton {\it et al.} 2003),  
Be05 (Besnby {\it et al.} 2005), 
T98 (Th\'evenin 1998), 
RLA06 (Reddy, Lambert \& Allende Prieto 2006), 
LH05 (Luck \& Heiter 2005), 
EN03 (Erspamer \& North 2003), 
FG98 (Feltzing \& Gustafsson 1998), 
Ce02 (Caliskan {\it et al.} 2002), 
FK99 (Fulbright \& Kraft 1999), 
H02 (Heiter 2002), 
Ae01 (Adelman {\it et al.} 2001),  
Ae06 (Adelman {\it et al.} 2006), 
and 
Ce09 (Cenarro {\it et al.} 2009). 
The $-A/B$ and $1/B$ values (second and third columns)
are, respectively, the additive and multiplicative coefficients of the calibration expressions (Eq. 2).
When they are represented by integer numbers it means that no calibration of [Mg/Fe] was applied
because there are too few stars or none in common between the work and BM05 samples to compute a linear fit.
N$_{\rm c}$ (forth column) represents the number of stars in common between each work and BM05
after applying the excluding criterion to each $lsq$ fit.
N$_{\rm i}$ (fifth column) is the number of stars from each reference  
that are eligible to be included into the MILES [Mg/Fe] catalogue.
N$_{\rm r}$ (sixth column) gives the number of MILES stars that are repeated in other works (once in each case).
The [Fe/H] range of the stars to be included into our catalogue is shown in the seventh column 
and their [Mg/Fe] ranges (from the original values) are shown in the ninth column.
The [Fe/H] and [Mg/Fe] intervals of the work samples in common with, respectively, the MILES and BM05 catalogues
can be read in the plots of Figs. 1 and A1. 
The typical uncertainties of [Fe/H] and [Mg/Fe] for each work are presented, respectively, in the eighth and tenth columns. 
In the eleventh column, the main physical constrains on measuring the Mg abundances 
are cited (LTE, non-LTE, and the ionization stage of the Mg lines used in the abundance determination). 
The propagated uncertainty of [Mg/Fe] over the calibration process for each data sample is written in the last column.
The weighted averages of [Mg/Fe] errors are given in the eleventh row 
for 103 stars whose data come from the first nine works. 
The weighted averages of [Mg/Fe] errors for 9 stars of the five last listed works 
(that do not have stars repeated in other works nor were their data calibrated to the [Mg/Fe] uniform scale in the current work) 
is shown in the penultimate row.
} 
\label{lsq_params_hr_to_BM05} 
\begin{tabular}{@{}lrrrrrcccclc} 
\hline 
\hline 
  Ref.   &$-A/B$  & $1/B$ & N$_{\rm c}$ & N$_{\rm i}$ & N$_{\rm r}$ & [Fe/H]$^{\rm i}_{\rm l,u}$ & $\delta$[Fe/H] & [Mg/Fe]$^{\rm i}_{\rm l,u}$ & $\delta$[Mg/Fe]
& Notes            & $\sigma$[Mg/Fe]$_{}$ \\ 

\hline 
         &   (dex)&       &         &         &         &               (dex)&           (dex) &               (dex) &            (dex)
&                  & (dex) \\ 
\hline 
\hline 
CGS00     &  0.000 & 1.000 &       9 &     5   &       3 &       $-$2.63,$+$0.13 &           0.08 &        $+$0.04,$+$0.64 &           0.09
& non-LTE, Mg I      & 0.09 \\ 
F00      &  0.000 & 1.000 &      20 &    18   &       7 &       $-$2.64,$-$0.99 &           0.04 &        $+$0.25,$+$0.61 &           0.07
&   LTE, Mg I      & 0.07 \\ 
Ge03     & $-$0.076 & 0.997 &     132 &     2   &       1 &       $-$1.49,$-$0.75 &           0.05 &        $+$0.19,$+$0.63 &           0.09
&   LTE, Mg I      & 0.09 \\ 
Be05     &  0.000 & 1.000 &      84 &     1   &       1 &          $-$0.75     &           0.10 &           $+$0.42     &           0.06
&   LTE, Mg I      & 0.06 \\ 
T98      &  0.029 & 0.974 &     224 &    44   &       9 &       $-$2.63,$+$0.60 &        $<$0.20 &        $-$0.58,$+$0.80 &        $<$0.20
&   LTE, Mg I      & 0.20 \\ 
RLA06     & $-$0.065 & 1.392 &      59 &     1   &       1 &          $-$1.01     &           0.08 &           $+$0.35     &           0.05
&   LTE, Mg I      & 0.07 \\ 
LH05     & $-$0.103 & 1.000 &      56 &    16   &       5 &       $-$0.60,$+$0.16 &           0.06 &        $-$0.02,$+$0.57 &           0.13 
&   LTE, Mg I      & 0.13 \\ 
EN03     &  0.000 & 1.000 &       7 &    10   &       3 &       $-$1.19,$+$0.28 &           0.18 &        $-$0.34,$+$0.17 &           0.10 
&   LTE, Mg I      & 0.18 \\ 
FG98     &  0.000 & 1.000 &       7 &     6   &       1 &       $+$0.02,$+$0.26 &           0.18 &        $-$0.06,$+$0.18 &           0.13
&   LTE, Mg I      & 0.13 \\ 
\hline 
         &        &       &         &   103   &      31 &                    &                &                     &       $<$0.15$>$
&                  & $<$0.17$>$     \\ 
\hline
Ce02     &  0     &  1    &       2 &     1   &       1 &           0.00     &           0.20 &            0.00     &           0.21
&    LTE, Mg I     & 0.21 \\        
\hline 
         &        &       &         &   104   &      32 &                    &                &                     &
&                  &      \\ 
         &        &       &         &   $-$16   &         &                    &                &                     &  
&                  &      \\
Sum      &        &       &         &    88   &         &                    &                &                     &  
&                  &      \\ 
\hline 
FK99     &  0     &  1    &       1 &      1  &       0 &         $-$2.55      &           0.06 &           $+$0.60     &          0.09
&    LTE, Mg I     & 0.09  \\ 
H02      &  0     &  1    &       0 &      1  &       0 &         $-$1.02      &           0.10 &           $+$0.32     &          0.10 
& LTE, Mg I, Mg II & 0.10 \\ 
Ae01     &  0     &  1    &       0 &      3  &       0 &       $-$0.56,$+$0.20 &           0.16 &        $+$0.02,$+$0.20 &          0.16 
& LTE, Mg I, Mg II & 0.16 \\ 
Ae06     &  0     &  1    &       0 &      1  &       0 &         $-$0.74      &           0.14 &           $+$0.28     &          0.14 
&   LTE, Mg II     & 0.14 \\ 
Ce09     &  0     &  1    &       0 &      3  &       0 &       $-$2.59,$-$1.73 &           0.10 &        $+$0.27,$+$0.47 &          0.14 
&   LTE, Mg I      & 0.14 \\ 
\hline  
Sum      &        &       &         &      9  &         &                    &                &                     &    $<$0.14$>$ 
&                  & $<$0.14$>$ \\ 
\hline 
Total    &        &       &         &     97  &         &                    &                &                     & 
&                  &      \\ 
\hline 
\hline 
\end{tabular} 
\end{table*}

Figure 1 shows the comparison of [Mg/Fe] for the stars in common between BM05 and other HR works.
As can be seen, the relations are usually well described by an offset or a linear transformation.
We derive these linear transformations using a 3-$\sigma$ clipping least-square ($lsq$) method
minimizing the distance in both axis (as the uncertainties in different studies are of comparable order).
\begin{equation} 
[{\rm Mg/Fe}]_{\rm work} = A + B [{\rm Mg/Fe}]_{\rm BM05}
\label{Eq1}
\end{equation}
where $[{\rm Mg/Fe}]_{\rm work}$ represents the values computed in those HR works different from BM05.
In BM05 and Cenarro {\it et al.} (2007), the comparison sample was gradually increased
as each set of stellar parameters, like [Fe/H], was calibrated to a uniform scale.
We, however, calibrate the [Mg/Fe] values separately for each work,
basically because the comparison sample adopted here is large enough (218 stars from the BM05 compilation)
and because this avoids the error propagation through the transformations.
The calibrated values are obtained, then, inverting the Eq. 1 as following:
\begin{equation} 
[{\rm Mg/Fe}]_{\rm HR} = (-A/B) + (1/B)[{\rm Mg/Fe}]_{\rm work} 
\label{Eq2}
\end{equation}
The transformation was performed only when $A$ and $B$ were significantly different from 0 and 1 respectively,
based on the student t-test with a 95\% confidence level.

\begin{figure*} 
\begin{center} 
\includegraphics[width=73mm]{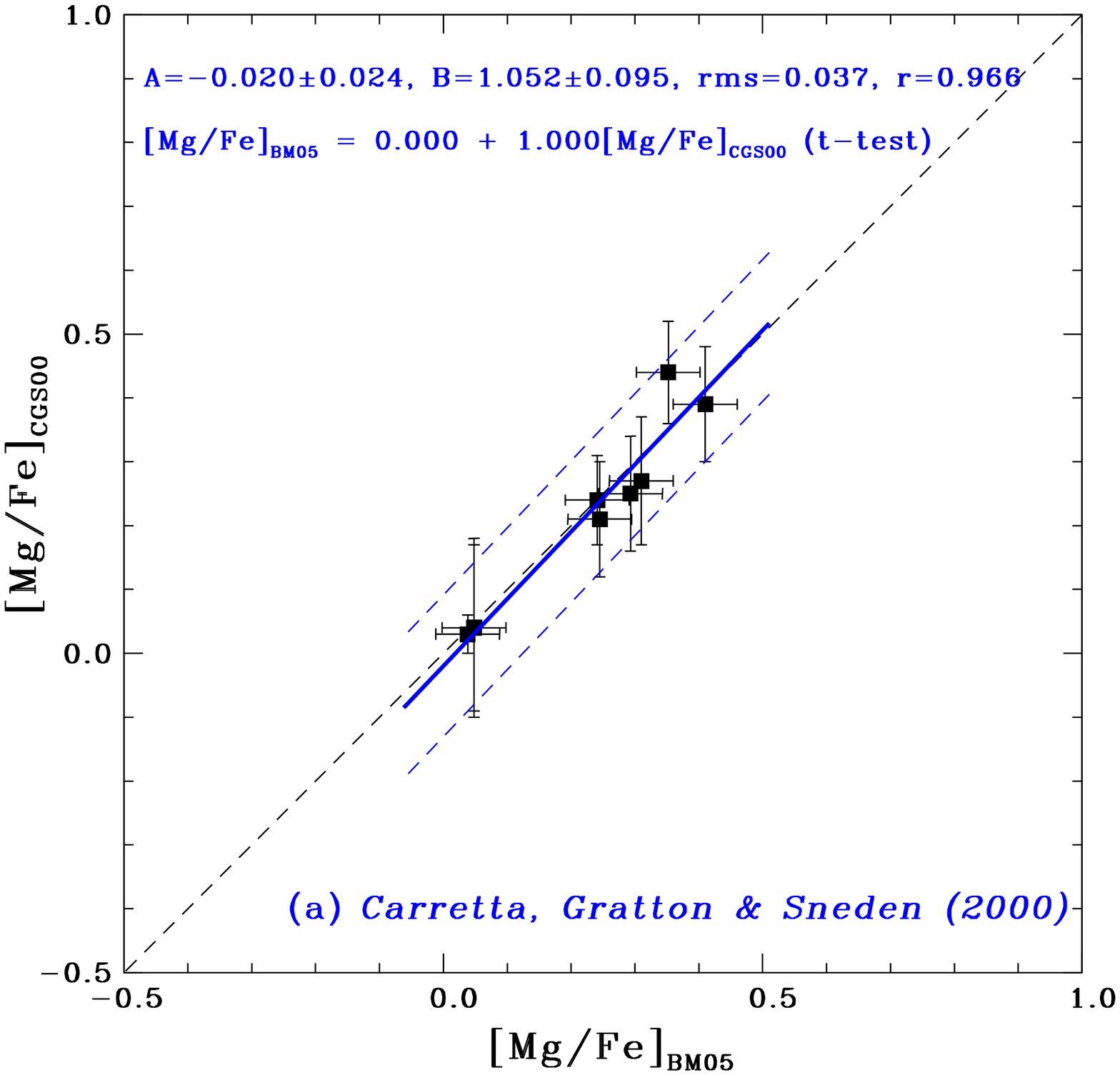} 
\includegraphics[width=73mm]{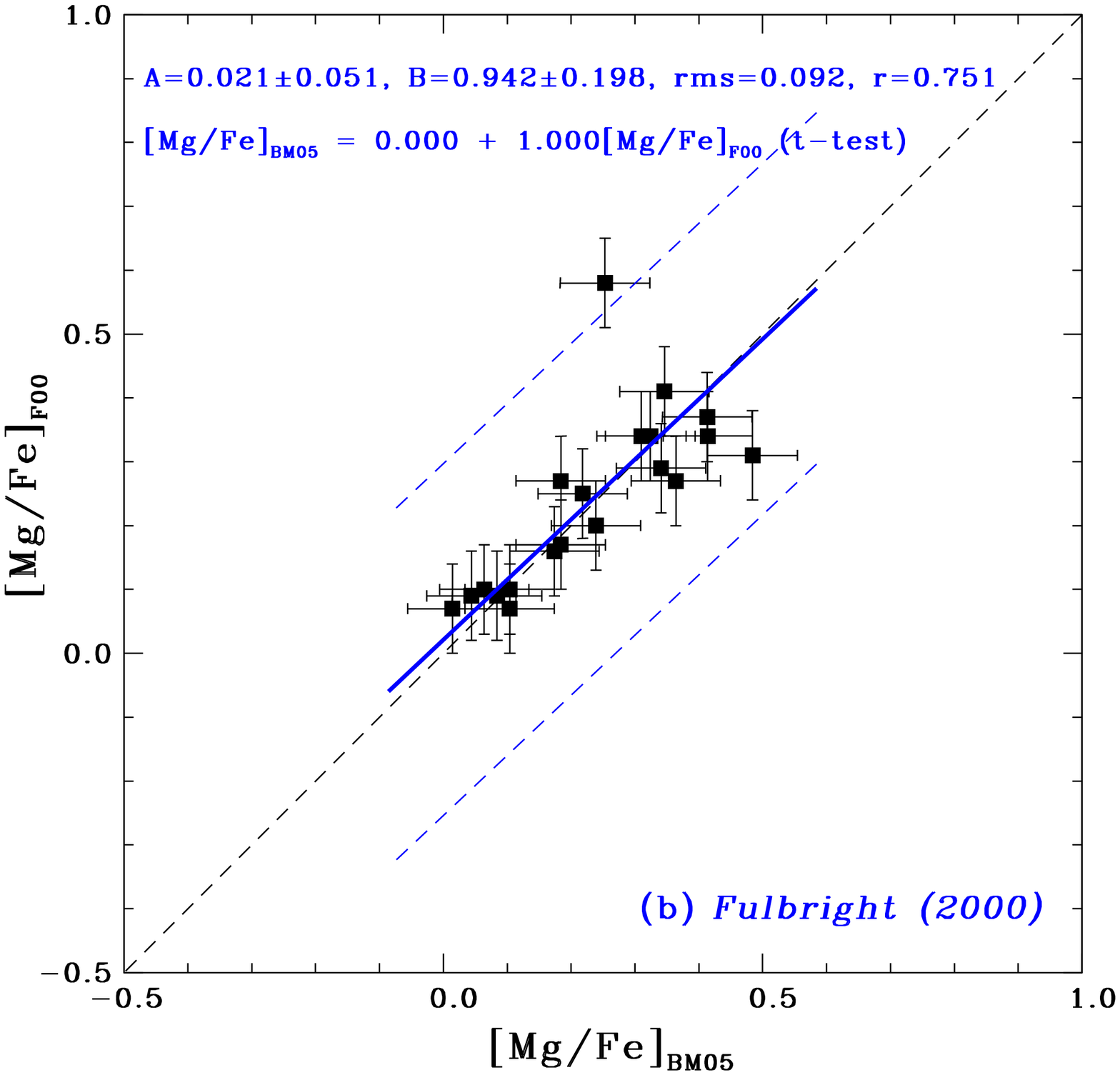} 
\includegraphics[width=73mm]{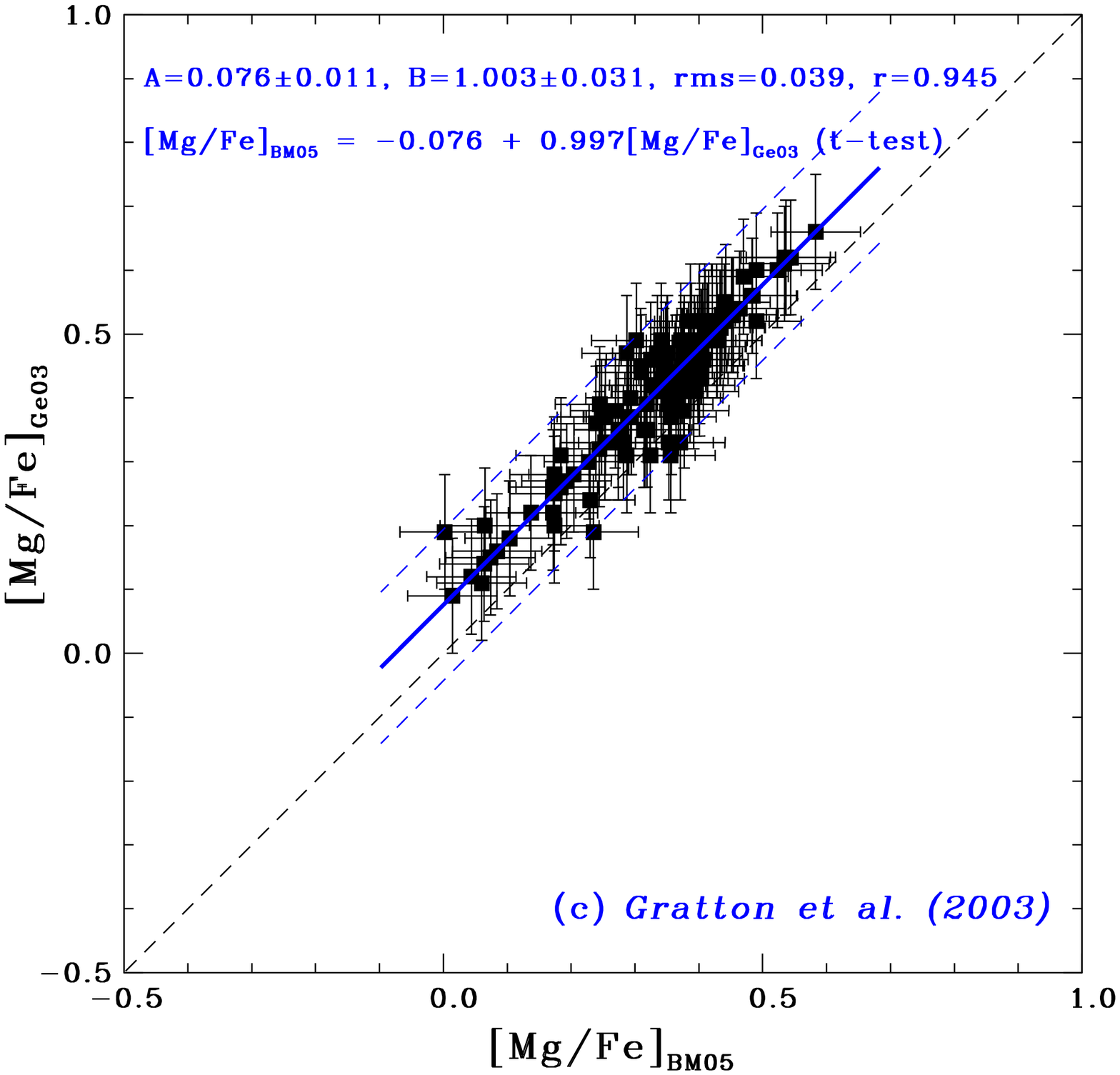} 
\includegraphics[width=73mm]{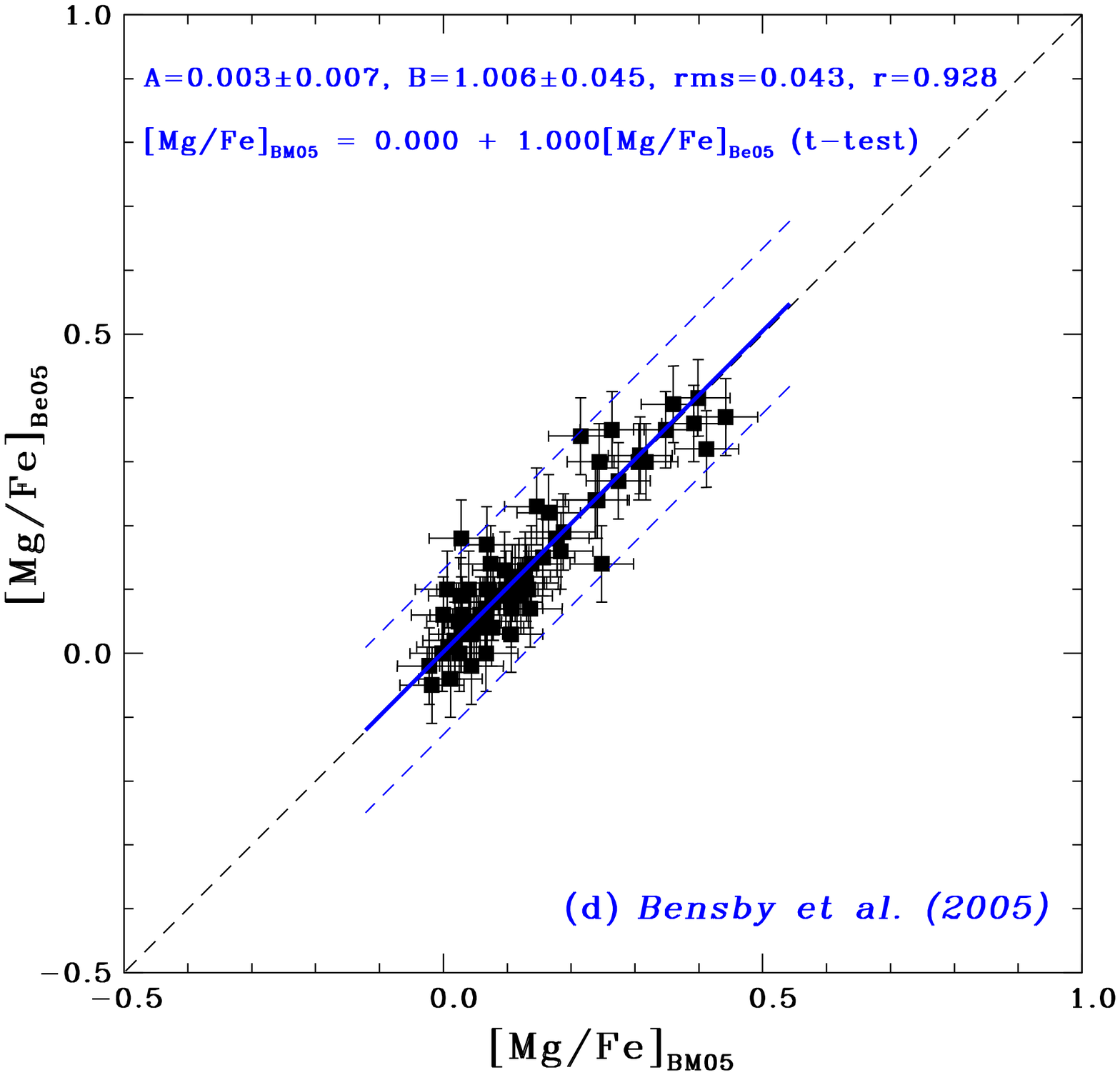} 
\includegraphics[width=73mm]{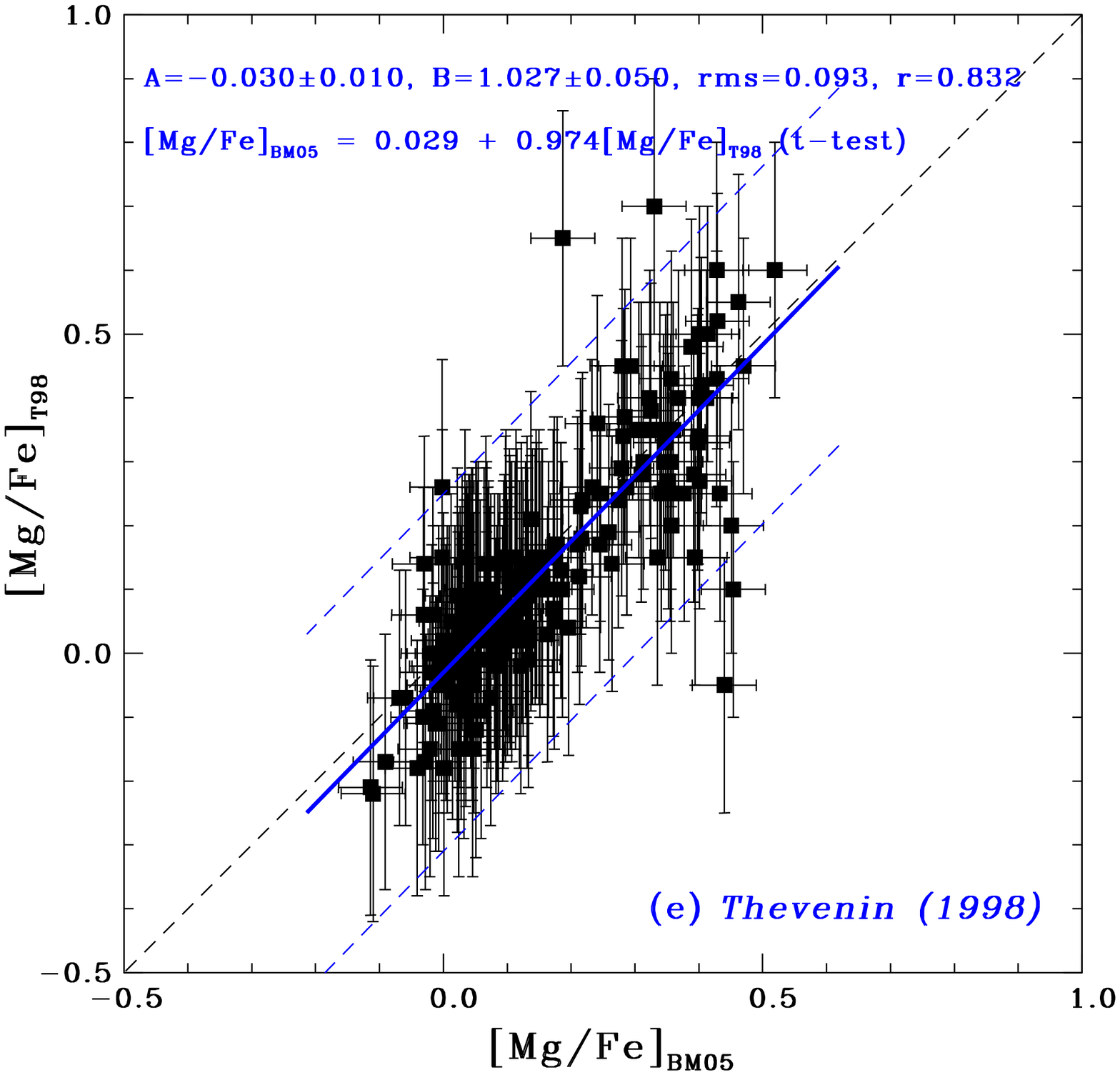} 
\includegraphics[width=73mm]{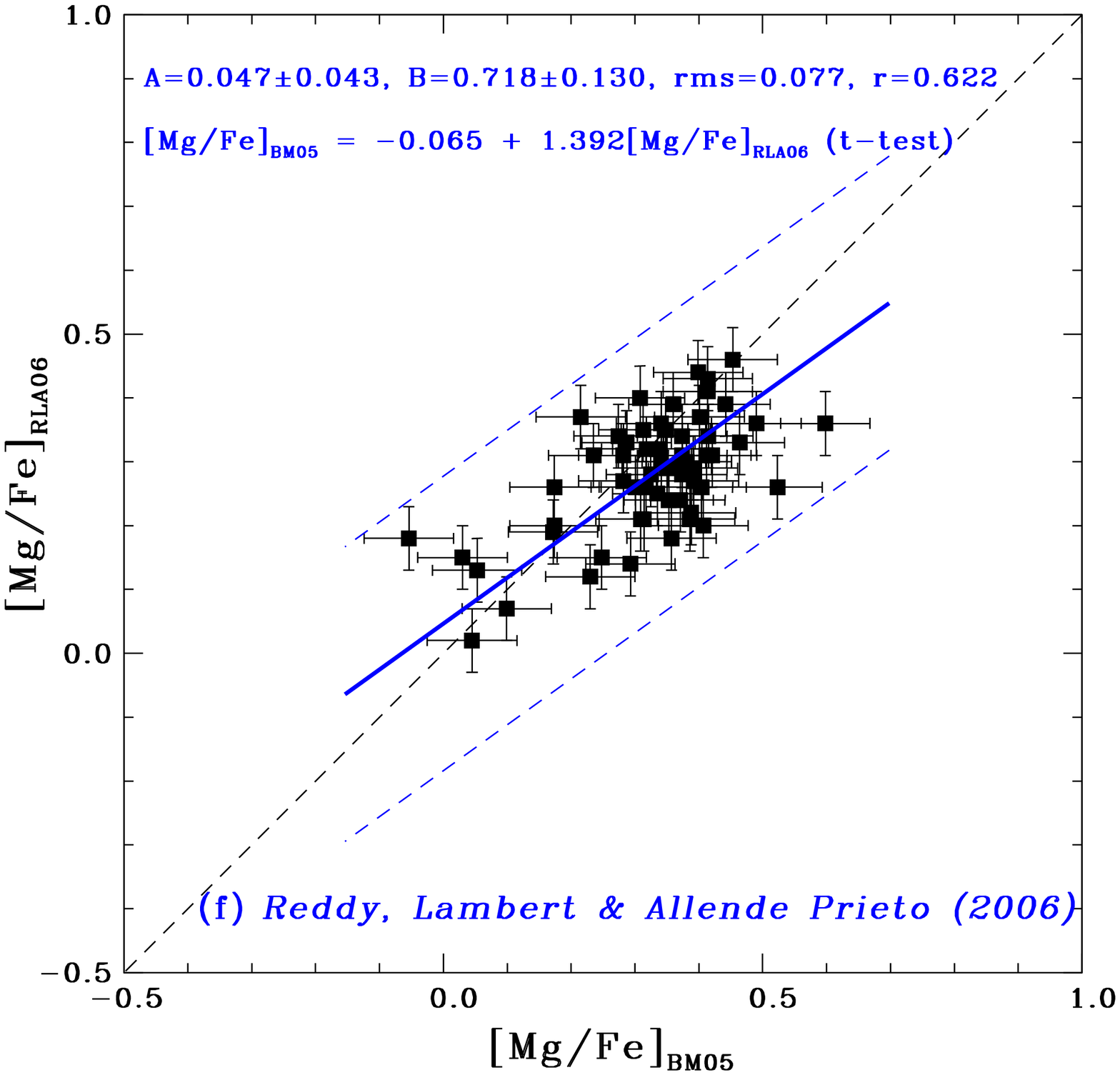}
\end{center} 
\caption{ 
[Mg/Fe]$_{\rm work}$ as a function of [Mg/Fe]$_{\rm BM05}$: 
9 panels ({\bf a} to {\bf i}) showing comparisons between different samples 
with the reference set from Borkova \& Marsakov (2005). 
The statistically representative, 3-$\sigma$-clipped (illustrated by parallel dashed blue lines),
linear $lsq$ fittings are presented by the thick blue lines
(clipped data is represented by red symbols, occurring in the last panel only).
The constants $A$ and $B$ and parameters $rms$ and $r$ (correlation coefficient) of the linear fits (Eq. 1)
are listed at the top of each panel.
The calibration expressions of [Mg/Fe]$_{\rm work}$ to [Mg/Fe]$_{\rm BM05}$ (Eq. 2) are also shown
after applying the 95\% t-test.
The work designation is cited at the bottom of each panel.
} 
\label{Fig1} 
\end{figure*}

Table 1 presents the list of works and the number of stars in common between each sample and the BM05 catalogue,
as well as the number of MILES stars to be included into our catalogue from each HR study
and the number of MILES stars of each work duplicated in other source(s), whose total is sixteen.
The calibration constants and the typical work uncertainties of [Fe/H] and [Mg/Fe] are shown in this table.
We also show the [Fe/H] and [Mg/Fe] ranges of the stars included from these works.
Comments about the LTE assumptions and spectral lines employed in each work are given in the table too.
Statistically reliable linear calibrations of [Mg/Fe] were applied for the first nine works listed  in Table 1.
The data from
Carretta, Gratton \& Sneden (2000),
Fulbright (2000),
Bensby {\it et al.} (2005),
Erspamer \& North (2003), and
Feltzing \& Gustafsson (1998)
were not altered because there is no detectable difference
between the [Mg/Fe] scales of these works and the BM05 scale.
No calibration was applied, either, to the data from the 6 last entries of Table 1:
Caliskan {\it et al.} (2002)
and Fulbright \& Kraft (1999),
because there are very few stars in common with the BM05 sample; 
Heiter (2002)
and Adelman {\it et al.} (2001, 2006)
because there is no star in common; and
Cenarro {\it et al.} (2009)
because their data are already on the same system we have adopted too
(see Appendix A).

Sixteen stars have [Mg/Fe] values from duplicated sources.
When the difference between distinct sources is larger than 4$\sigma$,
then the values with smaller uncertainties were adopted (3 cases only).
The final abundance ratios
for the 13 remaining duplicated cases were computed
as simple averages after separately calibrating to the BM05 scale.
These stars were also used to evaluate the calibration process as a whole and helped us 
to compare the uncertainties of [Mg/Fe] when there are duplicated or single data sources (see Sect. 2.3).

Summarizing, the [Mg/Fe] catalogue of MILES contains 315 stars with HR measurements
(see Table 1):
\begin{enumerate} 
\item 218 stars with [Mg/Fe] collected directly from BM05, 
\item 91 stars with [Mg/Fe] obtained from other published works and calibrated to the same single uniform scale, and 
\item 6 stars whose [Mg/Fe] ratios were collected from other works and inserted into it without any transformation. 
\end{enumerate}

\subsection[2.3]{Precision of calibrated [Mg/Fe]} 
 
The uncertainties of calibrated [Mg/Fe] to the BM05 scale were estimated through the propagation of
their original errors taking also into account the precision of calibration parameters
(see Table 1 and plots of Fig. 1).
The error propagation through the linear expressions of calibration process was based on adding variances.
The [Mg/Fe] uncertainties can be summarized as: 
\begin{enumerate} 
\item 0.07 and 0.05 dex for the 218 stars that define the base uniform scale, for 
respectively, [Fe/H] $\leq$ $-$1.0 dex (47 stars) and [Fe/H] $>$ $-$1.0 dex (171 stars) 
(as described in BM05); 
\item 0.17 dex for 75 stars (weighted average)
whose abundance ratios were calibrated from single values;
\item 0.10 dex for 13 stars based on average calibrated abundance ratios (data from duplicated sources);
\item 0.14 dex for 3 stars from Ce09; and
\item 0.13 dex for 6 other stars whose [Mg/Fe] were not transformed onto the BM05 scale.
\end{enumerate}

\begin{figure} 
\begin{center} 
\includegraphics[width=73mm]{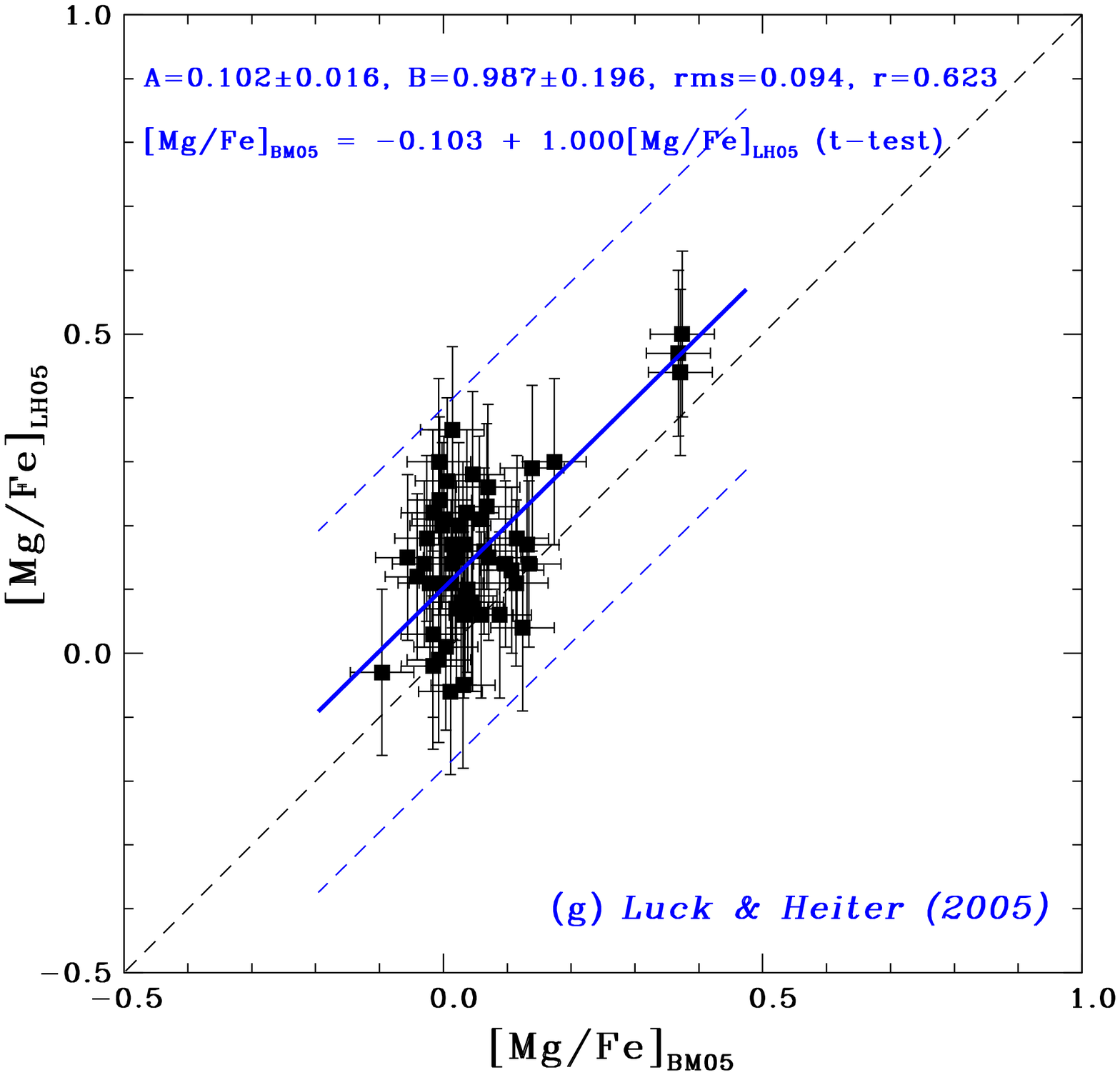} 
\includegraphics[width=73mm]{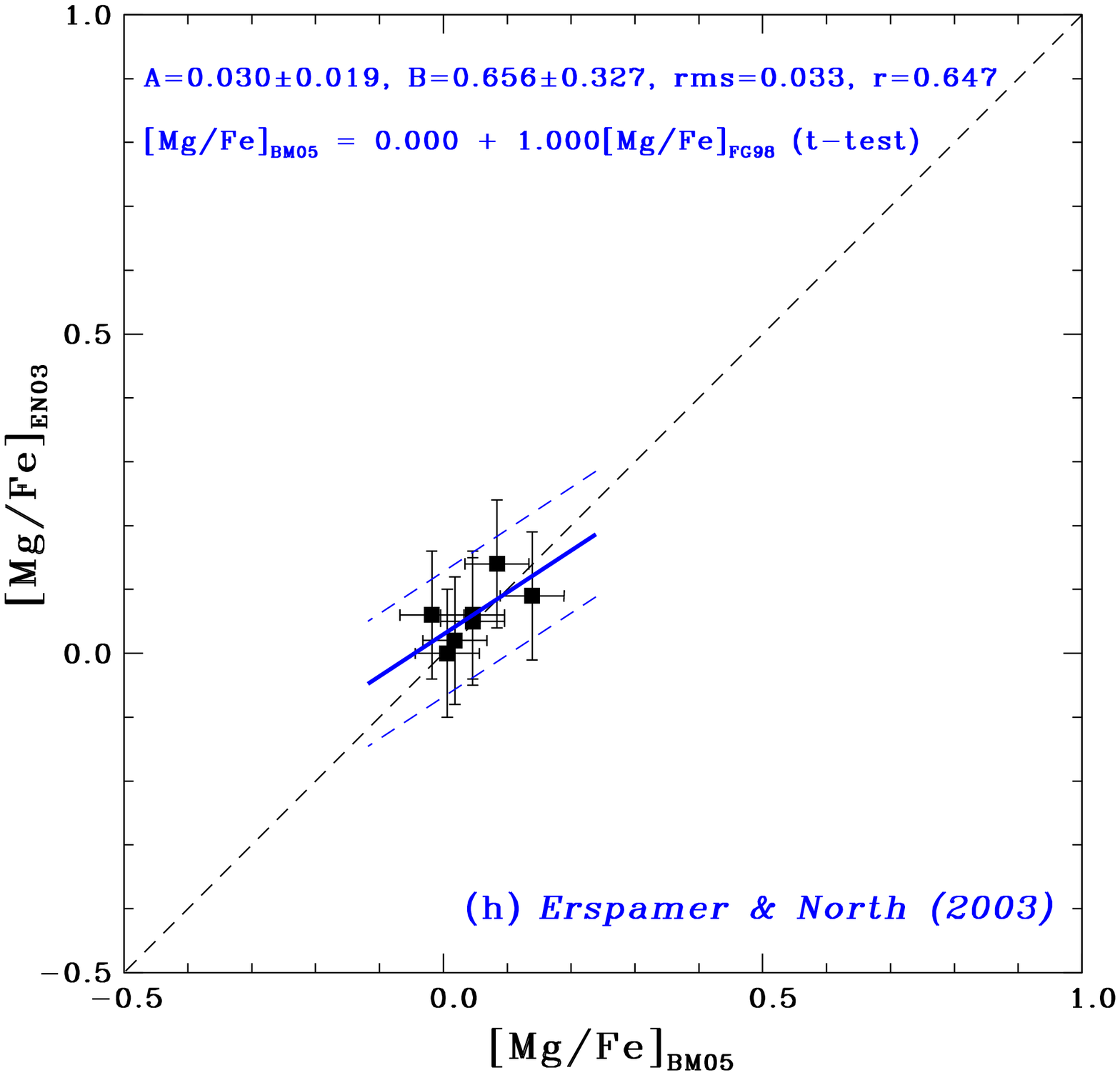} 
\includegraphics[width=73mm]{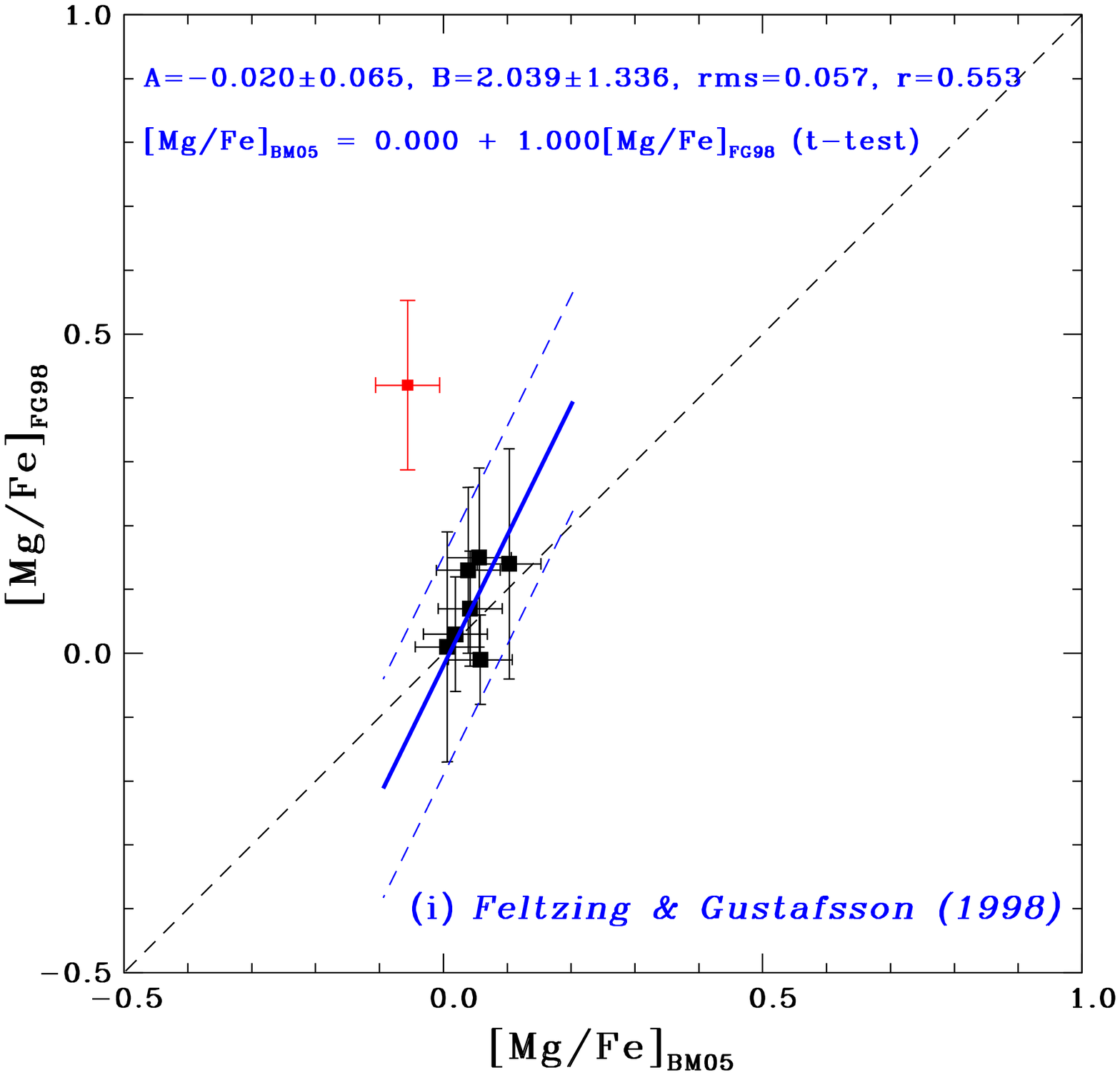} 
\end{center} 
\contcaption{} 
\label{Fig1cont} 
\end{figure}

The averaged [Mg/Fe] values compiled from duplicates are statistically more precise (1$\sigma$ = 0.10 dex)
than the calibrated abundance ratios that have been computed from single sources (1$\sigma$ = 0.17 dex).
The final weighted average uncertainty of [Mg/Fe]
is around 0.09 dex over all compiled HR data (315 stars). 
It is 0.10 dex for 88 metal-poor stars ([Fe/H] $\leq$ $-$1.0 dex) 
and 0.08 dex for 227 metal-rich ones ([Fe/H] $>$ $-$1.0 dex). 
The mean uncertainty of [Mg/H] over whole range of metallicity is  
0.13 dex, estimated by the quadratic sum of $\sigma$[Fe/H] and $\sigma$[Mg/Fe].

\begin{figure*} 
\begin{center} 
\includegraphics[width=120mm, angle=-90]{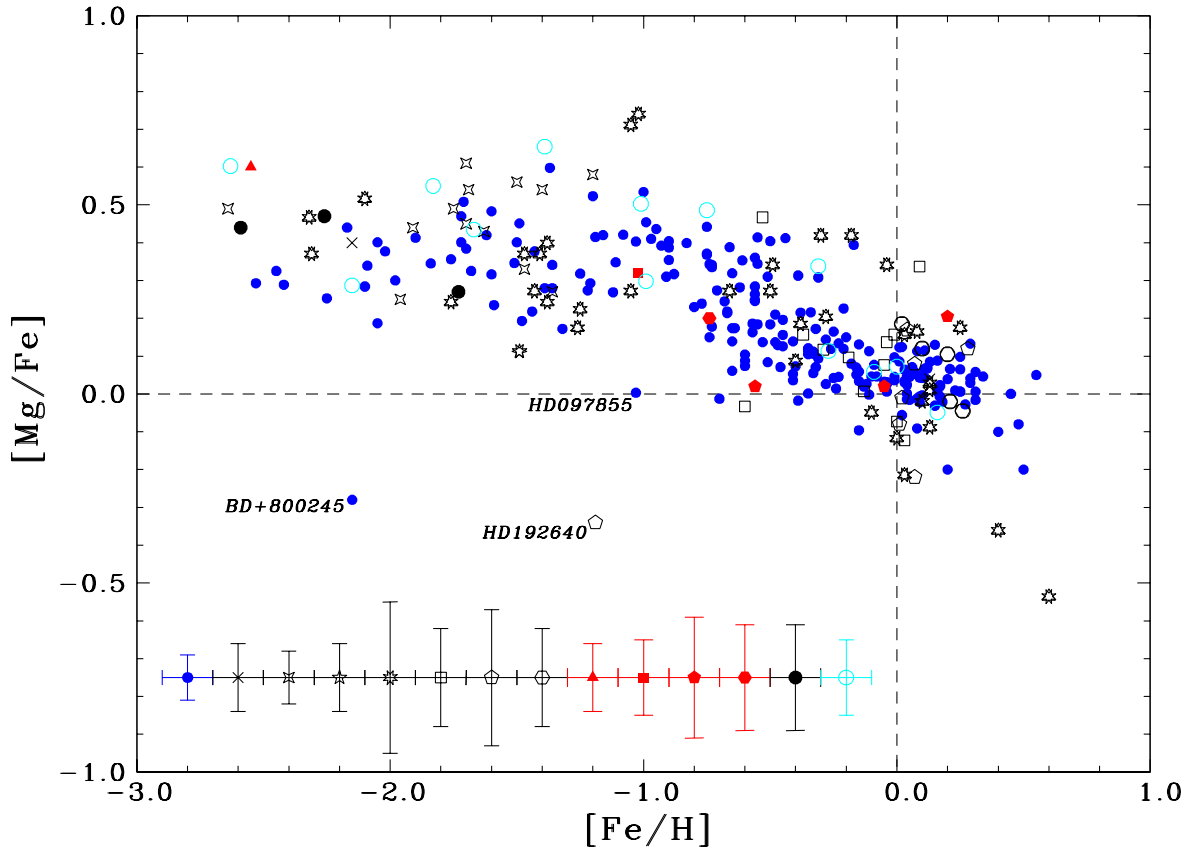} 
\end{center} 
\caption{ 
[Mg/Fe] vs. [Fe/H] for the HR data of MILES [Mg/Fe] catalogue (summing 315 stars).
[Fe/H] is on the MILES scale (Cenarro {\it et al.} 2007). 
The sources are represented by different symbols: 
BM05 catalogue as filled blue circles (218 stars), 
CGS00 as diagonal crosses (2 objects), 
F00 as four-pointed stars (13 objects), 
Ge03 as five-pointed stars (1 object),  
T98 as eight-pointed stars (35 objects), 
LH05 as open squares (12 objects), 
EN03 as open pentagons (7 objects), 
FG98 as open hexagons (5 objects), 
FK99 as filled red triangles (1 object),
H02 as filled red squares (1 object),
Ae01 as filled red pentagons (3 objects),
Ae06 as filled red hexagons (1 object), and
Ce09 as filled black circles (3 objects).
The data for 13 stars, whose sources are duplicated, are represented
by open cyan circles.
The red symbols represent stars whose [Mg/Fe] were not calibrated to the uniform scale (6 cases).
The designations of works are cited in the caption of Table 1.
Three chemically peculiar stars are identified by their names.
}
\label{Fig2} 
\end{figure*}

\subsection[2.4]{Extended control sample for calibrating [Mg/Fe]} 

In total, we compiled [Mg/Fe] for 315 MILES stars covering about 1/3 of the library 
(263 dwarfs and 52 giants, respectively 49\% and 12\% of them),
which was used to define an extensive control sample for calibrating our own Mg abundance measurements at mid-resolution (Sect. 3).
Dwarfs as designated when log $g$ $\geq$ 3.0 whilst giants
when log $g$ $<$ 3.0, as in BM05 and the MILES database itself.

Figure 2 shows [Mg/Fe] versus [Fe/H]
for the HR part of the MILES [Mg/Fe] catalogue
and its parametric coverage is shown in Table 6 (Sect. 4).
Specifically, the control sample owns 306 stars covering nearly 31\% of MILES (255 dwarfs, and 51 giants)
and practically presenting the same coverage.
Besides six stars whose [Mg/Fe] has not been transformed onto the reference scale,
three chemically peculiar metal-poor stars (BD+800245 confirmed by
Ivans {\it et al.} 2003,
HD097855, and HD192640)
have also been excluded from the control sample because their [Mg/Fe] lay far from the 
[Mg/Fe] vs. [Fe/H] trend described by the stars in the solar neighbourhood (see Fig. 2).
Recently,
Nissen \& Schuster (2010)
have classified a kind of galactic objects named low-$\alpha$(-enhancement) stars
distributed over two distinct nearby halo populations based on their kinematics,
whose origins might due to the accretion from dwarf galaxies
and some of them from the Omega Centauri globular cluster (denominated as a progenitor galaxy).
HD097855 might belong to the proposed group.
All of those nine stars have been incorporated into our catalogue.

\section[3]{Magnesium abundances measured at mid-resolution} 

To extend the magnesium abundance characterization of the MILES stars,
a spectroscopic analysis based on a LTE spectral synthesis of Mg features
was carried out at mid-resolution.

Recovering element abundances in stellar photospheres at medium spectral resolution
(resolving power between 1,000 and 10,000)
has been a well-established and alternative approach for many decades.
For instance,
Pagel (1970) 
reported that low-resolution spectroscopic analysis was one way to find the metallicity of nearby stars,
by calibrating the results with a reference sample
which, indeed, has been highlighted as an important step involved
(Friel \& Janes 1993;
Kirby {\it et al.} 2009;
Marsteller {\it et al.} 2009).

Our present Mg abundances reported in the current section of the paper are based on the MILES spectra,
which have a resolution comparable to many previous studies
(e.g. Chavez, Malagnini \& Morossi 1995;
Terndrup, Sadler \& Rich 1995;
Cook {\it et al.} 2007).

\subsection[3.1]{Computation of the synthetic spectra}
 
Our analysis is based on a LTE spectral synthesis computed with the most recent stable version of the MOOG code
at the time of developing this work
(Sneden 2002).
The synthesis code was fed by linearly interpolated model atmospheres 
over the MARCS 2008 grid 
(Gustafsson {\it et al.} 2008), 
up-to-date atomic line lists from the VALD database 
(Vienna Atomic Line Database, 
Kupka {\it et al.} 2000, 
Kupka {\it et al.} 1999, 
Ryabchikova {\it et al.} 1997, and 
Piskunov {\it et al.} 1995) 
and a set of important molecular lines of C$_2$, CN and MgH 
(Kurucz 1995) 
in order to compute a series of model spectra for each MILES star ranging over five values of [$\alpha$/Fe]: 
$-$0.60, $-$0.30, 0.00, $+$0.30, $+$0.60 dex.

The abundance of all $\alpha$-particle-capture elements -- O, Ne, Mg, Si, S, Ar, Ca and Ti --
were equally and simultaneously modified to represent a global variation of their chemistry in a stellar photosphere.
Although these elements may have distinct individual abundance ratios relative to iron,
assuming homogeneity for them should be a reliable approximation among many computations
such as molecular dissociative equilibrium and partial pressure of different species,
especially when the element abundances are not individually and previously known in each star.
Moreover, changing globally these ratios for all $\alpha$-elements
is coherent with the model atmospheres adopted in our spectral synthesis.
We tested the alternative approach at medium spectral resolution
(i.e. varying the Mg abundance only, keeping the abundances of other $\alpha$-elements fixed) 
for some MILES stars and the effect was negligible.
Even adopting the homogeneous global trend for the $\alpha$-elements,
their individual abundances can be independently measured by separately analysing their own absorption features.
For instance, the Mg abundance is only consistently quantified as one or more magnesium lines are analysed
instead of computing and measuring features of other $\alpha$-elements.

We selected model atmospheres whose chemistry follows the general pattern of solar neighbourhood
for the $\alpha$-elements (the standard composition group of the MARCS 2008 models, Gustafsson {\it et al.} 2008),
i.e. [$\alpha$/Fe] = $+$0.40 dex when [Fe/H] $\leq$ $-$1.00 dex
and [$\alpha$/Fe] = 0.00 dex for [Fe/H] $\geq$ 0.00 dex
with intermediate variable values between these fiducial metallicities
(+0.30, +0.20 and +0.10 dex for [Fe/H] = $-$0.75, $-$0.50 and $-$0.25 dex respectively).
On this standard range of compositions, every other element (X) follows the iron abundance, [X/H] = [Fe/H] meaning [X/Fe] = 0 dex.  
A MARCS model presents a stratification over 56 plane-parallel layers for its photosphere.
In the case of giants, the models (with $-$0.5 $\leq$ log $g$ $\leq$ 3.5 in the MARCS grid)
have been originally generated under a three-dimensional geometry for different masses
and afterwards transformed onto one-dimensional representations by their own developers.
We chose 3-d models with one solar mass for giants (log $g$ $<$ 3.0 in the current work).
All chosen models for dwarfs (with 3.0 $\leq$ log $g$ $\leq$ 5.5 in the grid)
and giants have a micro-turbulence velocity of 2.0 km s$^{\rm -1}$.
The MARCS 2008 grid ranges are: 
\begin{enumerate} 
\item 2500 $\leq$ T$_{\rm eff}$ $\leq$ 8000 K with steps of 100 K for T$_{\rm eff}$ $<$ 4000 K 
and 250 K for T$_{\rm eff}$ $\geq$ 4000 K; 
\item $-$1.0 $\leq$  log $g$ $\leq$ 5.0 (or 5.5 in some cases) with 0.5 constant steps; and 
\item {[}Fe/H{]} = $-$5.00, $-$4.00, $-$3.00, $-$2.00, $-$1.50, $-$1.00, $-$0.75, $-$0.50, $-$0.25,
0.00, $+$0.25, $+$0.50, $+$0.75 and $+$1.00 dex. 
\end{enumerate}

Linear interpolations of the model atmospheres were automatically done
by using the user-friendly software of 
Masseron (2008),
which is publicly available from the MARCS models' web-site
(http://marcs.astro.uu.se/).
The layers of each input model to the MOOG code were represented by
the optical depth at 5000 {\AA}, 
the thermodynamic equilibrium temperature,
the total gas pressure,
and the electron numeric density
(all quantities in cgs units).
The micro-turbulence velocity was fixed at 2.0 km s$^{\rm -1}$ for all layers of each model.
The effective broadening is dominated by the spectroscopic instrumentation
and is suitably represented by a Gaussian convolution. 

Our spectral synthesis computations adopt the same solar abundance pattern
used to build up the MARCS 2008 model atmospheres
(Grevesse, Asplund \& Sauval 2007, hereafter GAS07),  
aiming at an important internal consistency regarding the reference solar chemistry.
The solar abundances in GAS07 of those $\alpha$-elements on a logarithm scale are:
log(O) = 8.66,
log(Ne) = 7.84, 
log(Mg) = 7.53, 
log(Si) = 7.51, 
log(S) = 7.14, 
log(Ar) = 6.18, 
log(Ca) = 6.31, and 
log(Ti) = 4.90.
The abundance of iron is given by log(Fe) = 7.45 on this scale. 

Many authors re-scale the oscillator strengths ($gf$) of atomic lines to reproduce the solar spectrum,
i.e. calibrate the line-strengths to the Sun's photospheric conditions.
However, we preferred to maintain their laboratory $gf$ values from VALD instead of normalizing them to the solar scale
because we intend to calibrate our MR measurements using our HR control sample.
The theoretical spectra were calculated at 0.02 {\AA} wavelength-steps
by assuming an opacity contribution for the stellar continuum at 0.50 {\AA} bins.
For the often strong Mg b lines (see Sect. 3.2 and Table 2),
we adopted the Uns\"{o}ld approximation for the interaction constant (C$_{\rm 6}$)
of the van der Waals dump parameter $\gamma_{\rm 6}$ multiplied by a 6.3 factor.
This dumping represents collisions among neutral atomic species mainly H I and He I as in cold photospheric layers.
On the other hand, the same constant multiplied by a MOOG internal factor was chosen for the other Mg feature.
The Uns\"{o}ld approximation (default in MOOG) means
that C$_{\rm 6}$ is basically due to the H I atoms as a function of the excitation potential of an electronic transition,
but $\gamma_{\rm 6}$ remains also dependent on the local gas pressure and temperature.

All our model spectra were computed at the MILES resolution,
which is slightly different in each Mg feature region
(FWHM = $\Delta\lambda$ =  2.40 {\AA} for Mg5183 and 2.35 {\AA} for Mg5528), 
by applying Gaussian smoothing to represent the (instrumental dominated) broadening.
At this resolution, no additional stellar rotational broadening needs to be considered
unless the line-of-sight rotational velocity v$_{\rm rot}$sin($i$) is greater than $\sim$130 km s$^{\rm -1}$.
The wavelength scales of all MILES observed spectra were carefully shifted to the rest wavelength
to match the theoretical model scales, i.e. wavelengths in air as available in the VALD database
for $\lambda$ from 2000 {\AA} up to the infrared. 
Spectral cross-correlations were applied for this purpose
through the cross-correlation $fxcor$ task
of the Radial Velocity Analysis Package of the NOAO Optical Astronomy Packages
of IRAF (Image Reduction and Analysis Facility)\footnote{
IRAF is distributed by the National Optical Astronomy Observatories,
which are operated by the Association of Universities for Research in Astronomy, Inc.,
under cooperative agreement with the National Science Foundation, USA.}
by adopting a given correspondent model spectrum for each star as a template.
The model spectra computed for either [Mg/Fe] = +0.60 dex or +0.30 dex were chosen in order to guarantee reliable
spectroscopic cross-correlations for all MILES spectra, from the hotest metal-poor to the coldest metal-rich ones,
i.e. with strong absorption lines in the template spectra.
Then the wavelength scales of all computed spectra were re-binned
to exactly agree with the sampling of the MILES spectra at each observed wavelength bin (0.90 \AA).
Finally, flux normalizations of the empirical spectra based on the local pseudo-continuum
were carefully applied at the regions around each Mg feature
in order to adequately match to the flux normalized scale of the theoretical spectra.
Fiorentin {\it et al.} (2007)
emphasized the importance of performing reliable comparisons on compatible flux and wavelength scales
between observed MR spectra and synthetic ones,
paying special attention to choosing useful absorption features
and accurately calibrating the abundance results.

\begin{figure*} 
\begin{center} 
\includegraphics[width=75mm]{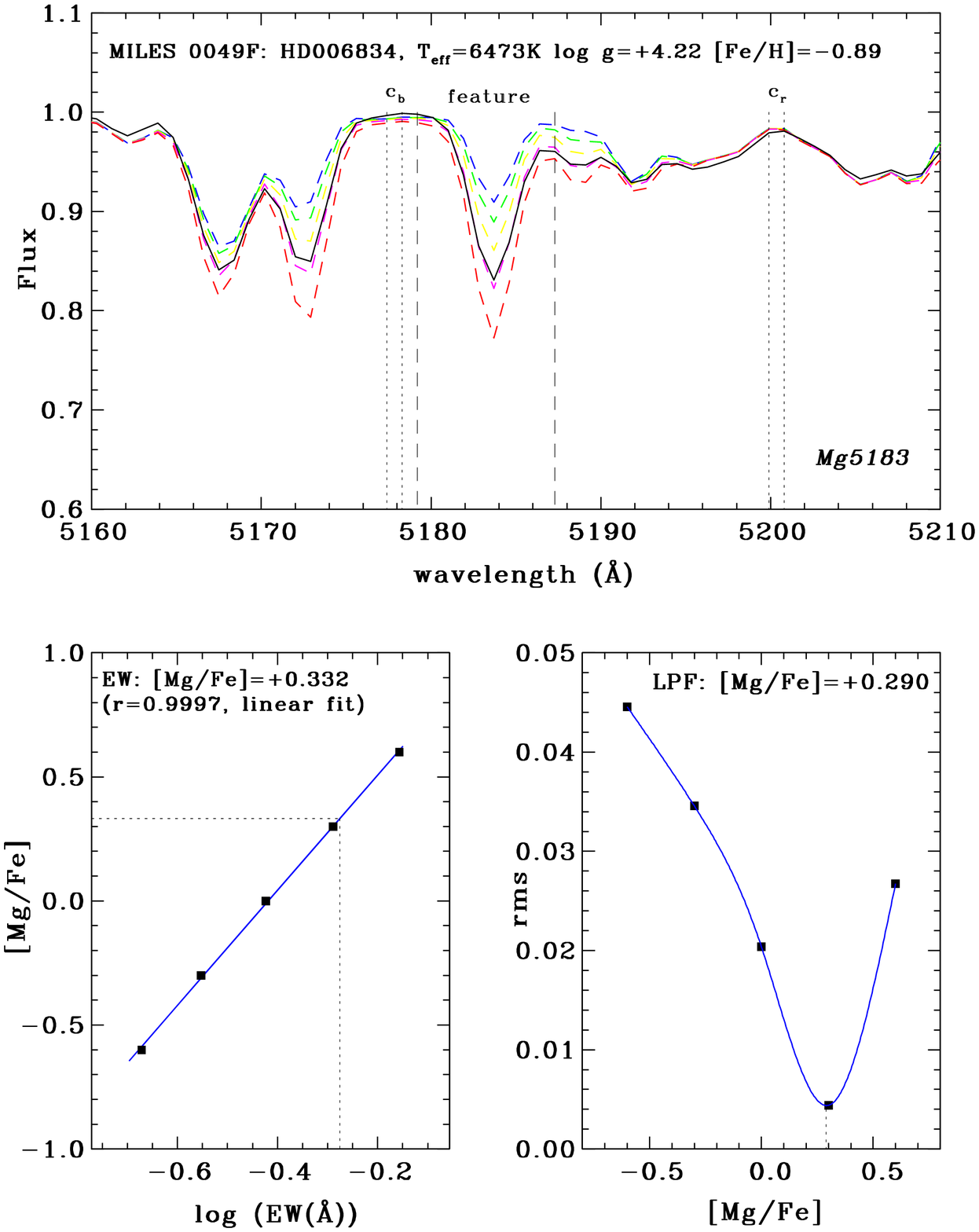}
\includegraphics[width=75mm]{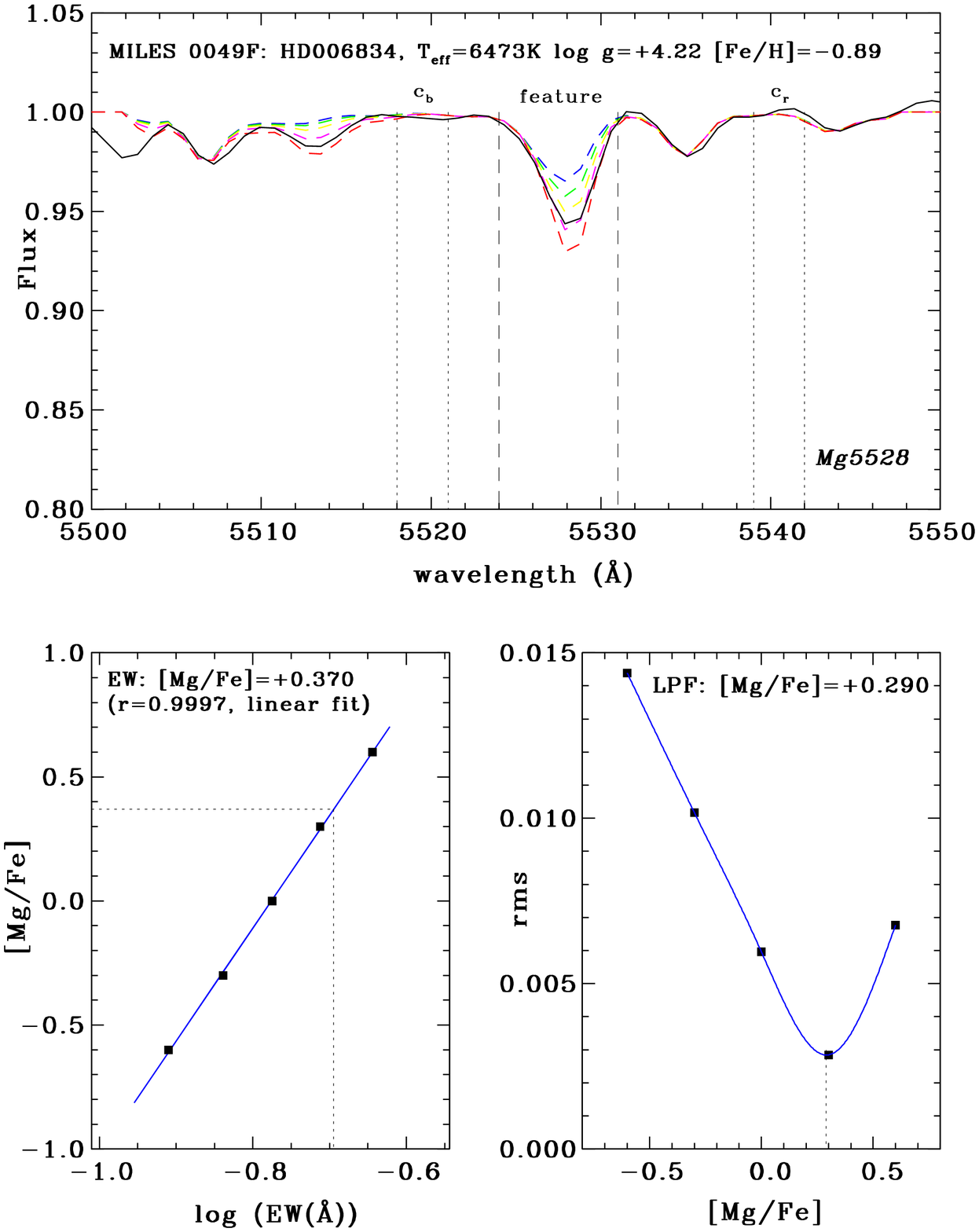} 
\includegraphics[width=75mm]{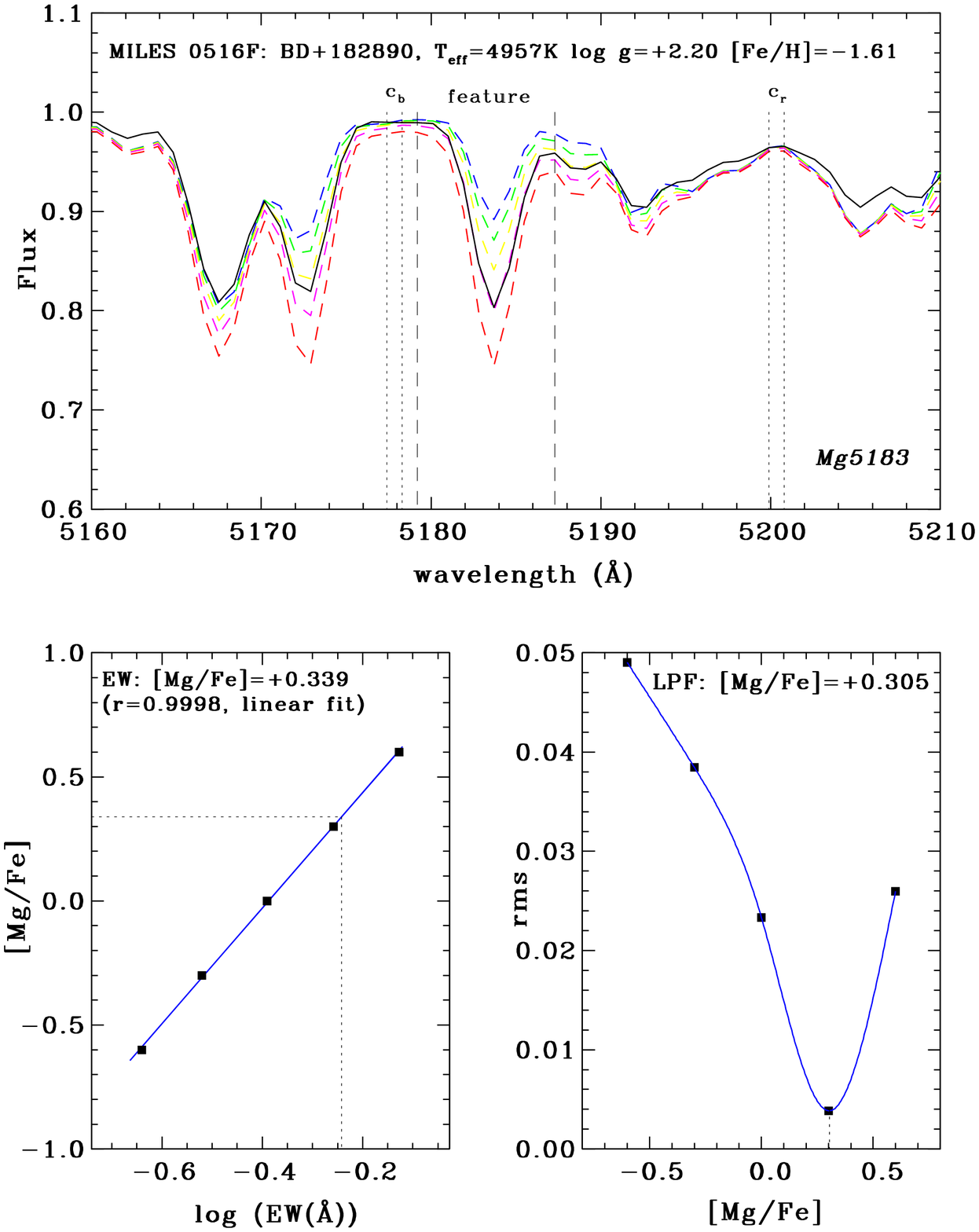} 
\includegraphics[width=75mm]{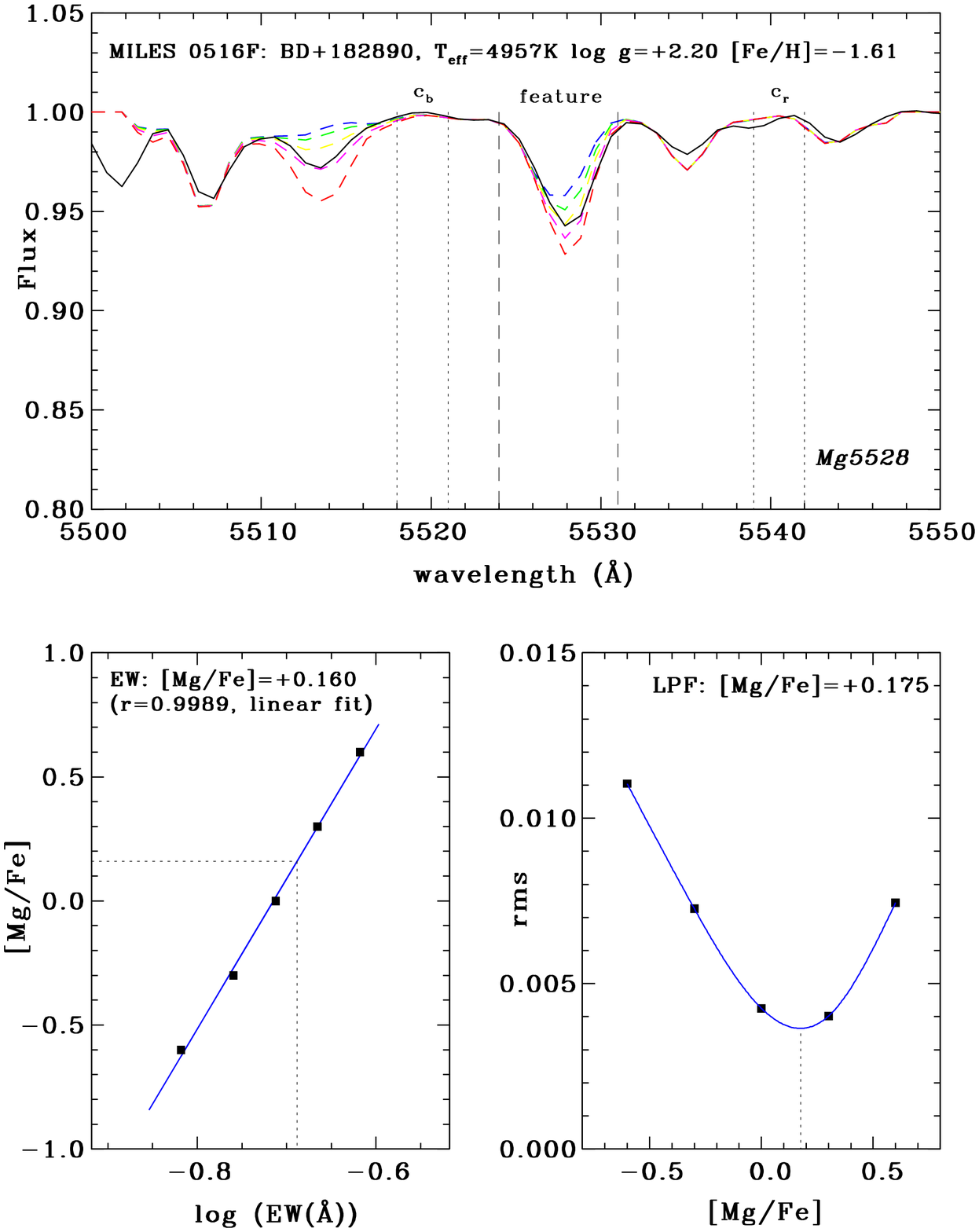} 
\end{center} 
\caption{ 
Examples of spectral synthesis of the Mg5183 and Mg5528 features
for a MILES' dwarf
(HD006834: T$_{\rm eff}$ = 6473 K, log $g$ = 4.22, [Fe/H] = $-$0.89, at the two top sub-figures)
and giant
(BD+182890: T$_{\rm eff}$ = 4957 K, log $g$ = 2.20, [Fe/H] = $-$1.61, at the two bottom sub-figures).
Their central bandpass and two pseudo-continuum windows, c$_{\rm b}$ and c$_{\rm r}$,
are shown in the main panel of each sub-figure.
The observed spectrum is represented by the solid black line 
and the synthetic ones by the colour dashed lines. 
Each set of theoretical spectra was computed for five $\alpha$-enhancements 
([$\alpha$/Fe] = $-$0.60, $-$0.30, 0.00, $+$0.30, $+$0.60 dex, respectively, on colours blue, green, yellow, magenta and red)
assuming the star's photospheric conditions fixed.
Two small graphs illustrate the two abundance determination methods in each sub-figure:
EW (bottom left corner) and LPF (bottom right corner) presenting their [Mg/Fe]$_{\rm feature}^{\rm method}$.
}
\label{Fig3} 
\end{figure*}

\subsection[3.2]{Mg features analysed: Mg5183 and Mg5528} 
 
Two strong Mg features were chosen and carefully tested to be measurable and useful
for recovering magnesium abundances at the MILES spectral resolution with acceptable precision
for generating new stellar population models with [Mg/Fe] constraints,
i.e. with comparable precision to that of [Fe/H] in MILES (0.10 dex).
These features are:
(i) the reddest line of the Mg b triplet ($\lambda$5183.604~{\AA}) named here Mg5183,
which is usually the strongest and the most Mg-sensitive of the three lines,
and 
(ii) MgI$\lambda$5528.405~{\AA}, hereafter Mg5528.

\begin{table} 
\caption{
The magnesium features: their central pass-bands Feature$_{\rm band}$
and two pseudo-continuum windows c$_{\rm b}$ and c$_{\rm r}$,
adopted for defining linear local continua for measuring their pseudo-equivalent widths.
}
\label{Mg_features} 
\begin{center} 
\begin{tabular}{@{}lccc} 
\hline 
\hline 
Feature & c$_{\rm b}$ & Feature$_{\rm band}$ & c$_{\rm r}$ \\ 
\hline 
        &     ({\AA}) &     ({\AA})          &    ({\AA})   \\ 
\hline 
\hline 
Mg5183  & 5177.4-5178.3  &  5179.2-5187.3  &  5199.9-5200.8  \\ 
Mg5528  & 5518.0-5521.0  &  5524.0-5531.0  &  5539.0-5542.0  \\ 
\hline 
\hline 
\end{tabular} 
\end{center} 
\end{table}

\subsection[3.3]{Methods applied: pseudo-equivalent width and line profile fit} 
 
Two methods were chosen to measure the magnesium abundances:
\begin{enumerate}
\item based on pseudo-equivalent widths, hereafter EW
(pseudo-ones in fact due to the extensive line blanketing at the MILES spectrum resolution),
and
\item applying line profile fittings, hereafter LPF. 
\end{enumerate}

These methods are usually adopted on HR and MR analyses
and both require a previous knowledge of the photospheric parameters.
When there are more than 2 lines of an element,
the modelling of their equivalent widths are preferable instead of fitting their profiles.
The fit of line profiles is commonly adopted to extract element abundances
from molecular absorptions where a myriad of lines from a single substance are very close to each other. 
When analysing atomic features, both methods generally need isolated lines,
however, they can work on composite lines that can be de-blended in single profiles
if the abundances of the others absorbers are already known.
Traditionally the EW method is applied in an automatic process to extensive sets of lines.
On the other hand, the second method is employed focusing on careful visual inspection
in a feature-by-feature and star-by-star base.
The abundance precision derived from each method depends on how many features are adopted
and how sensitive each absorption line is to a given elemental abundance variation.

At each chosen Mg feature region, we automatically applied these methods as explained below.

{\bf (i)}
The equivalent width of a weak line is proportional to the number of absorbers of element X (EW $\propto$ $n$(X)), 
whilst the equivalent widths of strong lines are dependent on $n$(X)$^{\rm 1/2}$.
Therefore direct comparisons were made on planes [Mg/Fe] vs. log(EW)
by adopting simple linear $lsq$ fittings for each set of theoretical equivalent widths.
Note that the iron abundance [Fe/H] is fixed in all model computations for each star
(like the other photospheric parameters assuming those compiled by
Cenarro {\it et al.} 2007)
and [Mg/Fe] = [Mg/H] $-$ [Fe/H].
Therefore we adopt that relationship instead of the direct log(EW) vs. log($n$(X)).
Consequently, [Mg/Fe] can be directly measured instead of [Mg/H]
and they were obtained through interpolation of the [Mg/Fe] vs. log(EW) relationship for each star and feature combination.
The equivalent width measurements of both Mg features were performed with
the LECTOR code
(A. Vazdekis' webpage, www.iac.es/galeria/vazdekis/SOFTWARE/)
and INDEXF software
(www.ucm.es/info/Astrof/software/indexf),
however, the uncertainties have been estimated through INDEXF only.
Both software provided the same results within an accuracy of milli-Angstroms.
The measurements were carried out within the central passband of each Mg feature
by adopting a linear local flux continuum that is defined
by the average fluxes and wavelengths of two pseudo-continuum windows held very near to the passband (at each side of the feature).
The central passband and pseudo-continuum windows of both Mg features were carefully chosen
in order to provide representative measurements for their equivalent widths
and are presented in Table 2.
We computed the equivalent width uncertainties as being dominated by photon statistics
based on the formalism from
Cardiel {\it et al.} (1998).
The relative uncertainties are distributed between few percents and 40\%,
whilst the equivalent widths range from 0.15 up to 2.60 {\AA}
and from 0.05 up to 0.95 {\AA} for the Mg5183 and Mg5528 features respectively.
We noted that the EW uncertainties are dependent on the spectral S/N, as expected,
and it is more evident for the Mg5183 feature.
The higher S/N, smaller the EW relative error is.
The S/N has been computed as an average between the blue and red regions of MILES stellar spectra,
and it ranges from 10 up to 550 per {\AA} with typical values around 235 per {\AA}.
It is also noted there is a strong inverse correlation between the relative errors and
the equivalent width values.

{\bf (ii)}
The line profile comparison between the observed Mg feature and each corresponding set of synthesized features
was made within the central passband through $rms$ statistics.
Optimal [Mg/Fe] from each feature were derived for each MILES' star through $rms$ minimization,
as is shown in Fig. 3 for a dwarf and giant spectra.
The pseudo-continuum windows were adopted to accurately match the continuum fluxes of the observed and model spectra.
Careful attention was paid to specific cases
for which the theoretical and observed spectra do not match each other simultaneously at both continuum windows.
Minor corrections were applied to the observed spectrum continuum flux for this purpose based on eye-trained inspections
(additive or multiplicative corrections).
Each curve of $rms$ as a function of [Mg/Fe] was fitted by a spline.

Figure 3 shows examples of spectral synthesis of the Mg5183 and Mg5528 regions for a dwarf and giant.
The main panels of each sub-figure (for each combination star-feature)
present the MILES spectrum compared with model spectra.
The sub-figures also include graphs illustrating the abundance measurement methods.

The global procedure for obtaining and calibrating [Mg/Fe] ratios from our MR spectra is as follows:
\begin{enumerate}
\item automatic measurements at each Mg feature by applying both methods to obtain [Mg/Fe]$_{\rm feature}^{\rm method}$;
\item for each feature, computation of simple averages from the two methods when both
show reliable results checked through visual inspections to get [Mg/Fe]$_{\rm feature}$
(in some cases, only one method results in a reliable value and in other cases, both methods fail);
\item linear calibration of [Mg/Fe]$_{\rm feature}$ to a uniform scale, separately for each feature, via
comparison with the HR star control sample (described at Sect. 2.4),
to compute [Mg/Fe]$_{\rm feature}^{\rm calib}$;
\item and simple averaging of the calibrated ratios obtained from both features combined when possible, 
in order to obtain the final calibrated abundance ratios [Mg/Fe]$^{\rm calib}$.
\end{enumerate}
When the spectral synthesis of a given Mg feature does not satisfactorily reproduce the observed stellar spectrum
(equivalent width and/or absorption line profile),
we call this a non-reproduced spectral synthesis and classify it
such as
(a) inadequate reproduction of the observed spectral continuum,
(b) saturation effect on the [Mg/Fe] vs. log(EW) relationship (non-linearity),
(c) extrapolation on the [Mg/Fe] vs. log(EW) relationship and/or on the $rms_{\rm LPF}$ vs. [Mg/Fe] one,
(d) possible inaccurate model atmosphere interpolation
(i.e. done around the borders of the MARCS 2008 grid),
(e) possible higher rotation velocity than 130 km s$^{\rm -1}$,
(f) absence of molecular lines to compute the model spectrum
(e.g. TiO bands),
(g) low quality observed spectrum (S/N below 50 per {\AA}),
(h) possible wrong photospheric parameters,
(i) suspect chemically peculiar star, and
(j) other unknown causes and effects.
Consequently, the abundance ratio provided by the correspondent method is not reliable.

Calibration of each [Mg/Fe]$_{\rm feature}$ was applied after averaging the measurements obtained from both methods.
We decided to compute simple averages because each method explores one aspect of a reliable spectral synthesis.
Whilst the EW comparison focuses on the reproduction of the total energy absorbed by the feature relative to the local continuum,
the line profile fit takes into account the line shape including the core and wings.

Moreover the absolute differences between EW and LPF methods are always smaller than 3$\sigma$[Mg/Fe]$_{\rm feature}$
(around 0.03 dex for Mg5183 and 0.07 dex for Mg5528). 
We checked for dependences of the differences between methods on the atmospheric parameters
T$_{\rm eff}$, log $g$ and [Fe/H] and the results do not show any dependence, for both features.
We also noted there is no parametric dependence of the differences
([Mg/Fe]$_{\rm feature}^{\rm EW}$ $-$ [Mg/Fe]$_{\rm HR}$) and ([Mg/Fe]$_{\rm feature}^{\rm LPF}$ $-$ [Mg/Fe]$_{\rm HR}$)
for both Mg features and methods.

\begin{figure} 
\begin{center} 
\includegraphics[width=80mm]{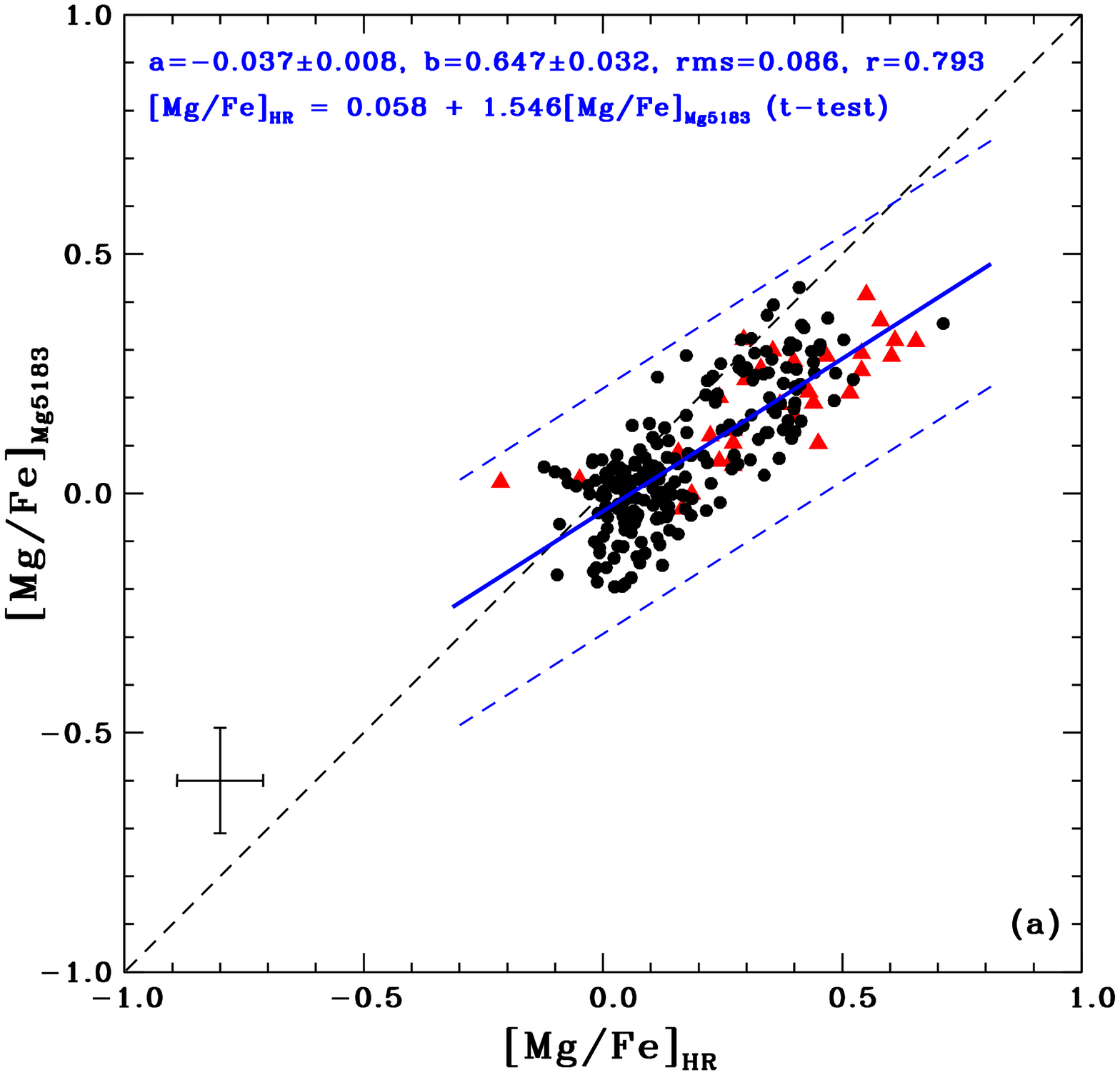}
\includegraphics[width=80mm]{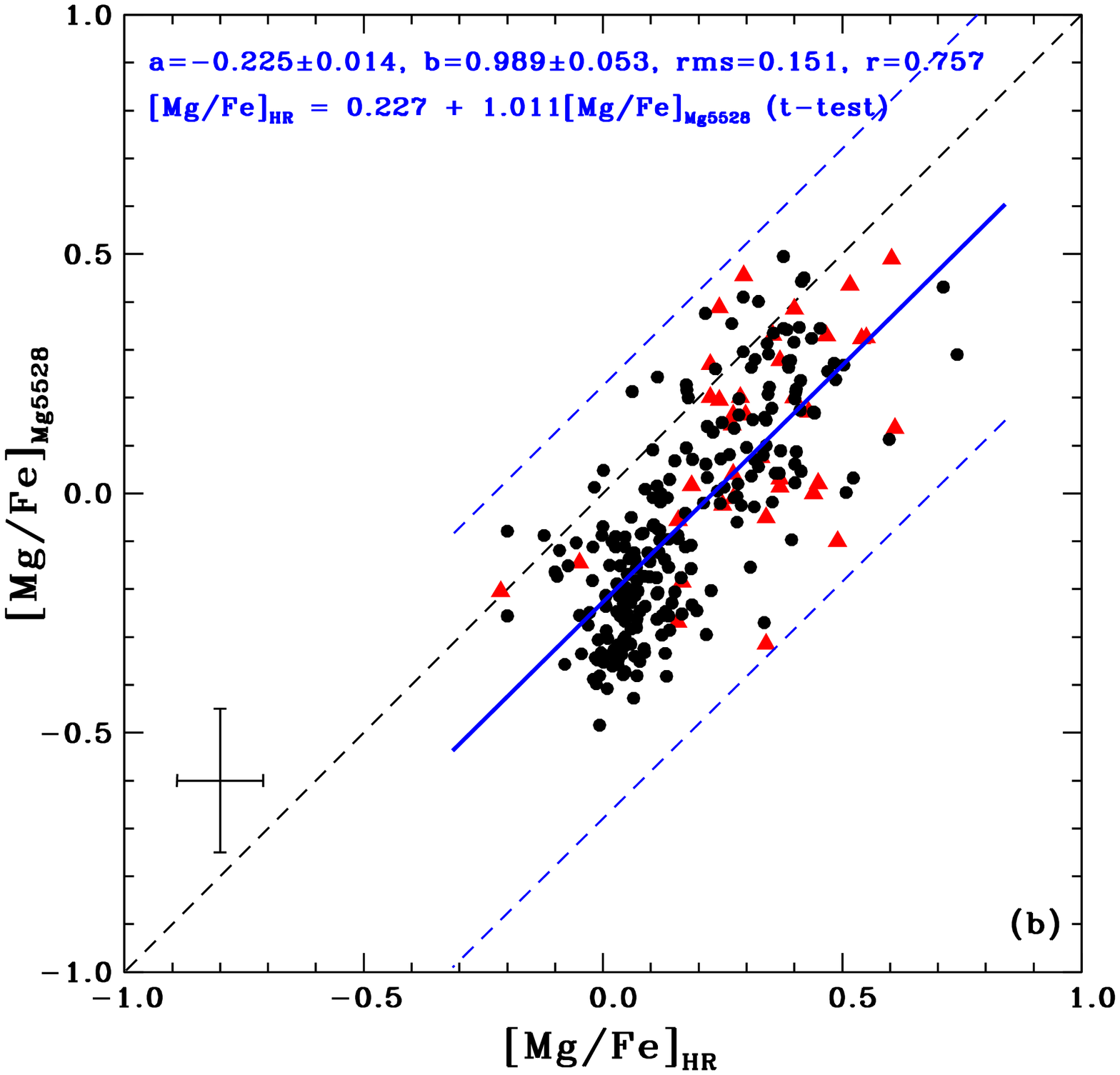} 
\end{center} 
\caption{ 
[Mg/Fe]$_{\rm Mg5183}$ vs. [Mg/Fe]$_{\rm HR}$, {\bf top panel (a)}, and
[Mg/Fe]$_{\rm Mg5528}$ vs. [Mg/Fe]$_{\rm HR}$, {\bf bottom panel (b)}, comparisons
for dwarfs and giants together in each,
respectively represented by black filled circles and red filled triangles.
The straight line inversely derived from the simple linear $lsq$ fit
[Mg/Fe]$_{feature}$ = $a$ + $b$[Mg/Fe]$_{HR}$
is also shown for each comparison (blue solid line)
with two parallel blue dashed lines illustrating the 3-$\sigma$ clipping procedure.
The parameters $a$ and $b$ are displayed on the top of graphs as well as
the calibration expressions themselves.
} 
\label{Fig4} 
\end{figure}

\subsection[3.4]{Calibration of the mid-resolution [Mg/Fe]}

The calibration to the previously adopted uniform scale relies on an inverse linear transformation,
\begin{equation} 
[{\rm Mg/Fe}]_{\rm feature}^{\rm calib} = (-a/b) + (1/b)[{\rm Mg/Fe}]_{\rm feature}
\label{Eq3}
\end{equation} 
that is obtained through a 3-$\sigma$ clipping simple linear $lsq$ fit of our MR measurements 
as a function of the HR calibrated values from the [Mg/Fe]$_{\rm HR}$ star control sample (Sect. 2.4).
We also applied a 95\% t-test to verify if the fit parameters $a$ and $b$ are distinguishable from zero and unity respectively.
In case they were not, we would adopt $a$ = 0 and $b$ = 1.

Figure 4 shows the comparisons of [Mg/Fe]$_{\rm Mg5183}$ and [Mg/Fe]$_{\rm Mg5528}$ against [Mg/Fe]$_{\rm HR}$,
in which the linear $lsq$ fittings are presented simultaneously for dwarfs and giants.
The computed calibration expressions are:
\begin{equation} 
[{\rm Mg/Fe}]_{\rm Mg5183}^{\rm calib} = 0.058~{\rm dex} + 1.546[{\rm Mg/Fe}]_{\rm Mg5183}
\label{Eq4}
\end{equation} 
\begin{equation} 
[{\rm Mg/Fe}]_{\rm Mg5528}^{\rm calib} = 0.227~{\rm dex} + 1.011[{\rm Mg/Fe}]_{\rm Mg5528}
\label{Eq5}
\end{equation} 
The calibrated abundance ratios [Mg/Fe]$^{\rm calib}$ from the Mg5183 and Mg5528 features
that respectively lay outside the intervals [$-$0.42 dex, +0.92 dex] and [$-$0.41 dex, +0.88 dex]
are based on extrapolations over the calibrations.
The intervals were estimated taking into account the [Mg/Fe]$_{\rm feature}$ scale coverages
of the stellar common sample between the HR compilation data and our MR measurements 
as well as the internal uncertainties of
[Mg/Fe]$_{\rm Mg5183}$ and [Mg/Fe]$_{\rm Mg5528}$ (described in Sect. 3.5).
Therefore the abundance ratios become more uncertain outside those intervals (see Fig. 4).

The stars' samples adopted in the calibrations are extensive enough and exhibit wide ranges of photospheric parameters,
which are as extensive as those of the MILES control sample, including uncertainties.
Parameter intervals covered can be seen in plots of Fig. 5 (described ahead).

\vspace{5mm}

We verified the atmospheric parameter sensitivity of the [Mg/Fe] differences between MR and HR results.
Figure 5 presents them as a function of T$_{\rm eff}$, log $g$ and [Fe/H] for both Mg features.
There is no significant dependence on the photospheric parameters,
as indicated by the linear fits (tiny linear correlation coefficients).
Furthermore the differences are always smaller or comparable to $rms$ scatter along the whole parameter scales.

Therefore, where possible, we averaged the calibrated ratios obtained from both features,
in order to obtain calibrated abundance ratios as follows:
\begin{equation} 
\overline{[{\rm Mg/Fe}]^{\rm calib}} = ([{\rm Mg/Fe}]_{\rm Mg5183}^{\rm calib} + [{\rm Mg/Fe}]_{\rm Mg5528}^{\rm calib})/2
\label{Eq6}
\end{equation} 

In Fig. 6, the residuals between calibrated results from the two features are plotted
as a function of T$_{\rm eff}$, log $g$ and [Fe/H].
There is no systematic dependency on the stellar parameters,
as it is noticed from the small linear correlation coefficients $r$.
This supports our procedure of computing averages of [Mg/Fe]$_{\rm feature}^{\rm calib}$.
The medium data dispersion (0.15 dex $rms$) is comparable
with the typical internal error and systematic error for the calibrated ratio from Mg5528,
but is slightly greater for the other feature.
It is equal to the systematic error of [Mg/Fe]$_{\rm Mg5528}^{\rm calib}$ ($\pm$0.15 dex),
but it is slightly greater than $\sigma$[Mg/Fe]$_{\rm Mg5183}^{\rm calib}$ (0.13 dex)
and $\sigma\overline{[{\rm Mg/Fe}]^{\rm calib}}$ (0.10 dex).

The criterion for deciding if the calibrated [Mg/Fe] measurements of both Mg features can be averaged
to obtain representative determinations
is fixed by the maximum deviation acceptable between them,
i.e. the condition is expressed by
$|$[Mg/Fe]$_{\rm Mg5528}^{\rm calib}$ $-$ [Mg/Fe]$_{\rm Mg5183}^{\rm calib}$$|$ $\leq$ 4$\sigma\overline{[{\rm Mg/Fe}]^{\rm calib}}$
or $\Delta$[Mg/Fe]$_{Mg5528-Mg5183}$ $\leq$ 0.40 dex, as illustrated in Fig. 6
and adopted for calibrated HR measurements collected from duplicated sources (Sect. 2.2).
See Sect. 3.5 to get details about how $\sigma\overline{[{\rm Mg/Fe}]^{\rm calib}}$ was estimated.
The Mg5183 determination was adopted for the stars that do not follow this condition
(5 dwarfs and 10 giants),
because this feature provides a better precision.

After separately calibrating [Mg/Fe]$_{\rm Mg5183}$ and [Mg/Fe]$_{\rm Mg5528}$, using the extensive control sample, 
we were able to compute average or individual feature values for a great number of dwarfs and giants.
In total we measured [Mg/Fe] at MR for 437 extra MILES stars (150 dwarfs and 287 giants). 
This represents around 44\% of the whole spectral library.

\begin{figure*} 
\begin{center} 
\includegraphics[width=85mm]{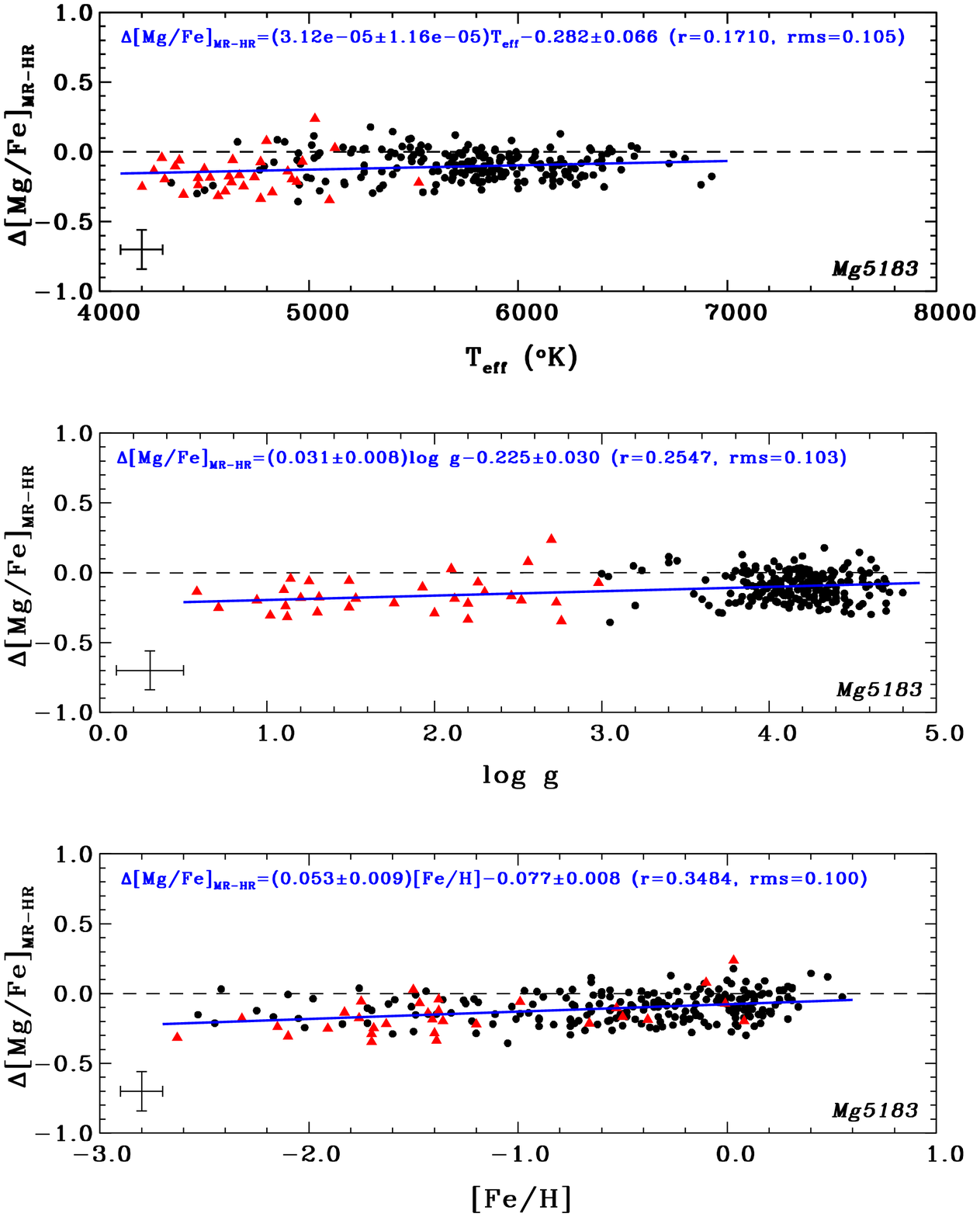}
\includegraphics[width=85mm]{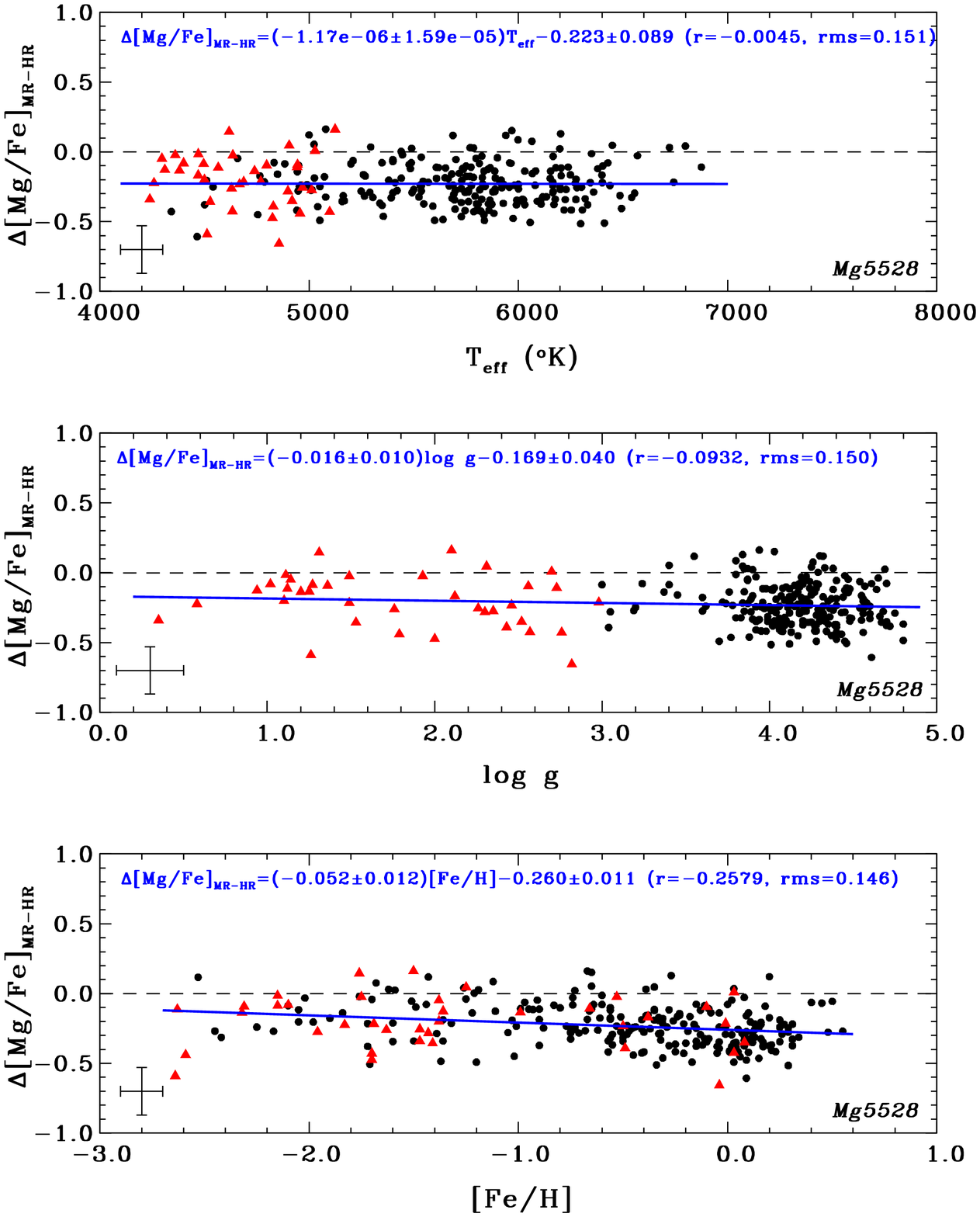} 
\end{center} 
\caption{ 
[Mg/Fe] differences between non-calibrated mid-resolution (MR) and calibrated high-resolution (HR) measurements
($\Delta$[Mg/Fe]$_{\rm MR-HR}$ = [Mg/Fe]$_{\rm feature}$ $-$ [Mg/Fe]$_{\rm HR}$)
computed for the features Mg5183 ({\bf left panel}) and Mg5528 ({\bf right panel})
as a function of T$_{\rm eff}$, log $g$ and [Fe/H] distinguishing dwarfs (black filled circles) and giants (red filled triangles). 
Simple linear $lsq$ fittings between the differences and each photospheric parameter are also shown (blue solid lines).
The resultant expressions, $rms$ and correlation coefficient of these linear fits are displayed on the top of each graph. 
} 
\label{Fig5} 
\end{figure*}

\begin{figure} 
\begin{center} 
\includegraphics[width=86mm]{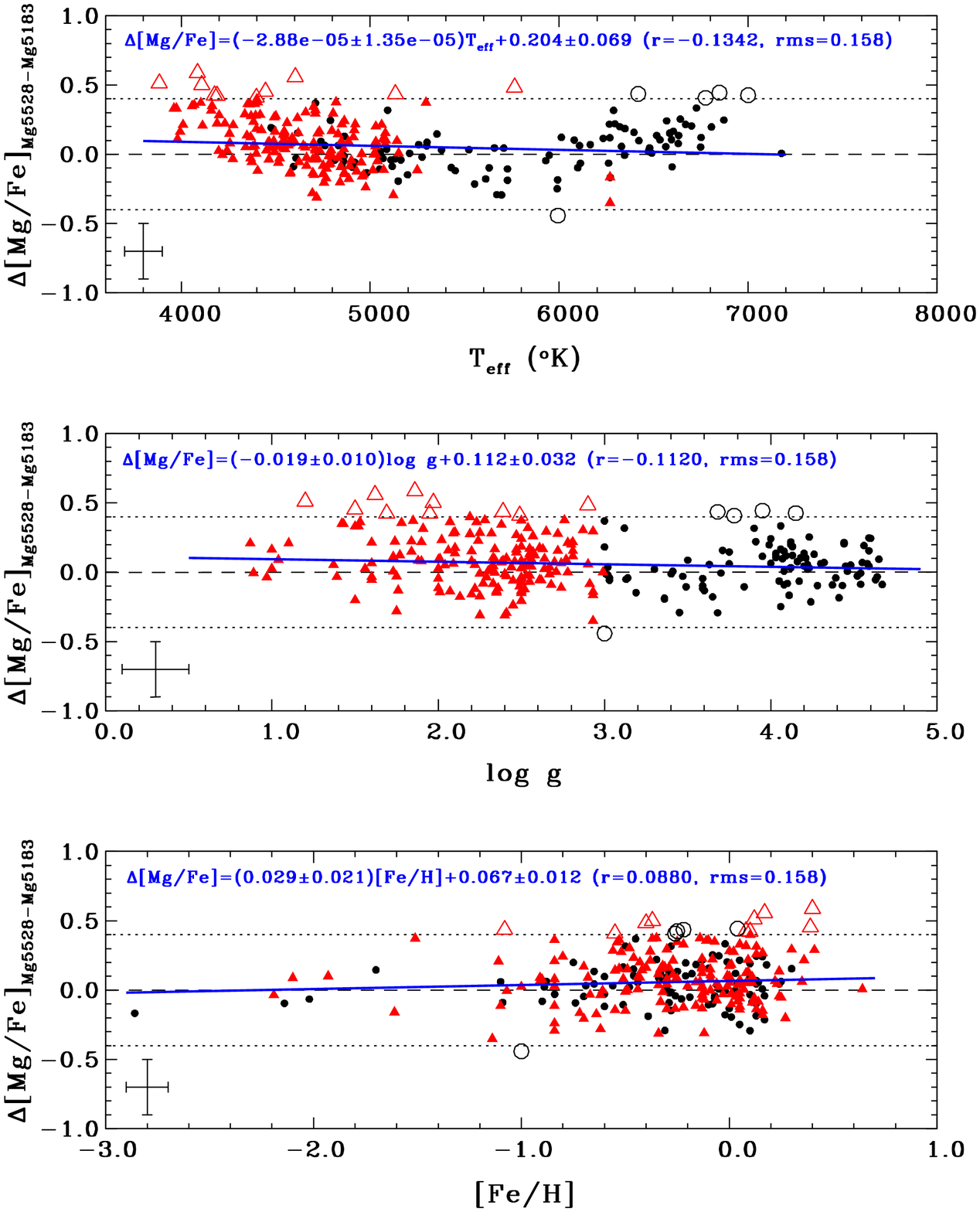}
\end{center} 
\caption{ 
[Mg/Fe] differences between the calibrated measurements obtained with the two Mg features 
($\Delta$[Mg/Fe]$_{\rm Mg5528-Mg5183}$ = [Mg/Fe]$_{\rm Mg5528}^{\rm calib}$ $-$ [Mg/Fe]$_{\rm Mg5183}^{calib}$)
as a function of T$_{\rm eff}$, log $g$ and [Fe/H]
for dwarfs (black filled circles) and giants (red filled triangles). 
Least-square linear fittings between the differences and each photospheric parameter are also shown
(solid blue lines) considering the data shown as filled symbols.
Two parallel dotted black lines are drawn to representing the 4$\sigma\overline{[{\rm Mg/Fe}]^{\rm calib}}$
as a maximum acceptable deviation for $\Delta$[Mg/Fe]$_{\rm Mg5528-Mg5183}$
(details in the end of Sect. 3.5).
The black open circles represent the dwarfs and red open triangles the giants
for which average abundance ratios were not computed based on the 4$\sigma$ criterion.
These stars are not considered in the $lsq$ fittings.
} 
\label{Fig6} 
\end{figure}

\subsection[3.5]{Uncertainty of the mid-resolution measurements} 

The internal uncertainties of [Mg/Fe] are due to the MILES photospheric parameter errors
and the abundance determination methods applied (EW and LPF).

We estimated the error propagation from the imprecision of T$_{\rm eff}$, log $g$ and [Fe/H]
by adopting model atmospheres directly collected from the MARCS 2008 grid,
whose parameters coincide with the photospheric parameters of
all MILES stars possible considering the typical errors (1$\sigma$).
This sample of MILES stars is composed of 135 dwarfs and 31 giants,
and it has an excellent coverage in the MARCS model parameter space
(distributing between 4200 and 6200 K in T$_{\rm eff}$,
1.0 and 4.7 in log $g$ and $-$2.10 and $+$0.40 dex in [Fe/H]).
Then, we computed their abundance ratios
based only on the EW method by using each Mg feature,
hereafter simply [Mg/Fe]$_{\rm atm}$.
Afterwards, we compared them with those measurements obtained by adopting interpolated model atmospheres,
whose parameters actually match T$_{\rm eff}$, log $g$ and [Fe/H] of the stars, naming them [Mg/Fe]$_{\rm interpol}$.
In this procedure, all [Mg/Fe] are not calibrated to the HR uniform scale, since only differences are needed for estimating errors.
The deviation ([Mg/Fe]$_{\rm atm}$ $-$ [Mg/Fe]$_{\rm interpol}$)
is well distributed around zero over all parameter scales
and it does not show any stellar parameter dependence.
These differences at 1$\sigma$ level
gave us a good estimation for the global internal uncertainty due to the star's parameter errors, $\sigma$[Mg/Fe]$_{\rm atm}$,
which is around 0.09 dex for the Mg5183 feature and 0.10 dex for Mg5528.

The internal uncertainties in [Mg/Fe] due to the EW method itself, $\sigma$[Mg/Fe]$_{\rm EW}$, for both Mg features were estimated
by computing the propagation of the observed equivalent width errors through the relationship [Mg/Fe] vs. log(EW).
We noted that the EW uncertainties dominate over the propagation error from the [Mg/Fe] vs. log(EW) theoretical fits.
The median of the $\sigma$[Mg/Fe]$_{\rm EW}$ asymmetric distribution for each Mg feature
characterises the typical uncertainty that it is around 0.09 dex for Mg5183 and 0.20 dex for Mg5528.

We also investigated the dependence of $\sigma$[Mg/Fe]$_{\rm EW}$ on the atmospheric parameters.
There is some correlation with [Fe/H] for Mg5183 only,
where $\sigma$[Mg/Fe]$_{\rm EW}$ is higher on average for metal-poor stars ($\sim$0.2 dex)
in comparison with metal-rich stars ($\sim$0.1 dex) with a transition limit around [Fe/H] = $-$1.0 dex.
Although a metallicity dependence is also noticed in HR measurements, as in BM05,
we decided not to estimate $\sigma$[Mg/Fe]$_{\rm EW}$ as a function of [Fe/H].

\vspace{2mm}

The uncertainty from the LPF method depends on how accurately the minimum of the curve is determined (see Fig. 3).
Typically this is estimated to be between 0.05 and 0.10 dex, giving a mean of $\sigma$[Mg/Fe]$_{\rm LPF}$ around 0.075 dex.
The final uncertainty due to the measurement process was computed
as mean value of the abundance uncertainties obtained with the two methods,
because, although they are not completely independent, the data is treated in different ways through them.
Typical value of $\sigma\overline{[{\rm Mg/Fe}]}_{\rm method}$ is 0.06 dex for Mg5183 and 0.11 dex for Mg5528.
The typical final internal error for each Mg feature, 0.11 and 0.15 dex for Mg5183 and Mg5528 respectively,
was obtained as the quadratic sum of $\sigma$[Mg/Fe]$_{\rm atm}$ and $\sigma\overline{[{\rm Mg/Fe}]}_{\rm method}$.

The systematic uncertainties of calibrated MR [Mg/Fe] were estimated
by comparing these values directly with HR values,
such that the $rms$ of deviation ([Mg/Fe]$_{\rm feature}^{\rm calib}$ $-$ [Mg/Fe]$_{\rm HR}$)
represent a good estimation, i.e. for each Mg feature we have
\begin{equation} 
\sigma[{\rm Mg/Fe}]^{\rm calib} = ( \frac{1}{N} \sum_{i=1}^{N} ([{\rm Mg/Fe}]^{\rm calib} - [{\rm Mg/Fe}]_{\rm HR})^{\rm 2} )^{\rm 1/2}
\label{Eq7}
\end{equation}
where $N$ means the number of stars in the MR vs. HR comparisons for each Mg feature determination.
Therefore [Mg/Fe] recovered by Mg5183 holds an uncertainty of 0.13 dex or
0.15 dex when is uniquely measured by Mg5528.
When it was possible to compute [Mg/Fe] as an average from both feature determinations,
its systematic error $\sigma\overline{[{\rm Mg/Fe}]^{\rm calib}}$ 
was estimated by quadratic mean reaching 0.10 dex.
Table 3 summarizes the [Mg/Fe] uncertainties. 

We also investigated the influence of the models atmosphere alpha-enhancement chemistry on the spectral synthesis carried out at MR 
by always keeping the model chemistry unchanged at each metallicity (see Sect. 3.1).
Besides the MARCS standard models adopted in our work,
there are other model classes: 
one named alpha-poor assuming $\alpha$/Fe solar for $-$2.00 $\leq$ [Fe/H] $\leq$ $-$0.25 dex, 
and another denominated alpha-enhanced with [$\alpha$/Fe] = +0.40 dex covering [Fe/H] from $-$0.75 to +0.50 dex.
Thus we performed spectral syntheses to cover four ([Fe/H], [$\alpha$/Fe]) (dex, dex) combinations as following:
(i) ($-$1.50, 0.00) and (ii) ($-$1.50, +0.40)
with the alpha-poor ([$\alpha$/Fe] = 0.00) and standard ([$\alpha$/Fe] = +0.40) models for this metallicity,
and (iii) (0.00, 0.00) and (iv) (0.00, +0.40)
with the standard ([$\alpha$/Fe] = 0.00) and alpha-enhanced ([$\alpha$/Fe] = +0.40) models for this metallicity.
The syntheses were applied to four stellar evolution stages:
main sequence at T$_{\rm eff}$ = 5000 K and log $g$ = 4.5,
turn-off main sequence at T$_{\rm eff}$ = 6000 K and log $g$ = 4.0,
sub-giant at T$_{\rm eff}$ = 5000 K and log $g$ = 3.5,
red giant branch at T$_{\rm eff}$ = 4500 K and log $g$ = 1.0.
However, just the red giant stage could be adopted for the solar metallicity
due to a limitation of the MARCS alpha-enhanced class coverage.
Afterwards, we directly compared the equivalent widths of each Mg feature measured on theoretical spectra
that have been computed for each metallicity-$\alpha$-enhancement combination but assuming two different model classes.

At [Fe/H] = $-$1.50 dex, 
it is noticed that the EW variations of both Mg features
are very acceptable within typical uncertainties,
i.e. $<$ 8\% for the main sequence and sub-giant stages and $<$ 3\% for the turn-off and red giant stages
that are translated into [Mg/Fe] abundance ratio changes smaller than around 0.10 and 0.05 dex respectively.
At solar metallicity, the EW variations of red giant stage  (the only one analysed)
are smaller than 7\% for Mg5183 (or about 0.09 dex in [Mg/Fe])
and 20\% for the Mg5528 feature (or $\sim$0.2 dex in [Mg/Fe]).
Therefore it is viable to perform spectral synthesis at MR by fixing the model atmosphere chemistry
and changing the $\alpha$-element abundances to cover a large range of [$\alpha$/Fe] values
as it was done in the current work.

However, a conservative variation of $\pm$0.2 dex for [$\alpha$/Fe]
in the spectral syntheses done under fixed model atmosphere chemistries
could be assumed in order to figure out which abundance ratio determinations
would be considered as extrapolations based on the $\alpha$-enhancement compatibility.
Concerning the MARCS standard models adopted and taking into account the uncertainties of [Mg/Fe]$^{\rm calib}$,
we have found 33 stars, which represent just 7.5\% of all MR determinations
and are identified in the catalogue (Sect. 4).
Consequently, the [Mg/Fe] of each case might have a less accuracy in the sense of it was based on a model atmosphere
whose chemistry does not exactly follow the abundances of $\alpha$-elements adopted in the spectral synthesis.
These cases include those extrapolations over the Mg feature calibrations themselves (Sect. 3.4).

\begin{table*} 
\caption{
Typical uncertainties of [Mg/Fe] (dex unity) from our mid-resolution measurements,
showing the internal errors for the Mg5183 and Mg5528 features
due to the photospheric methods adopted (pseudo-equivalent widths, EW, and line profile fitting, LPF)
and stellar atmospheric parameter errors.
The total uncertainty of non-calibrated [Mg/Fe]$_{\rm feature}$ and
the systematic error of calibrated [Mg/Fe]$^{\rm calib}$ for each Mg feature are presented respectively in the 
last and penultimate rows.}
\label{Special_cases} 
\begin{center} 
\begin{tabular}{@{}llllll}
\hline 
\hline 
\# &                                             & Mg5183 & Mg5528 & both  & Notes \\ 
\hline
   &                                             & (dex)  & (dex)  & (dex) &       \\ 
\hline
\hline
1  & $\sigma$[Mg/Fe]$_{\rm EW}$                  & 0.09   & 0.20   & ----- & from the EW method \\ 
2  & $\sigma$[Mg/Fe]$_{\rm LPF}$                 & 0.075  & 0.075  & ----- & from the LPF method \\ 
\hline 
3  & $\sigma$[Mg/Fe]$_{\rm method}$             & 0.06   & 0.11   & ----- & from rows 1 and 2 variance averaged \\ 
\hline 
4  & $\sigma$[Mg/Fe]$_{\rm atm}$                 & 0.09   & 0.10   & ----- & due to the photospheric parameter errors \\ 
\hline 
5  & $\sigma$[Mg/Fe]$_{\rm feature}$             & 0.11   & 0.15   & ----- & internal errors from rows 3 and 4 combined in quadrature \\ 
\hline    
6  & $\sigma$[Mg/Fe]$_{\rm feature}^{\rm calib}$ & 0.13   & 0.15   & 0.10  & systematic errors (from HR comparisons), variance averaged in the last column \\
\hline 
\hline 
\end{tabular} 
\end{center} 
\end{table*}

\subsection[3.6]{Coverage of the MR measurements}

There are 843 MILES stars with catalogued atmospheric parameters within the MARCS 2008 grid.
From those, 308 already had [Mg/Fe] measurements from HR studies (Sect. 2).
The 535 residual stars were spectroscopically analysed by us at MR.
The MR determinations cover wide ranges in atmospheric parameters (see Figs. 8 to 11 described in Sect. 4),
reaching 81.7\% efficiency or completeness level inside the MARCS parameter space when the stars with HR data are not considered.
Rather than be uniform over all scales, our measurements actually complement the HR data.
However, depending on the region, one Mg feature works better than the other.h
Section 4 presents a more extensive discussion about the parametric coverage of all MR and HR determinations.

\vspace{6mm}

In general, the Mg5183 feature does not work well on the coldest stars (T$_{\rm eff}$ $<$ 4000 K)
due to the presence of strong molecular absorptions of TiO and MgH as well.
Mg5528 is not satisfactorily applied for the hottest giants (T$_{\rm eff}$ $>$ 5500 K),
since this feature becomes too weak and it is practically insensitive to abundance variation in these cases.
In particular, Mg5183 gives reliable abundance measurements for dwarfs and giants with temperatures between 4000 and 8000 K,
while Mg5528 basically works on dwarfs with 3500 up to 8000 K.
However, for giants, Mg5528 can only be applied with great confidence for 3600 $\leq$ T$_{\rm eff}$ $\leq$ 5500 K.
Both Mg features can be used in the whole metallicity range.

In specified cases, the spectral synthesis does not work on
a single
or both features,
due to line saturation, non-reproduction of spectrum continuum,
extrapolation on the [Mg/Fe] vs. log (EW) and/or $rms_{\rm LPF}$ vs. [Mg/Fe] relationship,
incompleteness of line lists (mainly TiO bands), and low quality spectra in some cases.
The continuum and EW/LPF extrapolation cases do not exhibit any stellar parametric dependence for Mg5183.
On the other hand, we noted the cases of line saturation and incomplete line list
occur in metal-rich cold stars ([Fe/H] $>$ $-$1.0 dex with T$_{\rm eff}$ $<$ 4000 K) for both features.
A few of those non-reproduced spectral cases
(that sum in total 79 dwarfs and 115 giants for Mg5183 plus 60 dwarfs and 97 giants for Mg5528)
could be fixed, excepting those due to the incompleteness of line lists:
8 cases for Mg5183 (5 dwarfs and 3 giants),
and 15 cases for Mg5528 (5 dwarfs and 10 giants).
The line saturation effect was solved by adopting a spline fit on the EW method
and sometimes a smaller number of models (9 cases in total).

\section[4]{The MILES magnesium abundance catalogue} 

We have obtained [Mg/Fe] covering a bit more than 3/4 (more precisely 752 stars or 76.3\%)
of the MILES stellar spectrum library 
(411 dwarfs and 341 giants, respectively around 76\% and 77\% of their totals) that are suitable for SSP modelling,
i.e. the typical systematic uncertainty of [Mg/Fe] is 0.105 dex on average over our whole catalogue.
The stars' coverage in the four-dimensional parameter space of MILES T$_{\rm eff}$, log $g$, [Fe/H] and [Mg/Fe]
is extensive, as discussed in this section.
If we only consider the MILES stars with complete sets of photospheric parameters
that sum 946 objects
(Cenarro {\it et al.} 2007),
the coverage reaches 79.5\%.

The compiled [Mg/Fe] catalogue of MILES is presented in Tables 4 and 5 for field and cluster stars respectively.
The whole catalogue is only published in electronic form.
Basically, the catalogue tables provide the [Mg/Fe] values with their individual errors
together with the sources from where they have been obtained 
(i.e. the reference in case it has been compiled from HR works 
or the Mg feature(s) when it is a MR measurement).

\begin{table*} 
\caption{
The MILES [Mg/Fe] catalogue for field stars.
First column presents the star identification in the MILES database.
The identification in the CaT library
(Cenarro {\it et al.} 2002)
is in the second column.
The stars' names in other catalogues are shown in the third column.
The stellar photospheric parameters in MILES are listed from the 4$^{\rm th}$ to 6$^{\rm th}$ columns. 
[Mg/Fe] is shown in the seventh column together with its error in the eighth column.
Notes about the source of each [Mg/Fe] measurement are written in the last column
identifying its origin, i.e. from the high-resolution compilation (HR)
or our mid-resolution measurements (mr), as well as
the HR work, BM05 or other(s) listed in Table 1, 
the Mg feature(s) adopted in each stellar MR measurement,
and if the mid-resolution measurement represents an $\alpha$-enhancement model atmosphere extrapolation as described in Sect. 3.5
(designated by *).
The full table is only available in electronic form.
}
\begin{tabular}{@{}lllrrrrrl}
\hline 
\hline 
\#MILES&\#CaT& Star name       & T$_{\rm eff}$ & log $g$ & [Fe/H] & [Mg/Fe] & $\sigma$[Mg/Fe] & Notes           \\ 
\hline 
       &     &                 &     (K)       &         &  (dex) &   (dex) &           (dex) &                 \\ 
\hline 
\hline 
0081F &     & BD-010306        &  5650 &  4.40 & $-$0.90 &  +0.40 & 0.05 & HR BM05     \\  
0266F &     & BD-011792        &  4948 &  3.05 & $-$1.05 &  +0.71 & 0.20 & HR T98      \\  
0505F & 677 & BD+012916        &  4238 & +0.35 & $-$1.47 &  +0.37 & 0.20 & HR T98      \\  
0329F &     & BD-032525        &  5750 &  3.60 & $-$1.90 &  +0.41 & 0.07 & HR BM05     \\  
0777F &     & BD+044551        &  5770 &  3.87 & $-$1.62 &  +0.42 & 0.07 & HR BM05     \\  
0327F &     & BD-052678        &  5429 &  4.43 & $-$2.14 &  +0.41 & 0.10 & mr BothMg   \\  
0569F &     & BD+053080        &  4832 &  4.00 & $-$0.88 &  +0.55 & 0.10 & mr BothMg   \\  
0142F &     & BD+060648        &  4400 &  1.02 & $-$2.10 &  +0.52 & 0.20 & HR T98      \\  
0144F &     & BD-060855        &  5283 &  4.50 & $-$0.70 &$-$0.01 & 0.05 & HR BM05     \\ 
0537F &     & BD+062986        &  4450 &  4.80 & $-$0.30 &  +0.06 & 0.13 & mr Mg5183   \\  
\hline 
\hline 
\end{tabular} 
\label{catalogue} 
\end{table*}

\begin{table*} 
\caption{
The MILES [Mg/Fe] catalogue for star cluster stars (as in Table 4).
The cluster names and types are in 3$^{rd}$ and 4$^{th}$ columns.
The full table is only available in electronic form.
}
\begin{tabular}{@{}lllllrrrrrl}
\hline 
\#MILES&\#CaT& Cluster name & Type & Star name       & T$_{\rm eff}$ & log $g$ & [Fe/H] & [Mg/Fe] & $\sigma$[Mg/Fe] & Notes \\ 
\hline 
       &     &              &      &                 &     (K)       &         &  (dex) &   (dex) &           (dex) &       \\ 
\hline 
\hline 
0920C & 006 & Coma Ber & open     & HD107276         &  7972 &  4.21 & $-$0.05 &   +0.29 & 0.15 & mr Mg5528   \\
0921C & 007 & Coma Ber & open     & HD107513         &  7409 &  4.25 & $-$0.05 &   +0.32 & 0.15 & mr Mg5528   \\
0901C & 016 & Hyades   & open     & HD025825         &  5992 &  4.41 &   +0.13 & $-$0.09 & 0.10 & mr BothMg   \\
0902C & 017 & Hyades   & open     & HD026736         &  5657 &  4.45 &   +0.13 & $-$0.24 & 0.10 & mr BothMg*  \\
0904C &     & Hyades   & open     & HD027383         &  6098 &  4.28 &   +0.13 &   +0.03 & 0.10 & mr BothMg   \\
0905C & 023 & Hyades   & open     & HD027524         &  6622 &  4.28 &   +0.13 & $-$0.16 & 0.15 & mr Mg5528*  \\
0906C & 025 & Hyades   & open     & HD027561         &  6742 &  4.24 &   +0.13 &   +0.03 & 0.20 & HR T98      \\
0907C &     & Hyades   & open     & HD027962         &  8850 &  3.80 &   +0.13 & $-$0.09 & 0.20 & HR T98      \\
0908C & 029 & Hyades   & open     & HD028483         &  6486 &  4.30 &   +0.13 &   +0.01 & 0.20 & HR T98      \\
0909C &     & Hyades   & open     & HD028546         &  7626 &  4.11 &   +0.13 & $-$0.23 & 0.15 & mr Mg5528*  \\
\hline 
\hline 
\end{tabular} 
\label{catalogue} 
\end{table*}

\begin{figure} 
\begin{center} 
\includegraphics[width=65mm, angle=-90]{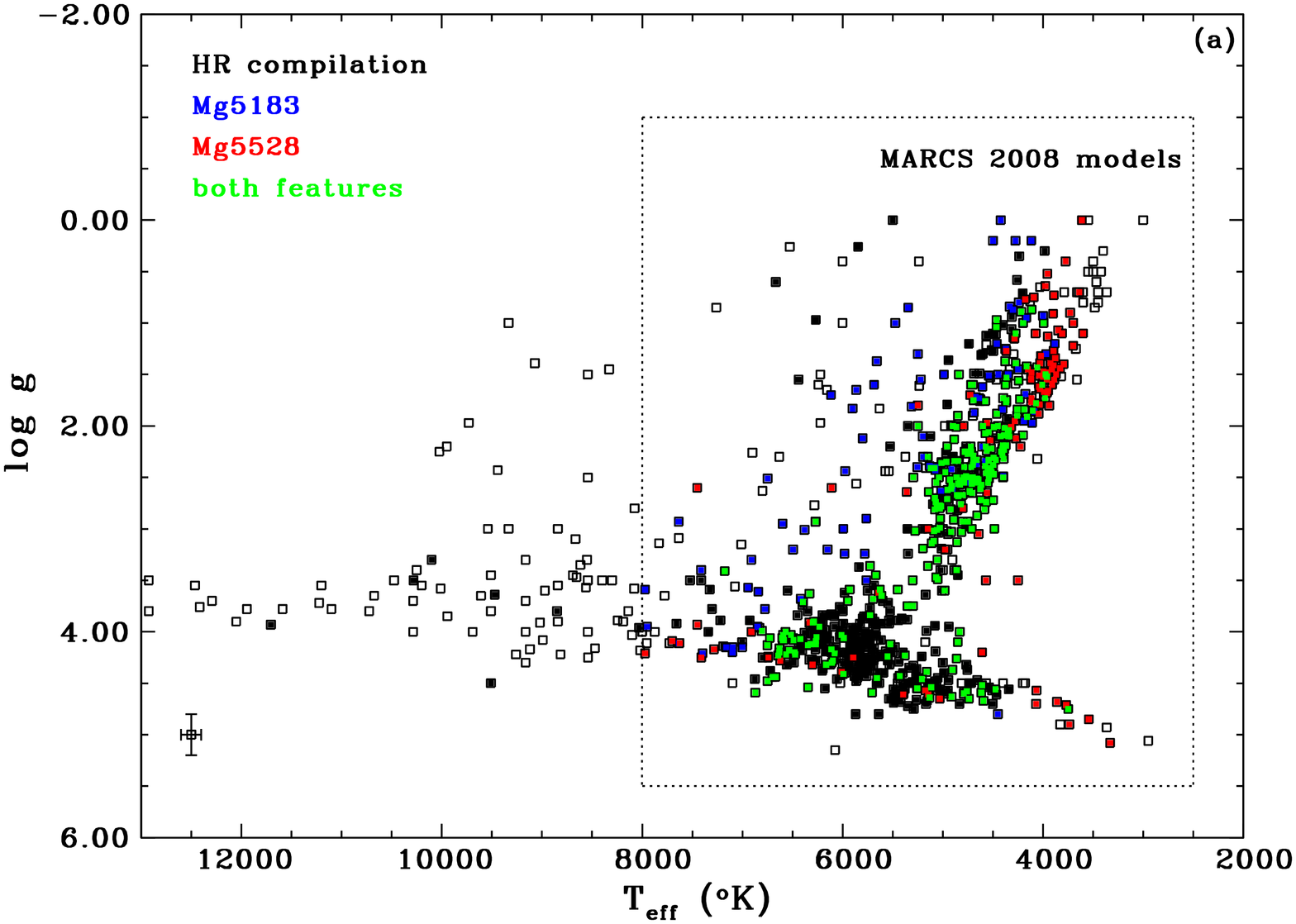} 
\includegraphics[width=65mm, angle=-90]{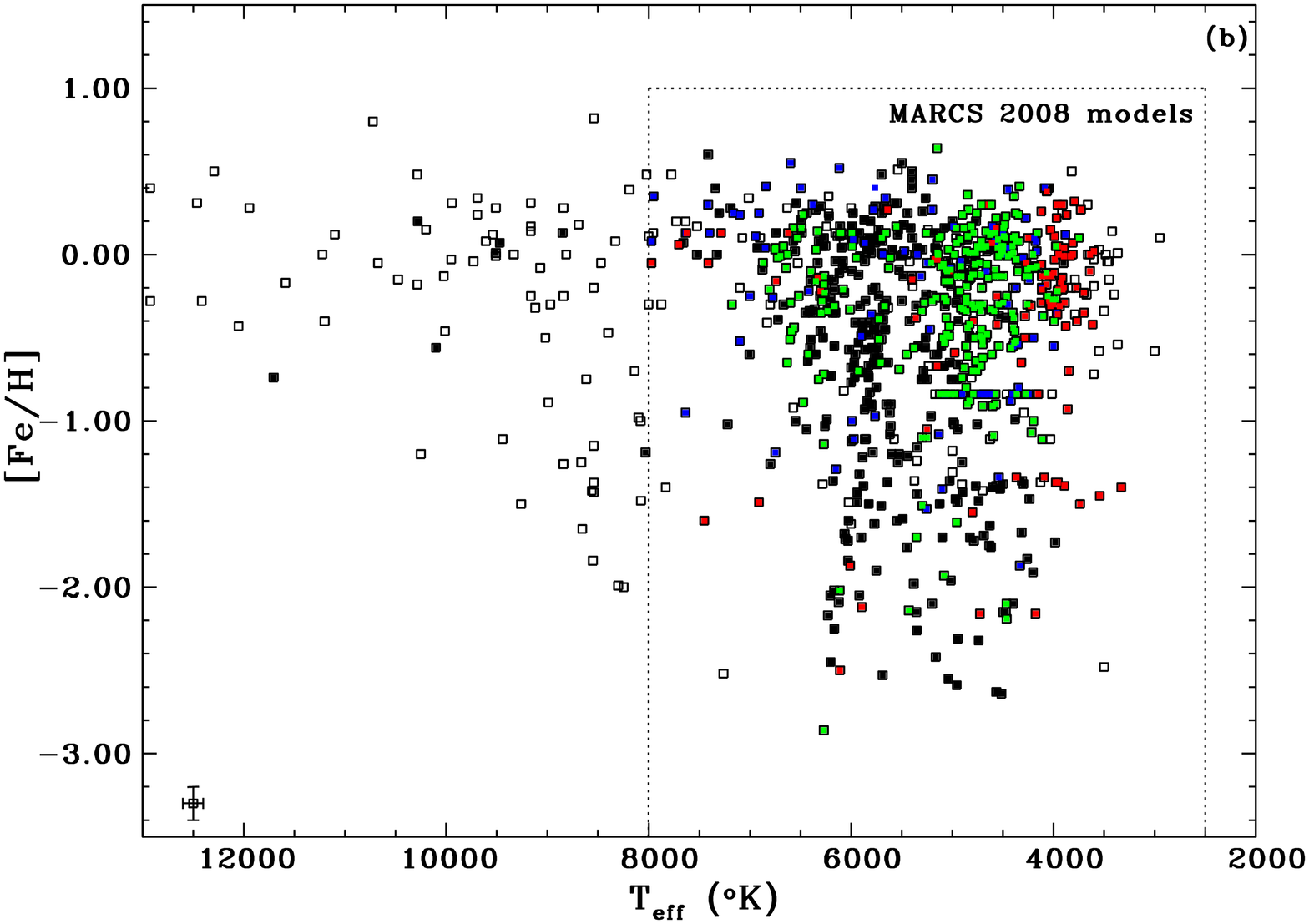} 
\includegraphics[width=65mm, angle=-90]{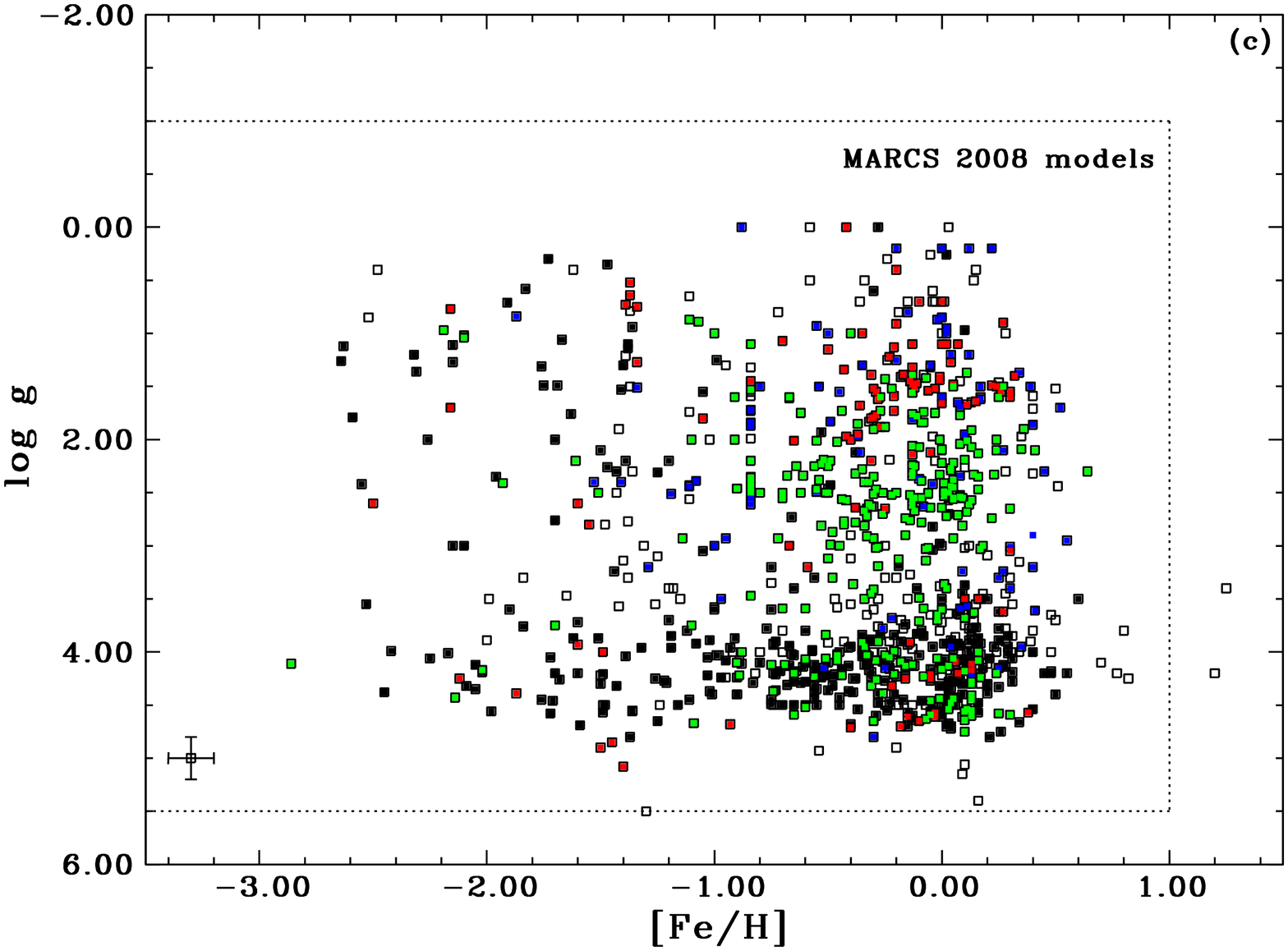} 
\end{center} 
\caption{ 
Coverage of the MILES stars with [Mg/Fe] in the photospheric parameter space:
{\bf (a)} modified H-R diagram log $g$ vs. T$_{\rm eff}$ ({\bf top panel}),
{\bf (b)} [Fe/H] vs. T$_{\rm eff}$ plane ({\bf middle panel}), and
{\bf (c)} log $g$ vs. [Fe/H] projection ({\bf bottom panel}).
The MILES stars with abundance ratios from HR studies are shown as black filled symbols,
and the stars with MR measurements from this work as colour filled symbols
(blue designating determinations based on the Mg5183 feature only, red on Mg5528 only and green on both Mg features combined).  
The library stars without [Mg/Fe] are drawn as open symbols.
The MARCS 2008 grid extension is also represented in each panel.
} 
\label{Fig7} 
\end{figure}

\begin{figure} 
\begin{center} 
\includegraphics[width=65mm, angle=-90]{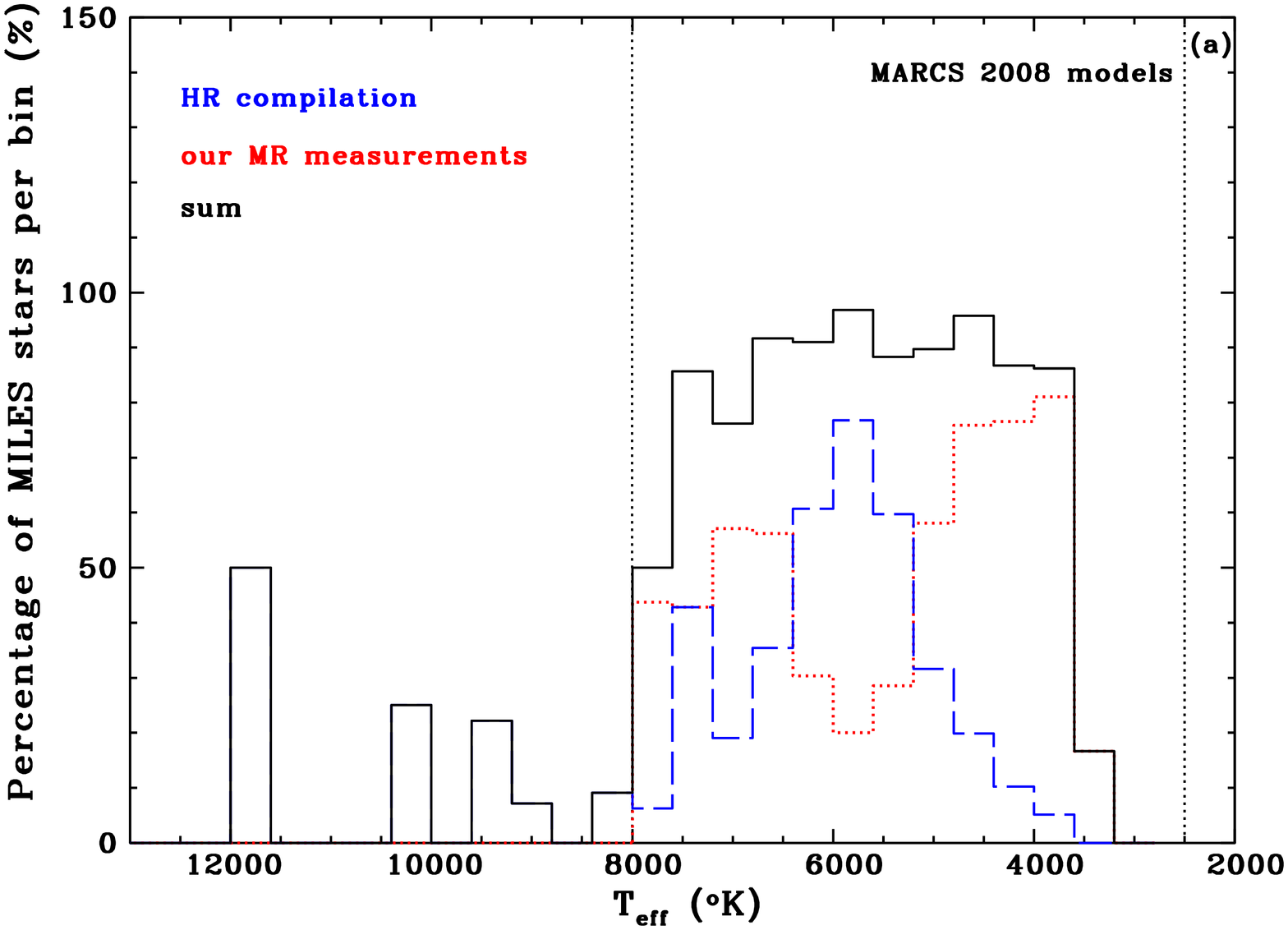} 
\includegraphics[width=65mm, angle=-90]{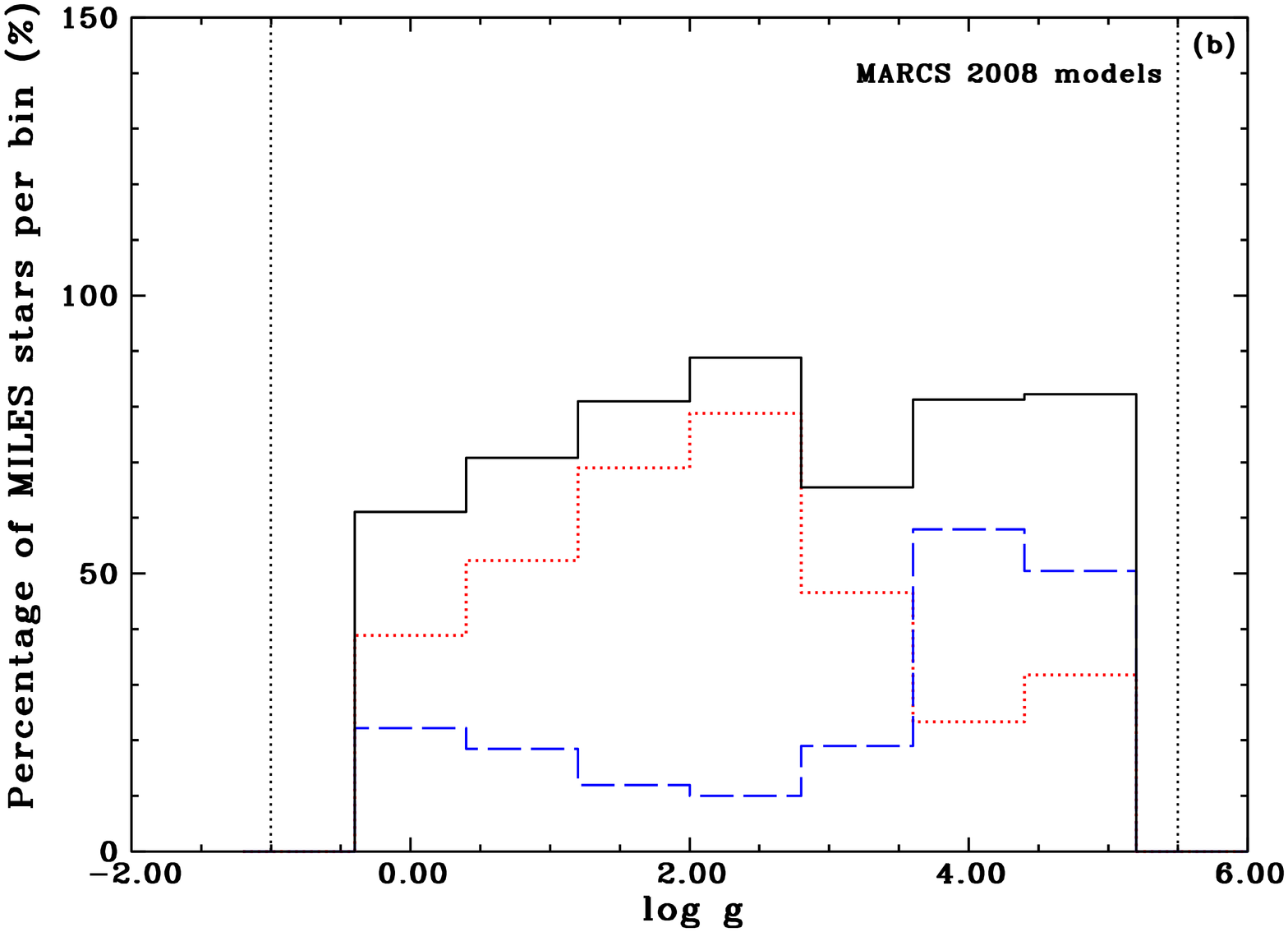} 
\includegraphics[width=65mm, angle=-90]{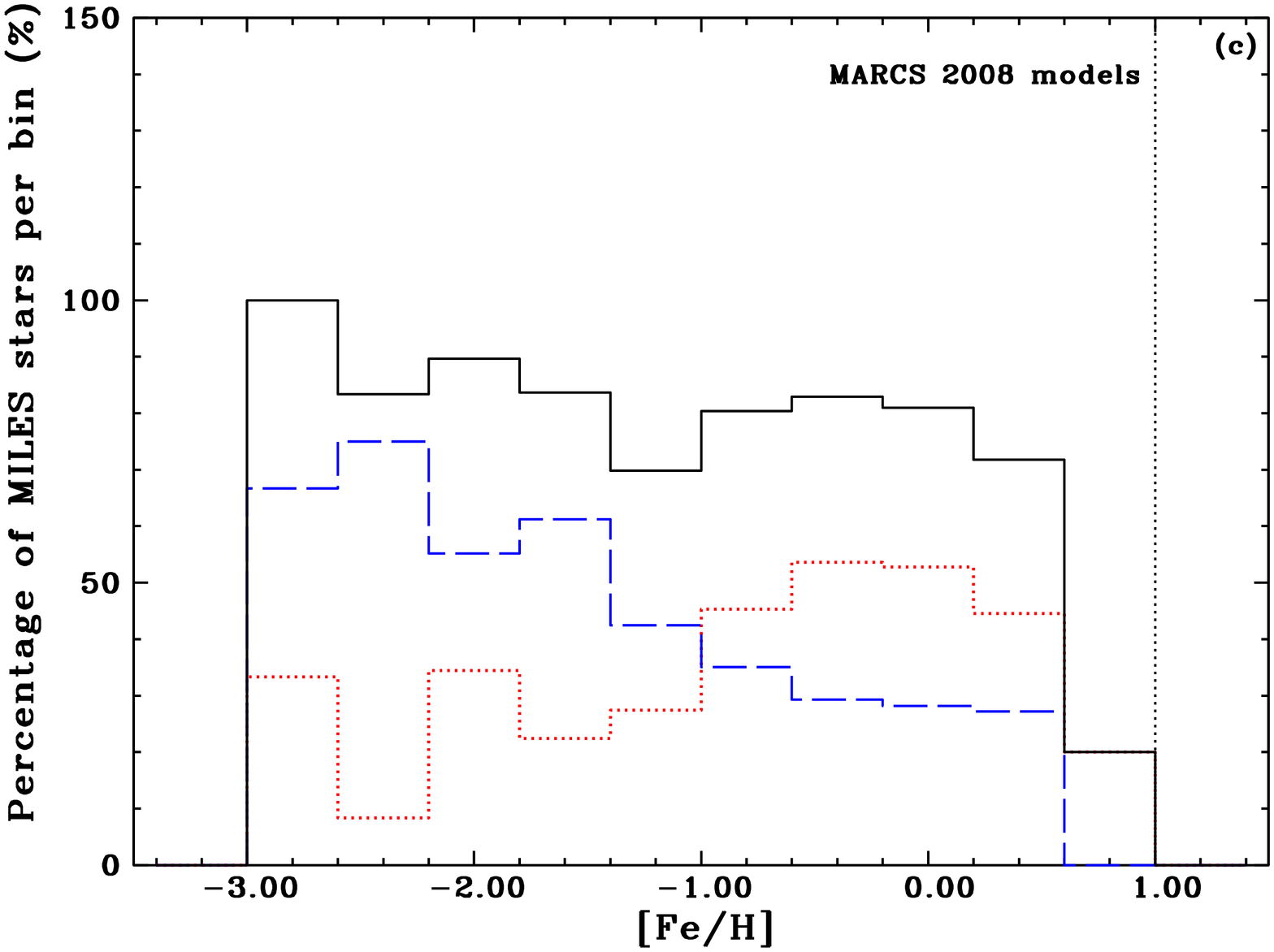} 
\end{center} 
\caption{ 
Distribution of the MILES stars with [Mg/Fe] over the photospheric parameter space:
{\bf (a)} effective temperature scale ({\bf top panel}),
{\bf (b)} surface gravity scale ({\bf middle panel}), and
{\bf (c)} metallicity escale ({\bf bottom panel}).
The percentage number per bin of the MILES stars with abundance ratios from HR studies is shown by blue dashed line,
and the percentage number of stars with our MR measurements by the red dotted line.
The sum of both is represented by the black solid line.
The adopted bins are equal to 4 times the parameter uncertainties.
The coverage of the MARCS 2008 grid of model atmospheres is also drawn in each panel.
} 
\label{Fig8} 
\end{figure}

\begin{figure} 
\begin{center} 
\includegraphics[width=65mm, angle=-90]{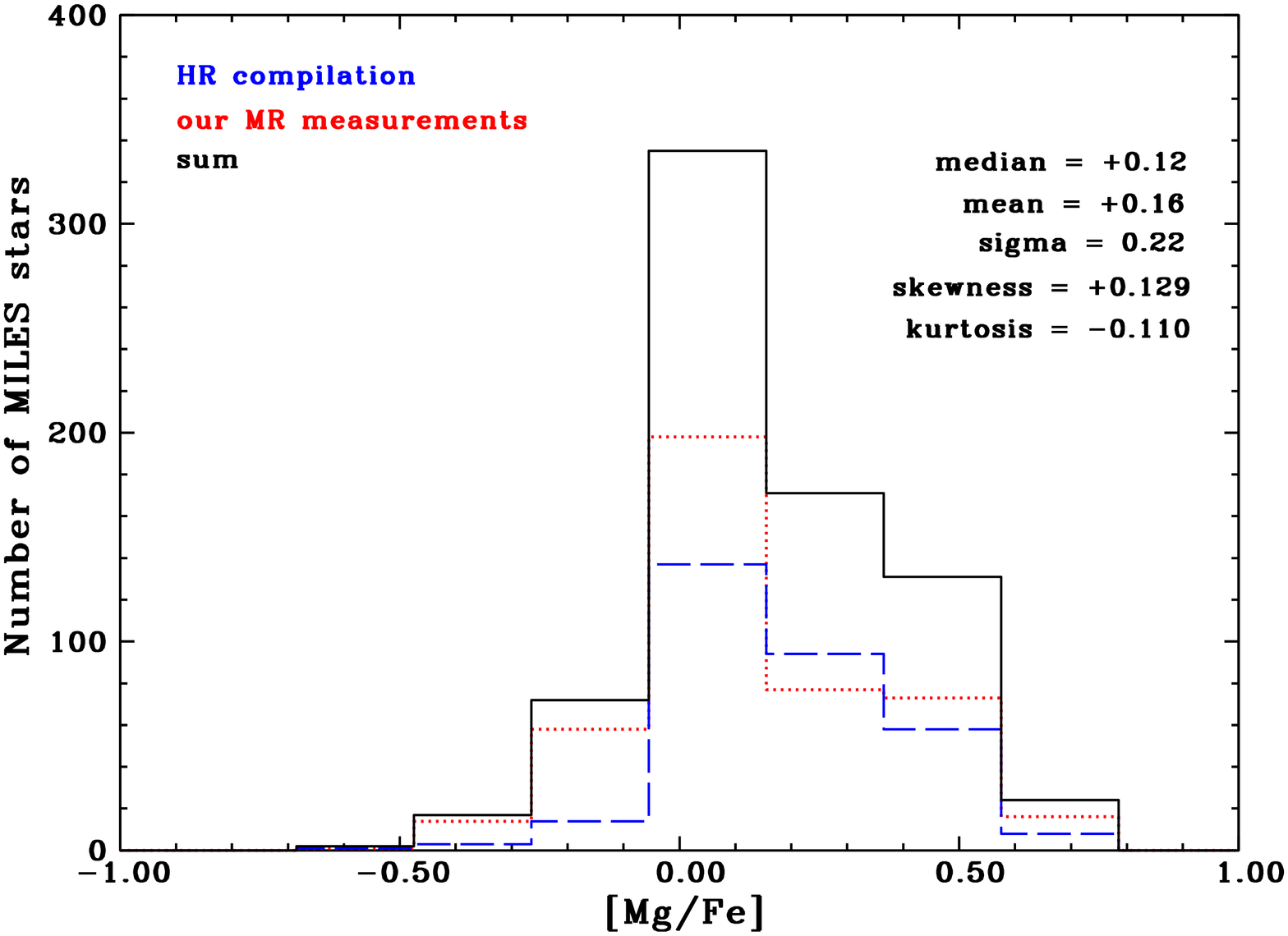} 
\end{center} 
\caption{ 
Distribution of abundance ratios over the MILES [Mg/Fe] catalogue built up in this work.
The histogram of stars with data from the HR compilation is shown by blue dashed line,
and the distribution of our MR measurements by the red dotted line.
The sum of both data sets is plotted by the black solid line.
The adopted bin is two times the average uncertainty of [Mg/Fe] over the whole catalogue
(2$\sigma$[Mg/Fe] = 0.21 dex).
The median, mean, sigma, skewness and kurtosis of the accumulated distribution are also informed.
} 
\label{Fig9} 
\end{figure}

Figure 7-(a) shows the stars' coverage over the modified H-R diagram log $g$ vs. T$_{\rm eff}$,
Figure 7-(b) explores the coverage around the projection [Fe/H] vs. T$_{\rm eff}$, and
Figure 7-(c) presents it on the plane log $g$ vs. [Fe/H].
The MILES stars with and without [Mg/Fe] are represented in all these plots,
in which the parametric extension of MARCS 2008 grid is also drawn.
The stars with [Mg/Fe] are distinguished according to the origin of their measurements.
Figure 8 exhibits the distribution of MILES [Mg/Fe] catalogue over the T$_{\rm eff}$, log $g$ and [Fe/H] scales.

The sum of all measurements (HR plus MR) show fairly flat distributions
across the photospheric parameter space covered by the MARCS models (histograms of Fig. 8).
The [Mg/Fe] HR data are distributed along the main sequence (MS) basically from T$_{\rm eff}$ about 4500 up to around 10000 K
and on the giant branch mainly from 4000 up to 5500 K, as seen in Fig. 7-(a).
Our MR measurements have a wide distribution over the plane log $g$ vs. T$_{\rm eff}$,
however there are some deficiencies such as in the low-MS, the red giant branch tip and the hottest giants. 
The histogram of T$_{\rm eff}$, Fig. 8-(a), shows a peak at 6200 K for the HR compilation data
while the MR measurements has a gradual increasing from 8000 to 4200 K.
There is a wide coverage over the whole MILES metallicity scale for both HR and MR data sets.
In the projection [Fe/H] vs. T$_{\rm eff}$, Fig. 7-(b),
we notice a predomination of the HR data between 5000 and 6000 K.
In the plane log $g$ vs. [Fe/H], Fig. 7-(c), the dwarfs are well covered by the HR data.
The MR distribution dominates in the metal-rich regime
and the HR data compilation dominates in the metal-poor regime,
complementing each other well (see also Figs. 9-(c) and 11).
Whilst the HR measurements provide mostly data for MILES dwarfs,
reaching the completeness maximum around log $g$ = 4.0 (Fig. 8-(b)),
our MR measurements contribute significantly to giants,
with a gradual decrease from log $g$ = 2.5 to 0.0.

Specifically for the mid-resolution measurements that were done within the limits of MARCS 2008 grid as shown in Figs. 8 and 9,
the highest completeness over the T$_{\rm eff}$ scale occurs around 4000 K whilst the smallest is at 6000 K. 
The maximum coverage in the gravity scale occurs at log $g$ = 2.5 and the minimum is around log $g$ = 4.0. 
The maximum of completeness over the metallicity scale is at $-$0.40 dex and the minimum occurs at [Fe/H] = $-$2.4 dex.

Figure 9 presents the MILES catalogue's [Mg/Fe] distribution itself, which is highly asymmetric around the solar ratio
showing a sharp decline towards negative values and a shallow decreasing towards over-enhanced ratios (skewness = +0.129).
The average of [Mg/Fe] is $+$0.16 dex having a standard deviation of 0.22 dex. The median of distribution is $+$0.12 dex.
The distribution of our MR measurements match well the HR data distribution (both sets on a same homogeneous scale).
The median and average of HR data are $+$0.16 dex and $+$0.19 dex (1$\sigma$ = 0.19 dex) respectively,
whilst they are $+$0.09 dex and $+$0.14 dex (1$\sigma$ = 0.24 dex) for the MR measurements.
The positive asymmetry also exists in both distributions:
the HR data have skewness equals to +0.081 and for the MR measurements it is +0.220.
The difference between them is that the HR data present a peaked distribution (kurtosis = +0.178)
whilst the MR values show a less peaked distribution (kurtosis = $-$0.279).
The MILES stars, therefore, now have [Mg/Fe] measurements covering a range
that is not restricted to the solar abundance ratio.

Figure 10 plots [Mg/Fe] as function of [Fe/H].
We can affirm that the MR measurements follow very well the solar neighbourhood global pattern of the HR data,
i.e. our determinations at medium spectral resolution statistically
recover with acceptable accuracy this abundance ratio along the whole metallicity scale.
At a given [Fe/H], the scatter of [Mg/Fe] is slightly larger
when it is measured by using Mg5183 than when Mg5528 or both features combined are adopted.
However, the systematic error on the [Mg/Fe] value is smaller when the Mg5183 feature is used
(see also Sect. 3.5).
In the MILES data set there is a mix of stars from different kinematic populations of our Galaxy
distributed over the thin and thick discs as well as the halo.
Several recent studies based on homogeneous HR spectroscopic analyses
have shown the intrinsic dispersion of [Mg/Fe] or [$\alpha$/Fe] at a fixed [Fe/H]
inside a particular disc population seems to be really small indeed
(e.g. Chen {\it et al.} 2000;
Mishenina {\it et al.} 2004;
Bensby {\it et al.} 2005;
Reddy, Lambert \& Allende Prieto 2006;
Bensby {\it et al.} 2010;
Nissen \& Schuster 2010)
but Neves {\it et al.} (2009) have found the opposite result.
On the other hand, halo stars ([Fe/H] $\leq$ $-$1.0 dex) exhibit great spread in the plane [Mg/Fe] vs. [Fe/H] 
(e.g. Stephens \& Ann Merchant 2002;
Borkova \& Marsakov 2005).
It is not the scope of the current work
to explore the details about the elemental abundances over the Galaxy's kinematic components.
These issues may be the central subject of a future work.

Figure 11 shows that there is good coverage of [Mg/Fe] over the T$_{\rm eff}$ and log $g$ scales
(uniformly from 4000 to 5500 K and nearly uniform along whole gravity scale),
with poorest completeness at the lowest and highest temperature ranges.
In addition there is a dearth of stars with sub-solar [Mg/Fe] around T$_{\rm eff}$ = 6000 K,
and for giant stars with low log g values.

Table 6 shows the stellar parameter coverage of the MILES [Mg/Fe] catalogue over the HR and MR data, and dwarfs and giants as well.
Table 7 summarizes the catalogued data presenting the number of HR and MR measurements
around dwarfs and giants together with their uncertainties.

\begin{table} 
\caption{
Stellar parameter coverage of the MILES [Mg/Fe] catalogue
for the high-resolution and mid-resolution data
distinguishing dwarfs (log $g$ $\geq$ 3.0) and giants (log $g$ $<$ 3.0).
} 
\label{catalogue_coverage} 
\begin{center} 
\begin{tabular}{@{}lcccc} 
\hline 
\hline 
  Class   &  T$_{\rm eff}$   &  log $g$  &  [Fe/H]      &  [Mg/Fe]      \\ 
\hline 
          &          (K)     &           &    (dex)     &     (dex)     \\ 
\hline 
\hline 
HR        &                  &           &              &               \\ 
\hline 
Dwarfs    & 4342,11704  &  3.0,4.80  & $-$2.53,$+$0.60 & $-$0.54,$+$0.74  \\ 
Giants    & 3902,6666   &  0.0,2.98  & $-$2.64,$+$0.10 & $-$0.21,$+$0.65  \\ 
\hline 
MR        &                  &           &              &               \\ 
\hline 
Dwarfs    & 3330,7972  &  3.0,5.08  & $-$2.86,$+$0.41 & $-$0.36,$+$0.73  \\ 
Giants    & 3600,7636  &  0.0,2.99  & $-$2.50,$+$0.64 & $-$0.47,$+$0.67  \\ 
\hline 
\hline 
\end{tabular} 
\end{center} 
\end{table}

\begin{table} 
\caption{
The distribution of HR and MR measurements around dwarfs (log $g$ $\geq$ 3.0) and giants (log $g$ $<$ 3.0)
in the MILES [Mg/Fe] catalogue.
The weighted average uncertainty $\sigma$[Mg/Fe] is also presented for each group of measurements.
}
\label{cat_summary} 
\begin{center} 
\begin{tabular}{@{}lrrrl}
\hline 
\hline 
Data source    & Dwarfs  & Giants  & Total  & $\sigma$[Mg/Fe]  \\ 
\hline 
               &         &         &        & dex              \\ 
\hline 
\hline 
HR             &    263  &     52  &   315  & 0.09             \\ 
\hline  
Mg5183         &     23  &     62  &    85  & 0.13             \\ 
Mg5528         &     31  &     69  &   100  & 0.15             \\ 
Both features  &     96  &    156  &   252  & 0.10             \\ 
\hline  
MR sum         &    150  &    287  &   437  & 0.12             \\ 
\hline  
Total sum      &    411  &    341  &   752  & 0.105             \\ 
\hline  
\hline 
\end{tabular} 
\end{center} 
\end{table}

\begin{figure*} 
\begin{center} 
\includegraphics[width=120mm, angle=-90]{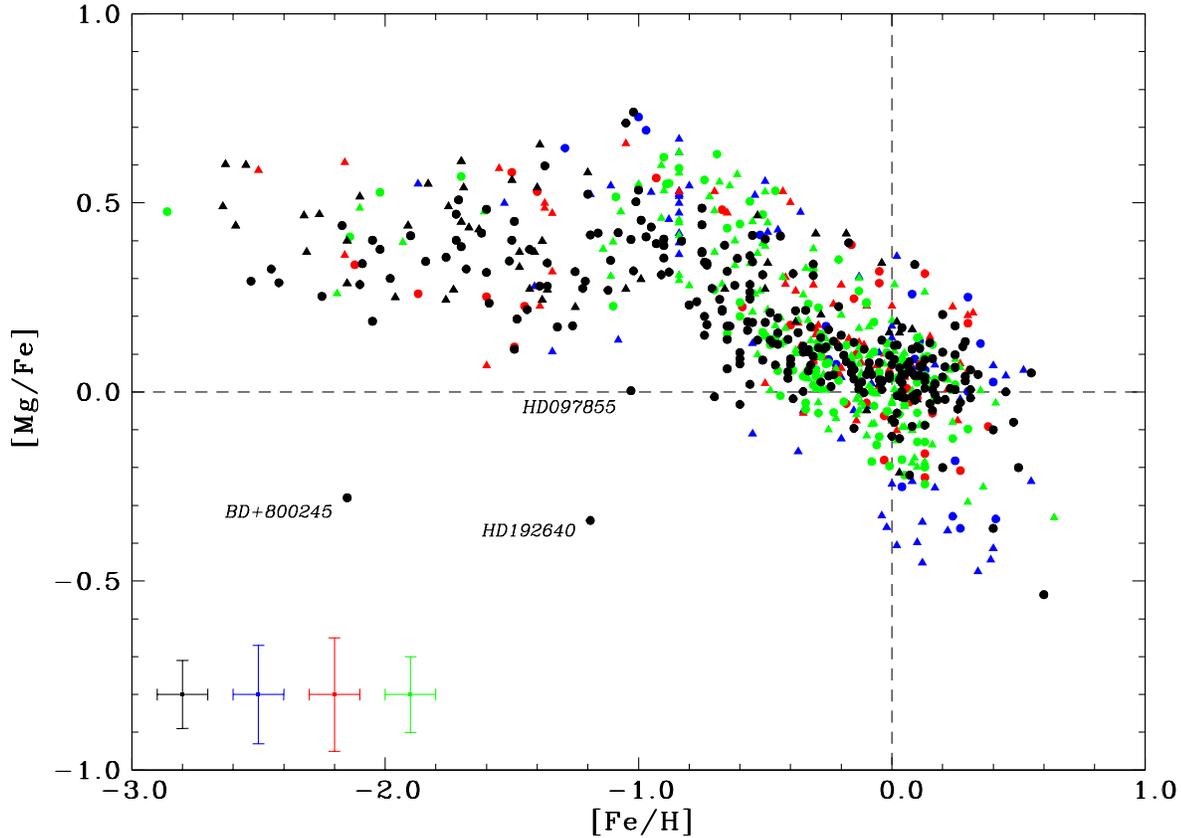} 
\end{center} 
\caption{ 
[Mg/Fe] as a function of [Fe/H] showing all MILES stars with [Mg/Fe]
from the HR compilation as black symbols and MR measurements as colour symbols
(blue designating determinations based on the Mg5183 feature only, red on Mg5528 only and green on both Mg features combined). 
Dwarfs (log $g$ $\leq$ 3.0) are shown as circles and giants as triangles.
The weighted average uncertainties for each data group (HR, Mg5183, Mg5528 and both features combined)
are illustrated on the bottom left corner. 
Three chemically peculiar stars are also identified.
}
\label{Fig10} 
\end{figure*}

\begin{figure} 
\begin{center} 
\includegraphics[width=65mm, angle=-90]{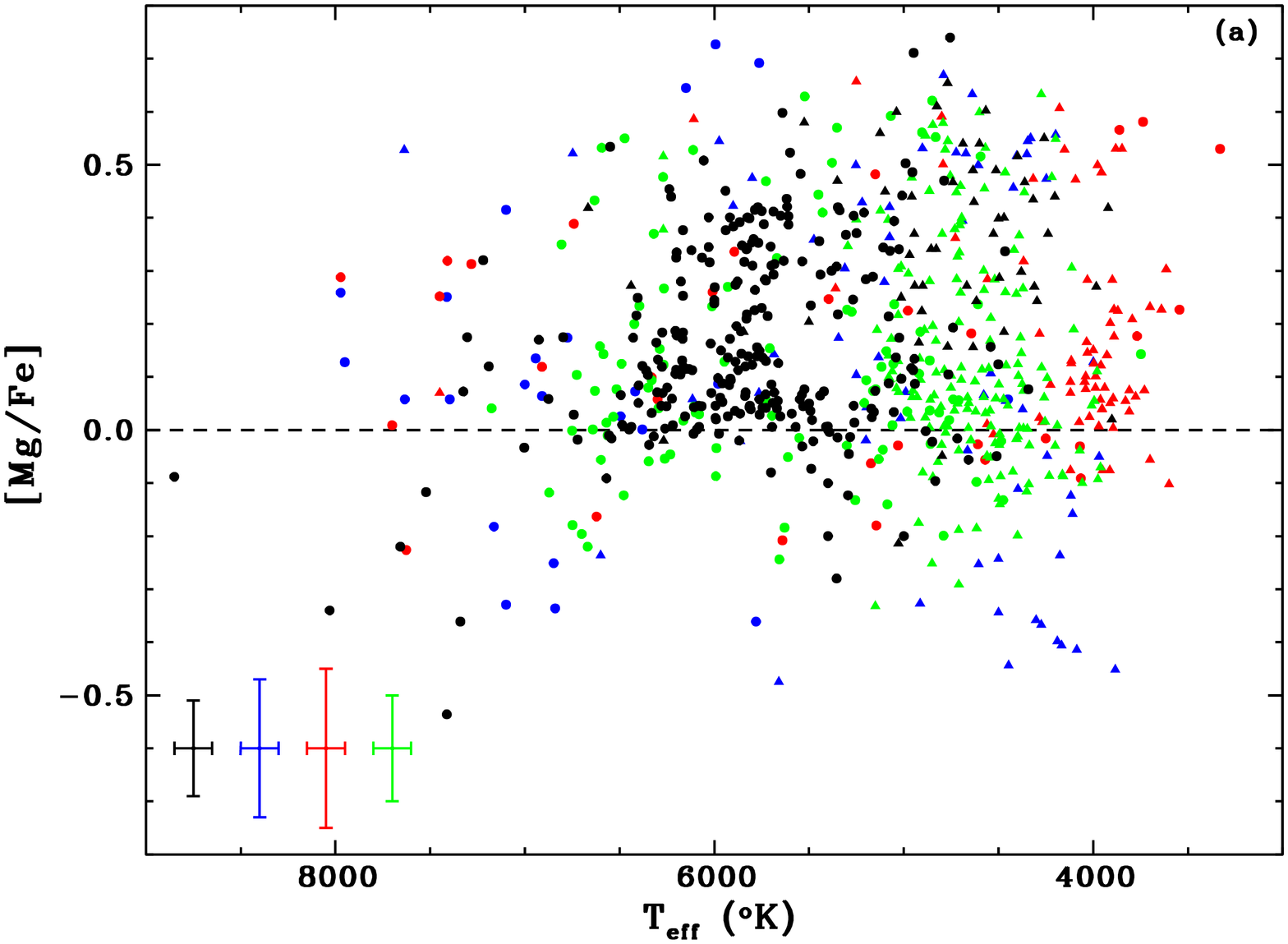} 
\includegraphics[width=65mm, angle=-90]{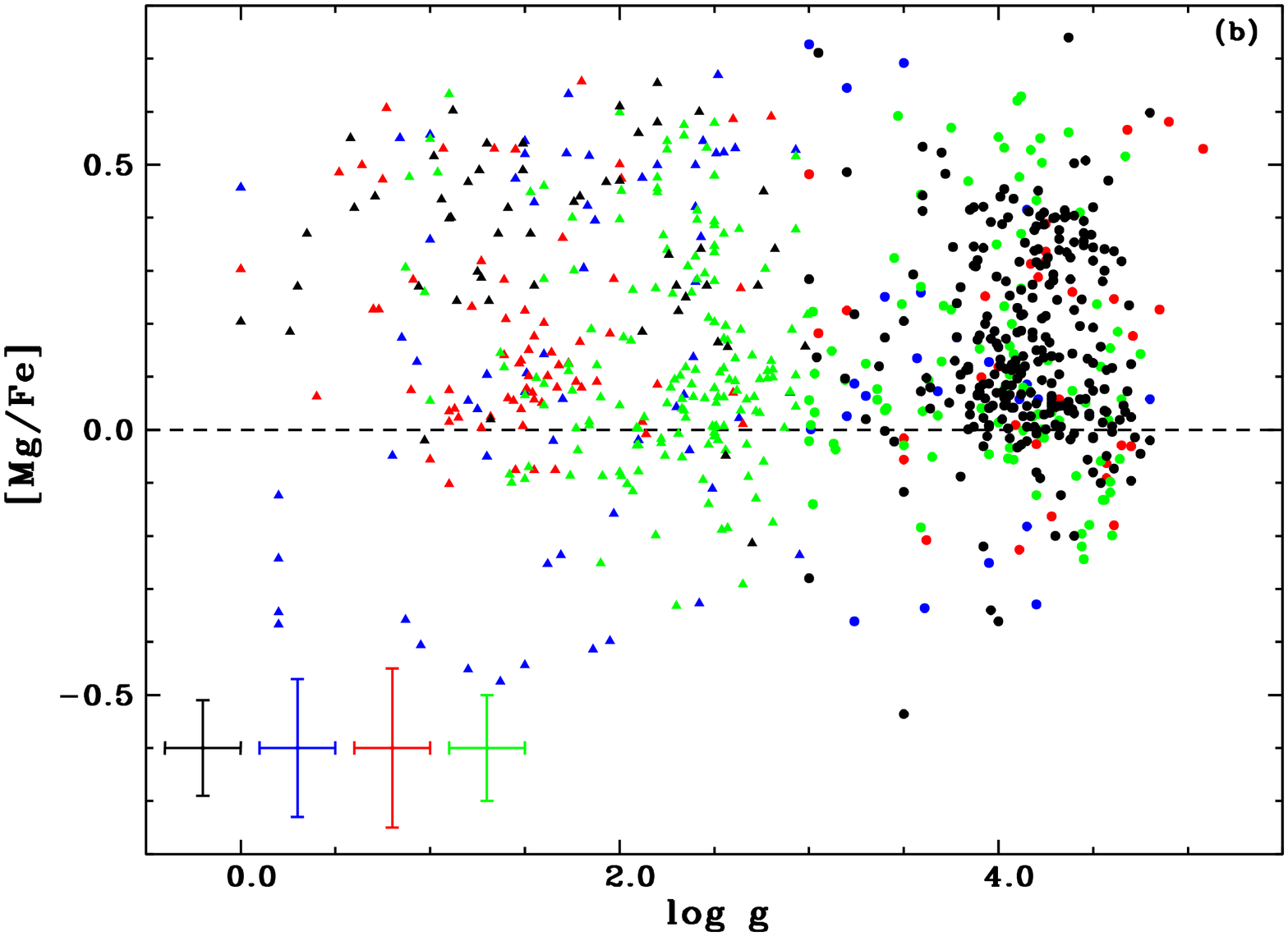} 
\end{center} 
\caption{
Distribution of [Mg/Fe] in our catalogue over the scales of
effective temperature (truncated at 9000 K) in the {\bf top panel (a)},
and surface gravity in the {\bf bottom panel (b)}.
The notations are those adopted in Fig. 10.
} 
\label{Fig11} 
\end{figure}

\section[5]{Comparison with theoretical model predictions} 

In previous stellar population studies, there have been attempts to account for variations in element abundance ratios
by the use of response functions that give the dependence of line-strength indices on just a single chemical element.
These response functions are calculated by adopting model atmospheres of, usually, just 3 or 4 different stars.
Examples include the models of
Weiss, Peletier \& Matteucci (1995),
Tripicco \& Bell (1995),
Korn, Maraston \& Thomas (2005) - hereafter K05,
Coelho {\it et al.} (2005), and
Serven, Worthey \& Briley (2005)
([$\alpha$/Fe] = 0.0 and +0.3 models, effects
on many spectral lines of many elements tested individually). 
Such models have been used in various studies to interpret the spectra of galaxies and globular clusters
(e.g. 
Vazdekis {\it et al.} 1997,
Trager {\it et al.} 2000b,
Proctor \& Sansom 2002,
Denicol\'o {\it et al.} 2005,
Lee \& Worthey 2005,
Schiavon 2007,
Coelho {\it et al.} 2007,
Pipino {\it et al.} 2009b, and
Smith, Lucey \& Hudson 2009).
The accuracy of those response functions, however, have not been calibrated or tested empirically.

In the present paper, variations in [$\alpha$/Fe] are characterised by measuring Mg abundances in {\it real stars}
and using this to represent $\alpha$-elements in general.
In this section, we compare the effects on Lick indices of these [$\alpha$/Fe] ratios derived from observations
with theoretical predictions for how Lick indices are expected to change with variations in [$\alpha$/Fe].
The models of K05 are used for this comparison, since they were used in many of the above referenced studies of galaxies.
Equation 7 from 
Thomas, Maraston \& Bender (2003)
shows how we can predict changes in spectral line indices with changing composition for lines
that tend to zero strength as the element abundance dominating that line tends to zero.
Other line indices (e.g. H$\gamma$, H$\delta$, G4300, Fe4383) take both negative and positive values, 
due to lack of a real continuum definition in complex star spectra.
This particularly affects indices in the blue part of the spectrum where the continuum level changes
rapidly with wavelength in long-lived, late-type stars.
For these indices we adopt the formalism given in equation 3 of K05,
which modifies fluxes rather than line strengths.    
For molecular bands and negative going lines differences in indices are compared,
whereas for positive absorption lines ratios are used for comparisons. 

The catalogue of MILES atmospheric parameters
(Cenarro {\it et al.} 2007)
was cross-correlated with the measured [Mg/Fe] cases and with the models of K05
to find samples of stars useful for comparing observations with models, in order to test their agreement.
There are then 31 stars in the MILES library whose surface temperature and 
gravity are the same as that of the turn-off star model (T$_{\rm eff}$ = 6200 K, log $g$ = 4.1) 
given in Table 13 of K05, within observational errors  ($\Delta$T$_{\rm eff}$ = $\pm$100 K, $\Delta$log$g$ = $\pm$0.2). 
Amongst these 31 stars are two that also have the same chemical composition as the model star ([Fe/H]=0, [$\alpha$/Fe]=0),
within observational errors ($\Delta$[Fe/H] = $\pm$0.1 dex, $\Delta$[$\alpha$/Fe] = $\pm$0.06 dex).
This allows us to use one of these two stars as a base with which to normalise the other stars
in order to show how changes in chemistry affect changes in indices, in a relative way.
Model stars can then be generated to match the 31 MILES stars and normalised by the model given in Table 13 of K05.
To generate the model star indices the equations were first applied to correct 
to a specific overall metallicity (using column 14 of K05),
then the equations were applied again to correct to a specific [$\alpha$/Fe] ratio,
modifying for all the $\alpha$-elements modelled by K05. 
Similarly, 13 cool giant stars and 7 cool dwarfs can be compared using Tables 14 and 12 from K05 respectively,
for solar composition models.
In this way we can compare normalised observations with normalised model predictions to see
if the observations agree with previously used methods of varying [$\alpha$/Fe] ratios in stellar population studies.

Examples of these comparisons are shown in 
Fig. 12
for Fe and Mg sensitive indices.
Fe sensitive features in general behave as expected in that the observed changes agree well with the predicted ones.
Mg sensitive features also show quite a good one-to-one agreement, but with scatter in excess of that expected from the observational errors.
There is some suggestion of a slight systematic deviation below the one-to-one line in the case
if Mg~b in turn-off and cool-dwarf stars.
These deviations will be explored in future work.
In general we see from 
Fig. 12
that the observed [Mg/Fe] abundances reported in this paper show the trends expected for Mg and Fe sensitive features,
when compared to models from K05. 
Other indices and comparisons with models will be discussed more extensively in a future paper.

\begin{figure*}
\begin{center} 
\includegraphics[width=115mm, angle=-90]{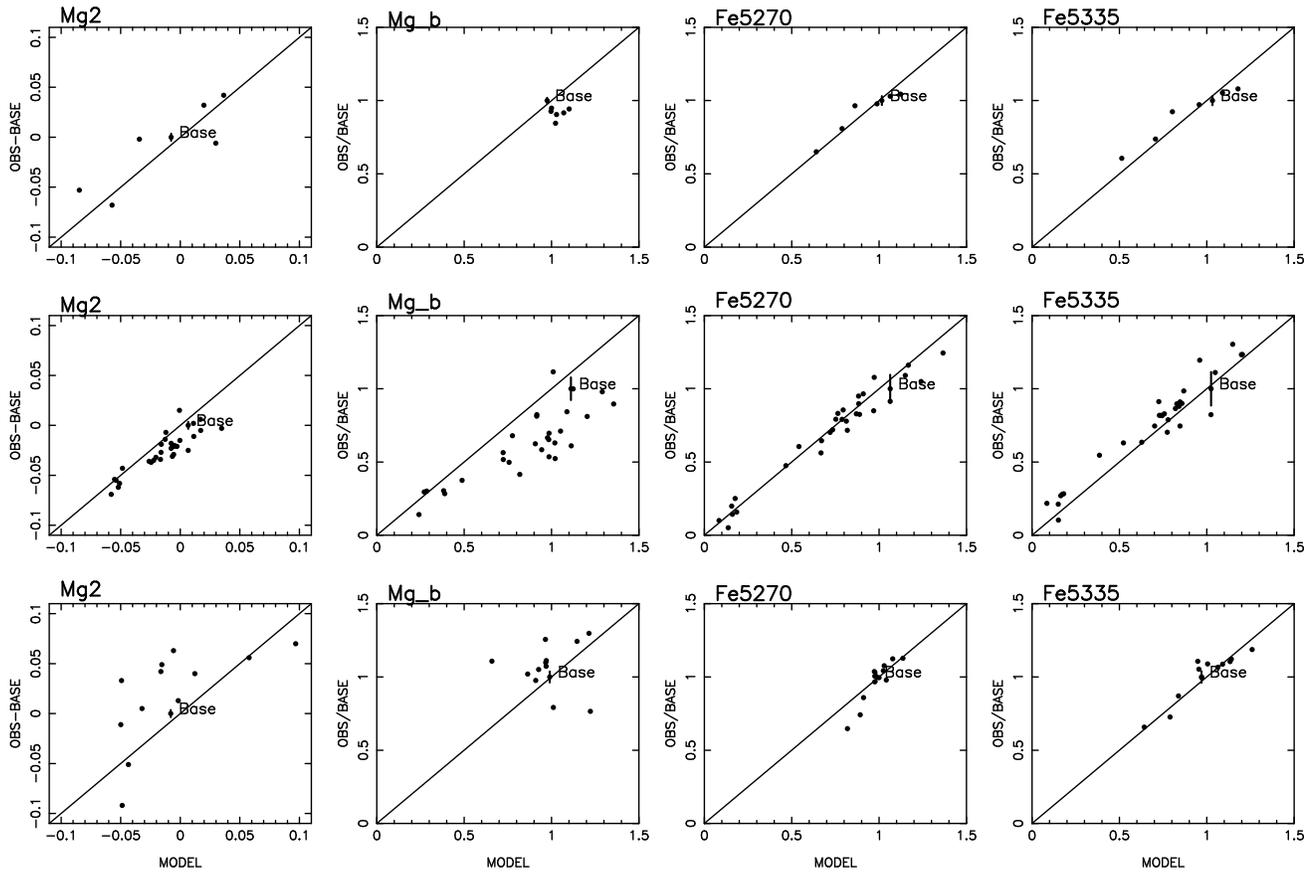}
\end{center} 
\caption{
Normalised observed versus normalised model indices for different stars
in the MILES library. The normalisation of the observations is achieved
using MILES stars with the same photospheric parameters as for the models
of K05 within errors. These normalising stars are labelled with 'base' in
the plots.
{\bf The top row}
shows MILES stars plotted against cool dwarf star models derived from
Table 12 of K05;
the {\bf central row} shows MILES stars plotted against
turn-off star models derived from Table 13 of K05 and
the {\bf lower row}
shows MILES stars plotted against cool giant star models derived from
Table 14 of K05. The models are normalised by the values given in tables
12 to 14 of K05. Models are normalised in the same way as the data,
as differences or ratios (see Sect. 5 for a description).
Stars in a given plot all have the same effective temperature and
surface gravity as that of the relevant model star, within the errors.
The chemistry ([Fe/H] and [$\alpha$/Fe]) is varied.
}
\label{Fig12}
\end{figure*}

\section[6]{Further applications to the analysis of stellar populations} 

From now on, it will be possible to build up new semi-empirical simple stellar population (SSP) models
by adopting the MILES star spectrum library
in order to more confidentially cover a range of values of metallicity and magnesium-to-iron abundance ratio
for some ages greater than 1 up to 14 Gyrs.
This is one of the important further applications based on the results of this work
that is scheduled by our group which has experience in studying stellar populations.
The MILES stars can be now selectively collected from the library
taking into account their characteristics in a more extensive parameter space,
i.e. [Fe/H], [Mg/Fe], log $g$ and T$_{\rm eff}$,
to represent different SSPs for given sets of age, [Fe/H] and [$\alpha$/Fe].
Magnesium may be considered a proxy of the $\alpha$-elements and
consequently [$\alpha$/Fe] might be represented by [Mg/Fe].

There are some caveats to be aware of for using the present catalogue results.
The measurements made and compiled in the present catalogue represent [Mg/Fe] well,
as our tests of the spectral measurements show, considering all $\alpha$-element or only Mg variations.
However, it is important to be aware that not all $\alpha$-elements may behave in exactly the same way in different populations.
This needs to be considered when applying the present catalogue to stellar population studies.
Other caveats are that [Mg/Fe] obtained through the MR calibrations applied for stars
whose parameters lay outside the control sample coverage might be more uncertain than the other determinations,
and that we warn the abundance ratios derived from the $\alpha$-enhancement model atmosphere extrapolations
should be used with certain precaution too.
Caveats aside, making the approximation that [Mg/Fe] can be used to represent [$\alpha$/Fe]
is a significant improvement over the scaled solar assumption only.
There is a great deal of interest in uncovering the information contained in non-solar abundance ratios.
Therefore we next illustrate how the catalogue may be used to generate SSPs
with empirically determined non-solar [$\alpha$/Fe] abundance ratios.

In this further step of our work, it will be necessary to take into account reliable cross-matching between
theoretical isochrones for non-solar ratios and real stars.
Basically, the dwarf and giant stars with known Mg/Fe ratio that are collected from MILES
in order to represent a given simple stellar population must be selected to precisely have on average the SSP's [Fe/H] and [Mg/Fe].
Moreover, the selected stars must be well sampled along the main evolutionary stages of an isochrone.
The stars must be sufficiently close to the isochrones taking into account the errors in log $g$ and T$_{\rm eff}$
in order to be included and weighted in the computation of SSP integrated colours and spectra.
It will be also necessary to generate additional stellar spectra for some non-completely-represented stages
by interpolating the MILES spectra in its four-dimensional parameter space.
Other approaches can be employed to extend the coverage of the SSP sets (age, [Fe/H], [$\alpha$/Fe])
such as evaluating the behaviour of integrated observational properties of the SSPs at specified metallicities
as a function of [$\alpha$/Fe] to computing and applying corrections to the SSP observables for a larger range of parameters.
Therefore we will be able to construct a large set of single-age single-metallicity single-$\alpha$-enhanced
stellar population models.
This will open new prospects for SSP modelling and evolutionary population synthesis.

Table 8 presents a set of age, [Fe/H] and [$\alpha$/Fe] combinations (36 in number)
for which there are sufficient number of dwarfs and giants in our MILES [Mg/Fe] catalogue
to be more consistently selected to construct SSPs by adopting scaled solar and $\alpha$-enhanced isochrones
with those properties.
Basically, in this approach, the selection of MILES stars for each metallicity value is done taking into account 1$\sigma$ or 2$\sigma$ variation
depending on the star's position in the H-R diagram.
The stars are also collected having [Mg/Fe] values around 1$\sigma$[Mg/Fe] and 2$\sigma$[Mg/Fe] of each isochrone's $\alpha$-enhancement.
The weighted uncertainty of the abundance ratio in the whole MILES [Mg/Fe] catalogue is 0.105 dex (1$\sigma$)
and $\sigma$[Fe/H] = 0.10 dex in MILES.
We also consider variations in the isochrones' ages,
which were estimated basically through plots of isochrones with distinct ages that are not shown in this section.

The Dartmouth isochrone models
(Dotter {\it et al.} 2008)
have been adopted as reference to match the stars' positions with isochrones in the HR diagram log $g$ versus T$_{\rm eff}$.
The Dartmouth models is a collection of scaled solar and $\alpha$-enhanced isochrones
that spans a range of [Fe/H] from $-$2.5 to +0.5 dex, [$\alpha$/Fe] from $-$0.2 to +0.8 dex (for [Fe/H] $\leq$ 0.0 dex)
or $-$0.2 to +0.2 dex (for [Fe/H] $>$ 0.0 dex), with 0.2 dex steps, and initial helium mass fractions from Y = 0.245 to 0.40.
Their stellar evolution tracks were computed for masses from 0.1 to 4 M$_{\odot}$,
allowing isochrones to be generated for ages as young as 250 Myr up to as old as 15 Gyr.

Figure 13 shows examples of isochrone-based plots by using Dartmouth
(Dotter {\it et al.} 2008)
and BaSTI
(Pietrinferni {\it et al.} 2004) models for 4 Gyr, [Fe/H] around zero and three distinct $\alpha$-enhancements.

The BaSTI scaled solar database
(Pietrinferni {\it et al.} 2004)
covers stellar evolution models for masses between 0.5 and 10 M$_{\odot}$ in a wide metallicity range
(10 values of [Fe/H] from $-$2.27 to +0.40 dex).
The initial He mass fraction ranges from Y = 0.245, for the more metal-poor composition, up to 0.303 for the more metal-rich one.
For each adopted chemical composition, the evolutionary models were computed without (called canonical models)
and with overshooting from the Schwarzschild boundary of the convective cores during the central H-burning phase.
The stellar models are used to compute isochrones in a wide age range, from 30 Myr up to 15 Gyr.
The overshooting models provide a better match to the observations at [Fe/H] around solar, and for ages equal and higher than 4 Gyr.
Besides these models, BaSTI $\alpha$-enhanced models were computed
for [$\alpha$/Fe] fixed at +0.4 dex and 11 values of iron metallicity between $-$2.62 and +0.05 dex
(Pietrinferni {\it et al.} 2006).

Figure 14 presents a good example of cross-matching between a group of MILES stars and an isochrone
to proceed to a reliable semi-empirical modelling of a single-age single-metallicity single-$\alpha$-enhanced stellar population.
The isochrone-based plot of this figure on the log $g$ vs. T$_{\rm eff}$ plane
is done for 6 Gyr, [Fe/H] = $-$0.4 dex and [$\alpha$/Fe] = $+$0.2 dex based on a Dartmouth theoretical model. 
The MILES stars have been carefully chosen with metallicity and [Mg/Fe] around the isochrone values within 1$\sigma$.
Furthermore, the restrictive matching of the stars' positions with the isochrone has considered
the MILES errors in log $g$ and T$_{\rm eff}$ within 1$\sigma$ too.
We have also computed the averages of [Fe/H] and [Mg/Fe] for the selected stars
to check if there is agreement with the model's values.
In this cross-matching SSP-MILES, 9 dwarfs (log $g$ $\geq$ 3.7) and 21 giants have been selected
whose averages [Fe/H] and [Mg/Fe] are $-$0.40 and +0.19 dex respectively,
in excellent accordance to the SSP's parameters
(the standard deviations are $\sigma$[Fe/H] = 0.07 dex and $\sigma$[Mg/Fe] = 0.06 dex).
The stars are roughly well sampled along the isochrone although there are some empty places
that might be filled through interpolations applied in the library parameter space.
This procedure must make the stars' distribution more uniform around the model chemistry
shown for instance in the plane [Fe/H] vs. [Mg/Fe],
refining in this way the computation of the SSP observables.

We have also begun a study of the dependence of absorption line indices on [Mg/Fe],
focusing on the observed behaviour of some indices of the Lick System as a function of the photospheric parameters, 
as described in Sect. 5. 
We intend to compute semi-empirical fitting functions for the main Lick indices 
in order to improve and extend the SSP models for different $\alpha$-enhancements.

The predictions of new semi-empirical SSP models will be compared with the observables
of distinct composite stellar systems such as globular clusters, dwarf galaxies, ellipticals and spiral bulges.
Consequently, the models will be very useful to understand their star formation histories and chemical evolutions.

Moreover, we can test and apply the same approach of this work 
to obtaining the calcium abundances for the MILES stars. 
Indeed, there are interesting questions about how the calcium-enhancement
behaves in several composite stellar systems relative to other $\alpha$-elements like magnesium.
For instance, 
Smith {\it et al.} (2009)
found for a sample of 147 red-sequence galaxies from the Coma cluster and the Shapley Supercluster
that the [Ca/Fe] ratio is positively correlated with the velocity dispersion, at fixed [Fe/H],
however its dependence is significantly less steep than that of [Mg/Fe].
On the other hand,
Pipino  {\it et al.} (2009a)
obtained that the [Ca/Fe]-mass relation is naturally explained by such a standard galactic chemical evolution model,
and explained that the observed under-abundance of Ca with respect to Mg
can be attributed to the different contributions from Type Ia and Type II supernovae to the nucleosynthesis of these two elements. 

Additional applications of the present [Mg/Fe] catalogue will 
potentially improve areas of our understanding of stellar atmospheres
and spectral flux distributions from different types of stars present
in the MILES library.

\begin{table} 
\caption{
Sets of age, metallicity [Fe/H] and [Mg/Fe] (representing $\alpha$-enhancement)
for which there are reasonable number of dwarf and giant stars in MILES
to build up semi-empirical SSP models.
The star counts are shown for the main sequence (MS), main sequence turn-off (TO),
sub-giants (SG) and red giant branch (RGB) stages.
The Dartmouth isochrones
(Dotter {\it et al.} 2008)
have been adopted as reference for this purpose.
The stars have been selected in the [Fe/H]-[Mg/Fe] parameter space
assuming specified ranges around these quantities,
whose criterion of choice was such that the sum of stars in 
the MS + TO stages $\geq$ 5 and in the SG + RGB stages $\geq$ 10.  
The weighted average $\sigma$[Mg/Fe] is 0.105 dex in our catalogue
and $\sigma$[Fe/H] is 0.10 dex in MILES.
Isochrone-based plots for the combinations assigned by asterisks are presented in Figs. 13 and 14.
}
\begin{center} 
\begin{tabular}{@{}rrrrrrr} 
\hline 
\hline 
Age        & [Fe/H]         & [Mg/Fe]              &  MS  &  TO  &  SG  &  RGB  \\
\hline 
(Gyr)      &  (dex)         &   (dex)              &      &      &      &       \\
\hline 
\hline 
 1$\pm$2   & $+$0.4$\pm$0.2 & $-$0.2$\pm$2$\sigma$ &    6 &    7 &    3 &     7 \\ 
 1$\pm$2   & $+$0.4$\pm$0.2 &    0.0$\pm$2$\sigma$ &   13 &    5 &    3 &    10 \\ 
\hline
 1$\pm$2   & $+$0.5$\pm$0.2 &    0.0$\pm$2$\sigma$ &    8 &    3 &    3 &      7 \\ 
\hline 
 2$\pm$2   &    0.0$\pm$0.1 & $-$0.2$\pm$1$\sigma$ &    5 &    4 &    1 &    13 \\ 
 2$\pm$2   &    0.0$\pm$0.1 &    0.0$\pm$1$\sigma$ &   24 &    9 &    4 &    35 \\ 
 2$\pm$2   &    0.0$\pm$0.1 & $+$0.2$\pm$1$\sigma$ &    3 &    3 &    3 &    15 \\ 
 2$\pm$2   &    0.0$\pm$0.2 & $+$0.4$\pm$2$\sigma$ &    3 &    3 &    1 &    13 \\ 
\hline
2$\pm$2   & $+$0.2$\pm$0.1 & $-$0.2$\pm$2$\sigma$ &    9 &    4 &    2 &    14 \\ 
2$\pm$2   & $+$0.2$\pm$0.1 &    0.0$\pm$1$\sigma$ &   12 &    8 &    3 &    15 \\ 
2$\pm$2   & $+$0.2$\pm$0.1 & $+$0.2$\pm$2$\sigma$ &    8 &    7 &    4 &    16 \\ 
\hline 
 4$\pm$2   & $-$0.2$\pm$0.1 & $-$0.2$\pm$2$\sigma$ &    4 &    4 &    0 &    17 \\ 
 4$\pm$2   & $-$0.2$\pm$0.1 &    0.0$\pm$1$\sigma$ &   11 &    8 &    1 &    30 \\ 
 4$\pm$2   & $-$0.2$\pm$0.1 & $+$0.2$\pm$1$\sigma$ &    4 &    7 &    0 &    16 \\ 
 4$\pm$2   & $-$0.2$\pm$0.2 & $+$0.4$\pm$2$\sigma$ &    5 &    2 &    1 &    22 \\ 
\hline 
$\ast$~4$\pm$2   &    0.0$\pm$0.1 & $-$0.2$\pm$2$\sigma$ &    7 &    2 &    3 &    33 \\ 
$\ast$~4$\pm$2   &    0.0$\pm$0.1 &    0.0$\pm$1$\sigma$ &   19 &    7 &    6 &    44 \\ 
$\ast$~4$\pm$2   &    0.0$\pm$0.1 & $+$0.2$\pm$2$\sigma$ &    8 &    3 &    4 &    32 \\ 
\hline 
~6$\pm$2 & $-$0.4$\pm$0.1 &    0.0$\pm$1$\sigma$ &    3 &    4 &    1 &     9 \\ 
$\ast$~6$\pm$2 & $-$0.4$\pm$0.1 & $+$0.2$\pm$1$\sigma$ &    4 &    3 &    2 &    21 \\ 
~6$\pm$2 & $-$0.4$\pm$0.1 & $+$0.4$\pm$2$\sigma$ &    4 &    1 &    2 &    16 \\

\hline 
 8$\pm$2   & $-$0.6$\pm$0.2 &    0.0$\pm$2$\sigma$ &    4 &    5 &    2 &    14 \\ 
 8$\pm$2   & $-$0.6$\pm$0.1 & $+$0.2$\pm$1$\sigma$ &    3 &    5 &    2 &     9 \\ 
 8$\pm$2   & $-$0.6$\pm$0.1 & $+$0.4$\pm$1$\sigma$ &    2 &    4 &    1 &    13 \\ 
\hline
10$\pm$2   & $-$0.8$\pm$0.1 & $+$0.4$\pm$2$\sigma$ &    4 &    4 &    2 &    14 \\ 
\hline 
10$\pm$2   & $-$0.6$\pm$0.2 &    0.0$\pm$2$\sigma$ &    3 &    4 &    2 &    14 \\ 
10$\pm$2   & $-$0.6$\pm$0.1 & $+$0.2$\pm$1$\sigma$ &    3 &    5 &    2 &     9 \\ 
10$\pm$2   & $-$0.6$\pm$0.1 & $+$0.4$\pm$1$\sigma$ &    2 &    3 &    2 &    10 \\ 
\hline 
12$\pm$2   & $-$2.0$\pm$0.2 & $+$0.4$\pm$2$\sigma$ &    4 &    2 &    2 &    11 \\ 
\hline 
12$\pm$2   & $-$1.0$\pm$0.2 & $+$0.4$\pm$2$\sigma$ &    5 &    8 &    4 &    20 \\ 
12$\pm$2   & $-$1.0$\pm$0.2 & $+$0.6$\pm$2$\sigma$ &    4 &    1 &    1 &    15 \\ 
\hline 
14$\pm$1   & $-$2.0$\pm$0.2 & $+$0.4$\pm$2$\sigma$ &    2 &    7 &    2 &    11 \\ 
\hline
14$\pm$1   & $-$1.8$\pm$0.2 & $+$0.4$\pm$2$\sigma$ &    3 &    6 &    3 &    11 \\ 
\hline
14$\pm$1   & $-$1.6$\pm$0.2 & $+$0.4$\pm$2$\sigma$ &    5 &    8 &    4 &    13 \\ 
\hline 
14$\pm$1   & $-$1.4$\pm$0.1 & $+$0.2$\pm$2$\sigma$ &    3 &    2 &    2 &    12 \\ 
14$\pm$1   & $-$1.4$\pm$0.1 & $+$0.4$\pm$2$\sigma$ &    4 &    4 &    1 &    13 \\ 
\hline 
14$\pm$1   & $-$1.2$\pm$0.2 & $+$0.2$\pm$2$\sigma$ &    2 &    5 &    1 &    12 \\
\hline 
\end{tabular} 
\end{center} 
\label{SSPs} 
\end{table}

\begin{figure} 
\begin{center} 
\includegraphics[width=64mm, angle=-90]{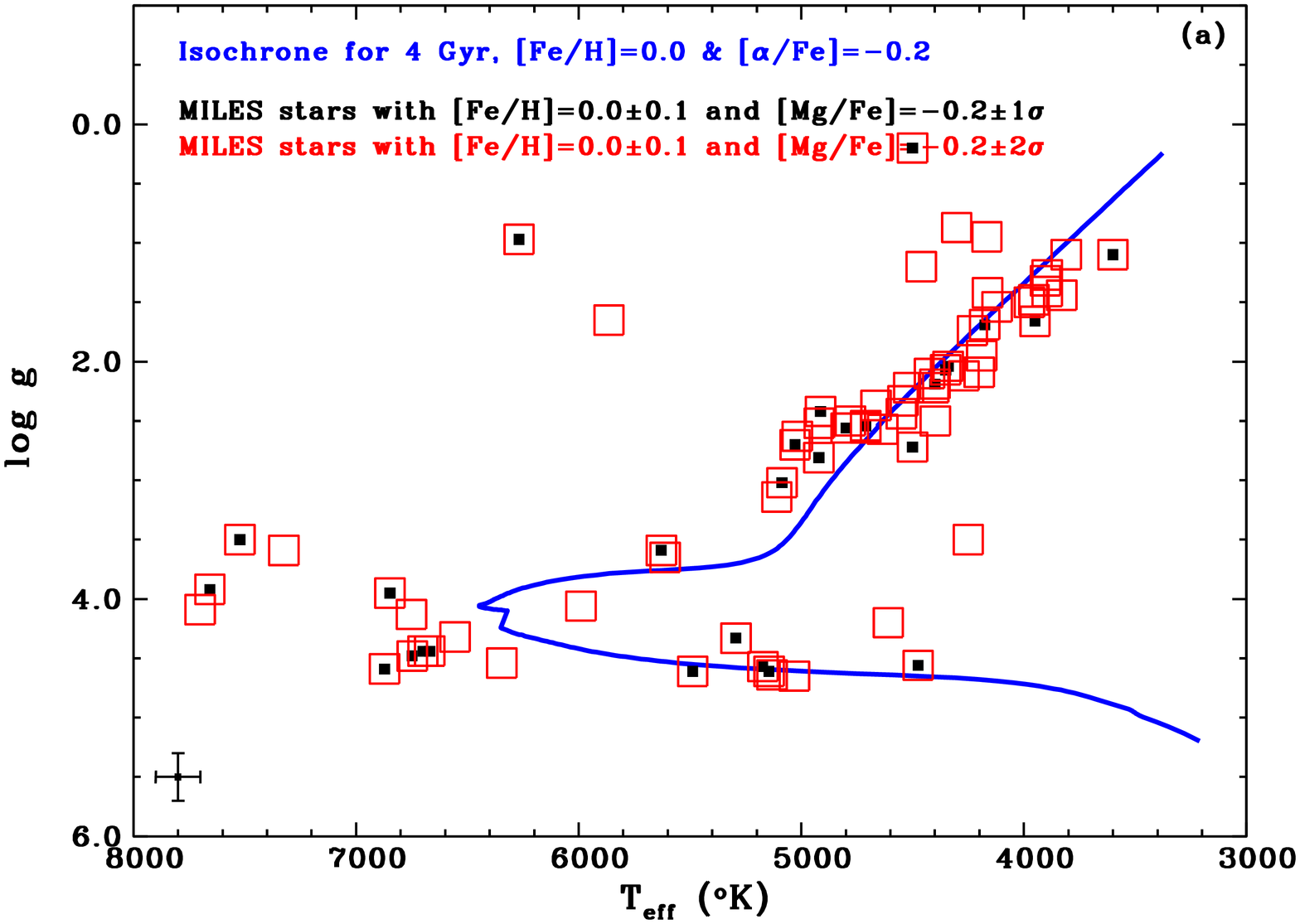} 
\includegraphics[width=64mm, angle=-90]{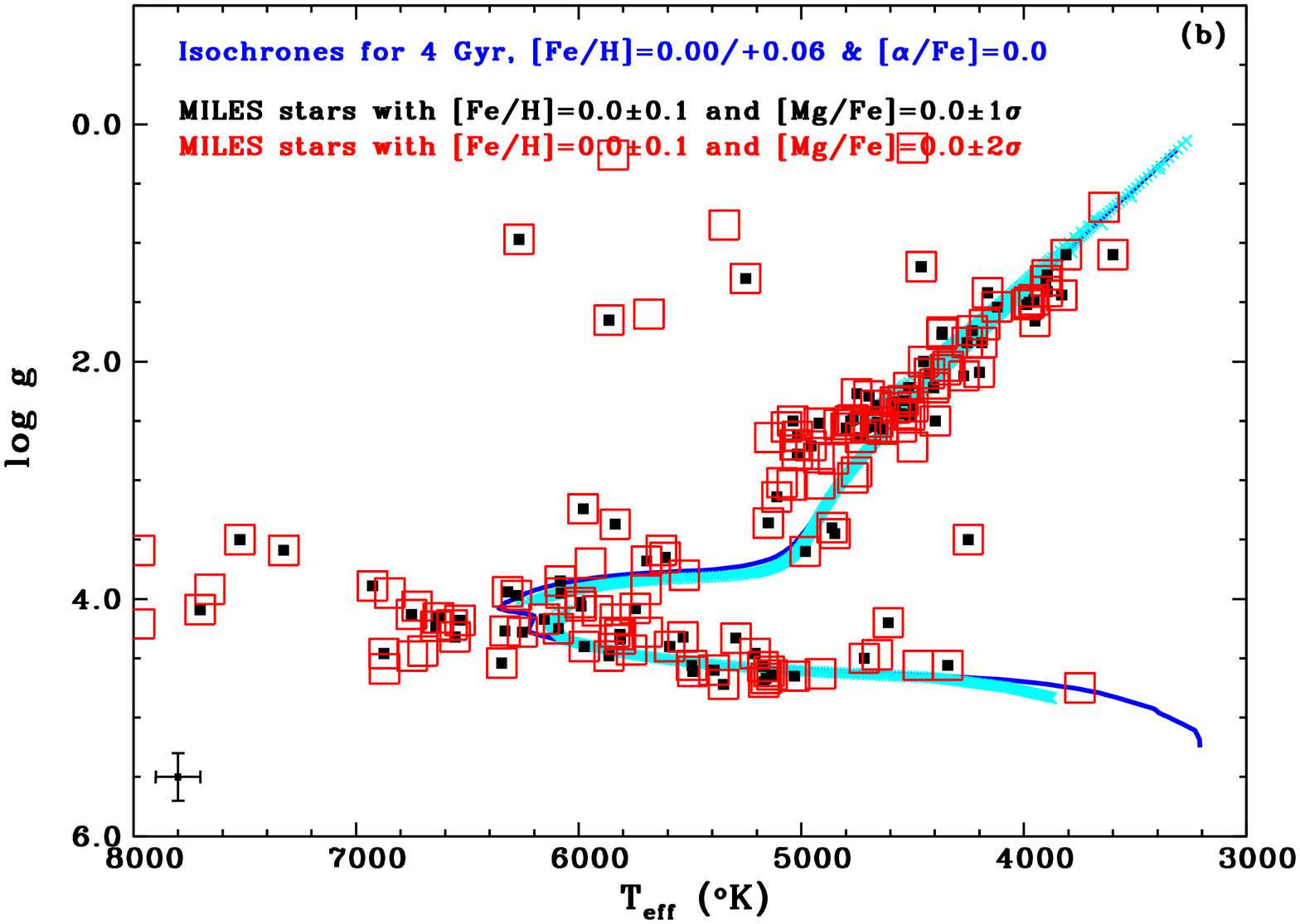}
\includegraphics[width=64mm, angle=-90]{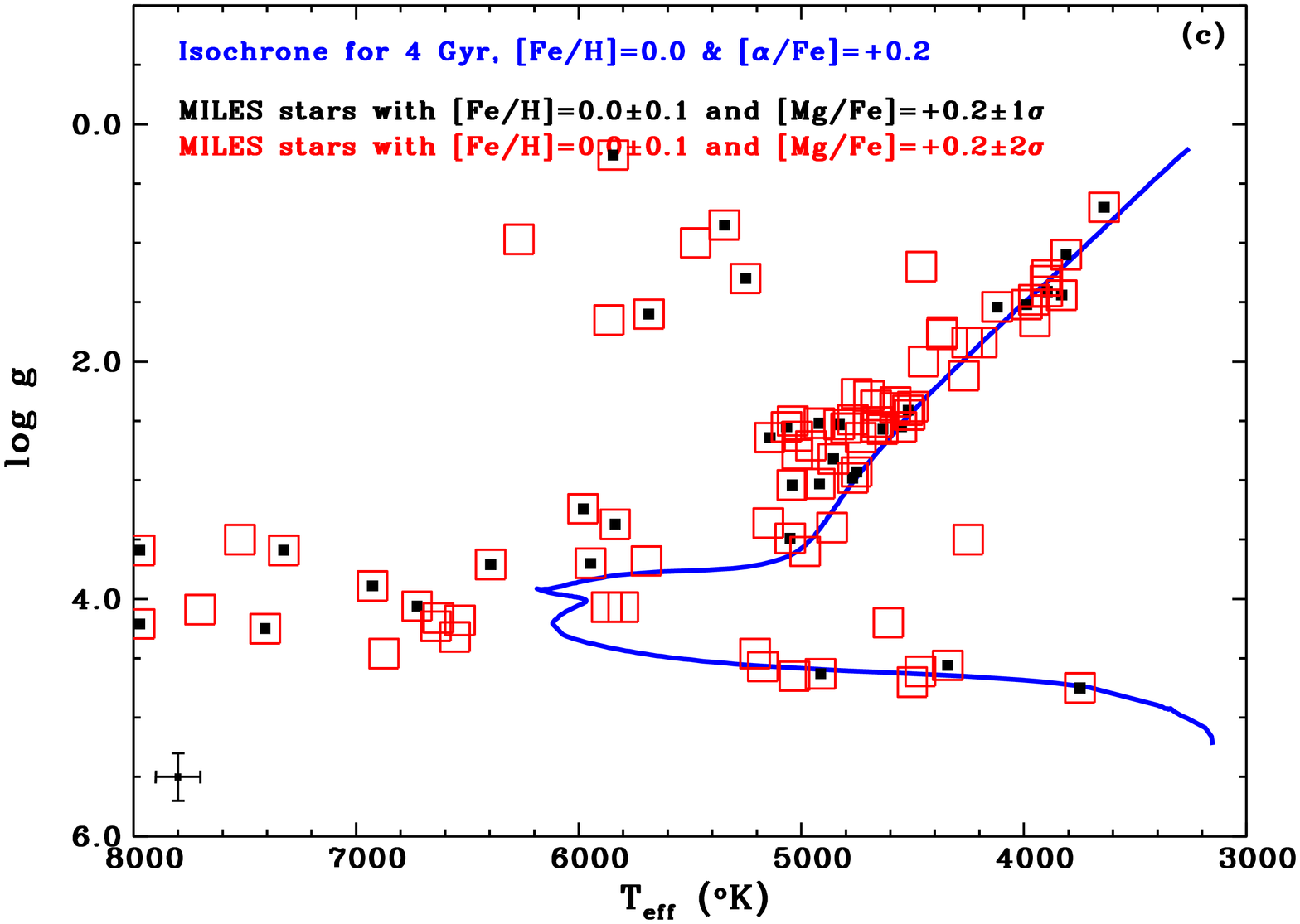} 
\end{center} 
\caption{ 
Isochrone-based plots on the HR diagram log $g$ vs. T$_{\rm eff}$ for age 4 Gyr, [Fe/H] = 0.0 dex
and three different [$\alpha$/Fe] ($-$0.2, 0.0 and $+$0.2 dex),
by adopting MILES stars with [Fe/H] = 0.0 dex and [Mg/Fe] around these $\alpha$-enhancements with 1 and 2 standard deviations ($\sigma$):
{\bf top panel (a)} for [Mg/Fe] = $-$0.2 dex,
{\bf middle panel (b)} for [Mg/Fe] = 0.0 dex, and
{\bf bottom panel (c)} for [Mg/Fe] = $+$0.2 dex.
The Dartmouth isochrones 
(Dotter {\it et al.} 2008)
are drawn as thick blue lines.
In the panel {\bf (b)},
a BaSTI
overshooting scaled solar model
for [Fe/H] = +0.06 dex
(Pietrinferni {\it et al.} 2006)
is also drawn (cyan crosses). 
The library stars are shown as filled black squares for which [Mg/Fe] has 1$\sigma$ precision
and as open red squares for 2$\sigma$.
}
\label{Fig13} 
\end{figure}

\begin{figure} 
\begin{center} 
\includegraphics[width=85mm]{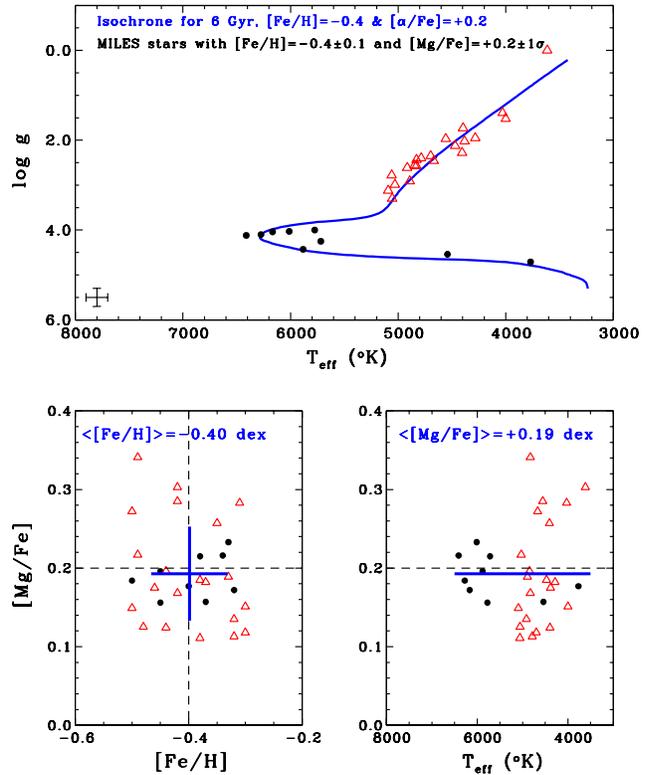}
\end{center} 
\caption{ 
Example of a cross-matching between a selected group of MILES stars and an isochrone for SSP modelling.
{\bf Top panel}:
isochrone-based plot on the H-R diagram log $g$ vs. T$_{\rm eff}$
for 6 Gyr, [Fe/H] = $-$0.4 dex and [$\alpha$/Fe] = $+$0.2 dex 
by collecting MILES stars with metallicity and [Mg/Fe] around these values with 1$\sigma$ deviation 
that restrictively match the correspondent Dartmouth isochrone
(Dotter {\it et al.} 2008)
taking into account the MILES errors in log $g$ and T$_{\rm eff}$ within 1$\sigma$ too
(error bars placed in the bottom left corner).
The dwarfs are represented as black filled circles and giants as red open triangles in all plots,
where dwarfs have log $g$ $\geq$ 3.7, covering the stages MS, TO and SG listed in Table 8, and giants log $g$ $<$ 3.7.
{\bf Bottom left panel}:
[Mg/Fe] vs. [Fe/H] plot for the selected dwarfs and giants,
in which the two blue thick lines show the stars' average values of [Fe/H] and [Mg/Fe]
(the line lengths are equal to the standard deviations of the averages),
{\bf Bottom right panel}:
[Mg/Fe] vs. T$_{\rm eff}$ plot for the selected dwarfs and giants,
in which the horizontal blue thick line shows the stars' average [Mg/Fe] over the whole stellar sample temperature scale.
The averages of [Fe/H] and [Mg/Fe] are written on the top of left and right panels respectively.
}
\label{Fig14} 
\end{figure}

\section[7]{Conclusions and summary}

We have obtained [Mg/Fe] for 76.3\% of the MILES stellar library (411 dwarfs and 341 giants,
76\% and 77\% of the total respectively), suitable for SSP modelling.
The weighted average uncertainty $\sigma$[Mg/Fe] is 0.105 dex over this MILES [Mg/Fe] catalogue.

Compilation of high spectral resolution [Mg/Fe] abundance ratios in the literature was extremely useful
in defining a uniform scale for [Mg/Fe] and in obtaining an extensive reference sample
for the calibration of abundance ratios measured in our work at medium resolution.
We emphasize that the calibration of mid-resolution measurements is an important step to be done
in the whole process to achieve reliable results in a homogeneous reference system.

A robust spectroscopic analysis was carried out using the MILES mid-resolution spectra and LTE spectral synthesis of two Mg features.
Two methods were applied through an automatic process: pseudo equivalent width and line profile fitting. 

The typical error of [Mg/Fe] from the collected and calibrated high-resolution measurements is 0.09 dex,
and the uncertainties from our MR analysis range from 0.10 to 0.15 dex, with a weighted average of 0.12 dex.
Thus we show that the accuracy of our measurements from MR spectra is quite acceptable,
but not better than those from HR analyses.
It is possible to measure element abundances in many more stars with such accuracy at MR
when a large control sample from HR measurements is adopted.
Hence this catalogue of [Mg/Fe] measurements will be useful for a range of applications for stellar population modelling and understanding stellar spectra.

The pattern of [Mg/Fe] versus [Fe/H] found for the stars in the MILES library is as expected for stars in the solar neighbourhood,
ranging from sub-solar [Mg/Fe] for high metallicity stars to super-solar [Mg/Fe] for low metallicity stars,
as shown in Fig. 10.

[Mg/Fe] measurements approximately characterise the alpha-to-iron ratios in stars.
Although not all types of stellar populations would be well sampled by the stars in this catalogue,
applications exploring the effects of non-solar abundance ratios will be possible
for such objects as globular clusters, spiral galaxies, various types of low luminosity galaxies and dwarf galaxies.
The importance of this is: (i) abundance patterns in stellar populations hold clues to their histories
and (ii) the accuracies of previously used characterisations of abundance patterns, based on theoretical models, can now be tested.
These applications will be followed up in future work. 

We also plan to use the MILES [Mg/Fe] catalogue to compute empirical and self consistent models of stellar populations
with Mg/Fe different from solar for certain values of age and metallicity.
These models will serve as a benchmark for other models based on theoretical libraries
as they will allow calibration of the effects of uncertainties in the final predictions
due to uncertainties in specific groups of stars.
We will study empirically the dependences of Lick System indices as a function
of the Mg/Fe ratio by adopting MILES stars with similar photospheric parameters but showing distinct [Mg/Fe]
in order to help computing robust stellar fitting functions of line-strengths.
Comparisons of empirical and theoretical line strengths will also be made
(Sansom {\it et al.}, in preparation).

\section*{Acknowledgements} 
 
A. Milone thanks the Brazilian foundations CAPES (abroad post-doctoral grant BEX 2895/07-2),
FAPESP (international congress support 2008/03161-7) and
CNPq (abroad short duration visit support 17.0018/2010-5 under the PCI/MCT/INPE program).
He is also grateful to the UCLan that provided a temporary visiting fellowship position for 12 months
in its Jeremiah Horrocks Institute.
PSB is supported by the Ministerio de Educacion y Ciencia through a Ram\'on y Cajal fellowship.
She also thanks support from the FP6 program of the EU through a ERC grant.
This work has been supported by the Programa Nacional de Astronom\'{\i}a y Astrof\'{\i}sica of the 
Spanish Ministry of Science and Innovation under the grant AYA2007-67752-C03-01.
This work has made use of BaSTI web tools. 
We thank Dr. A. Vazdekis, Dr. Paula R. T. Coelho and the anonymous referee
for their critical comments and suggestions that have improved the final version of this paper.

\appendix 

\section{}

\begin{figure} 
\begin{center} 
\includegraphics[width=85mm]{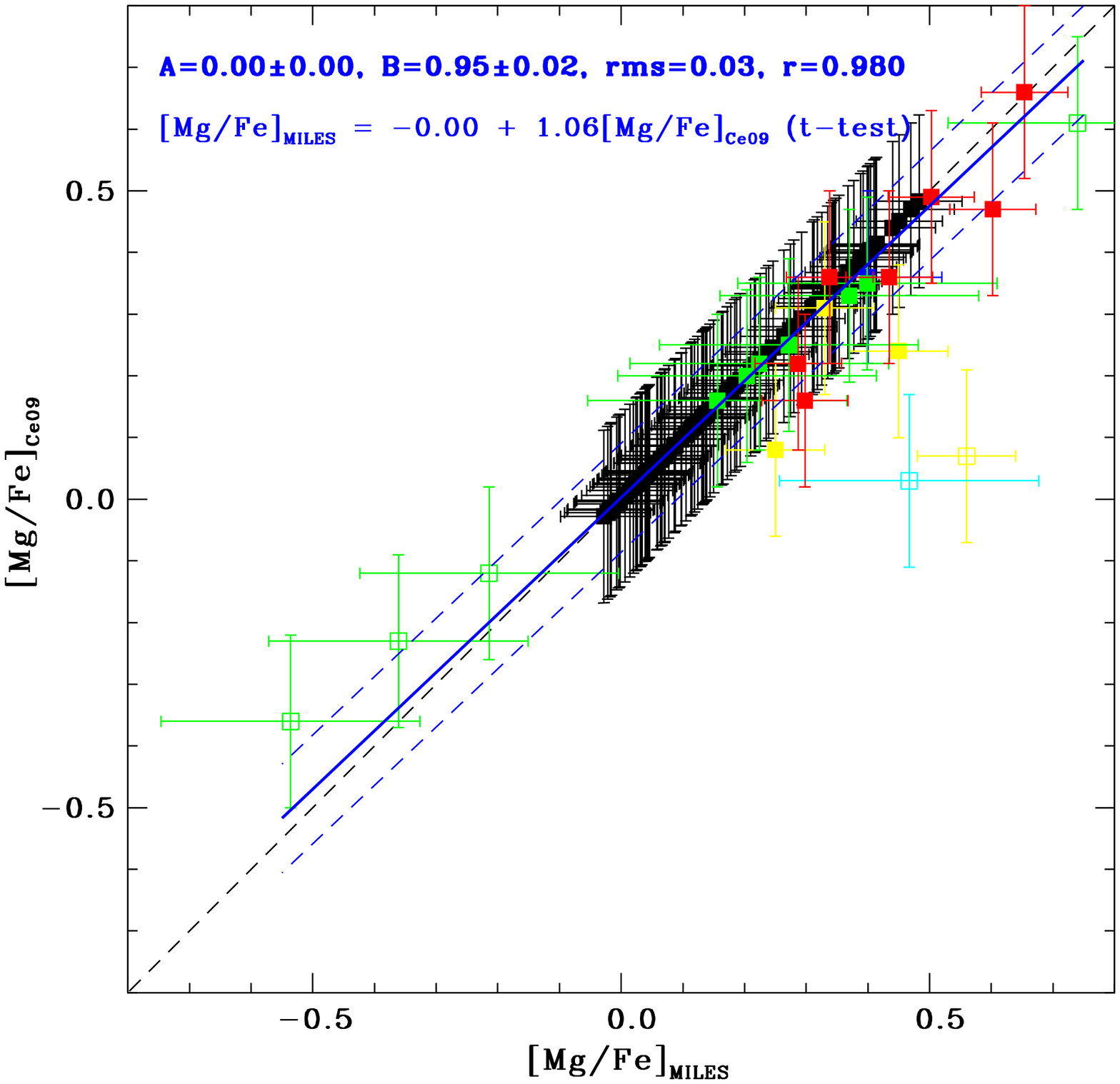} 
\end{center} 
\caption{ 
Comparison of our HR data with the [Mg/Fe] compilation for the CaT library
(Cenarro {\it et al.} 2009) that both have been independently calibrated to that scale of BM05
(whose measurements are represented by the filled black squares).
The filled colour squares represent different data sets: 
blue for CGS00, green for T98, yellow for F00, and
red for the stars with duplicated sources.
The linear $lsq$ fit [Fe/H]$_{\rm Cen09}$ = $A$ + $B$ [Fe/H]$_{\rm MILES}$ (solid blue line) is shown
with parallel dashed blue lines illustrating the 3$\sigma$ data clipping
(its $A$ and $B$, $rms$ and correlation coefficient $r$ are presented).
The statistically representative inverse expression
[Fe/H]$_{\rm MILES}$ = $-A$/$B$ + 1/$B$ [Fe/H]$_{\rm Ce09}$, based on a 95\% t-test, is also shown on the top.
The excluded data is represented by open colour squares (3-$\sigma$ criterion and outliers from T98).
} 
\label{FigA1} 
\end{figure} 

This appendix shows the comparison of our [Mg/Fe] reference scale with the CaT catalogue
(Cenarro {\it et al.} 2009).

Figure~A1 presents a direct comparison of the [Mg/Fe] high-resolution measurement compilation done by
Cenarro {\it et al.} (2009)
for their CaT library stars
with the HR part of MILES [Mg/Fe] catalogue.
A linear $lsq$ fitting between their results and ours
(also considering the errors in both variables and minimizing the distance along both directions)
gives [Mg/Fe]$_{\rm CaT}$ = [Mg/Fe]$_{\rm Ce09}$ = 0.00 + 0.95 [Mg/Fe]$_{\rm MILES}$
with a data spreading of 0.03 dex ($rms$).
This fit shows deviations, which are 0.035 dex at maximum, comfortably within the correspondent uncertainties
of [Mg/Fe] in CaT (typically 0.14 dex) and MILES (0.09 dex on average).
Two pairs of data were excluded by the 3-sigma clipping treatment taking into account their errors as well.
Few outliers with high and low [Mg/Fe] values (all from T98) were also excluded from this comparison.
The t-test with a 95\% confidence level appoints that the relationship [Mg/Fe]$_{\rm MILES}$ versus [Mg/Fe]$_{\rm Ce09}$
is not far from the 1:1 relation (see Fig. A1).
Therefore there is a quite good agreement between both scales.

\section{}

In this appendix we investigate possible systematic differences between the atmospheric parameters of the MILES catalogue
(Cenarro {\it et al.} 2007), that have been adopted all over in the current work,
and those of the BM05 sample.
Just concerning the metallicity scale, comparisons are also made with the data of those consulted HR works
(see Sect. 2).

Figure B1-(a)-(l) presents [Fe/H] from 12 consulted works compared with the MILES [Fe/H] scale.
Lines of least-square ($lsq$) linear fittings [Fe/H]$_{\rm work}$ = $A$ + $B$ [Fe/H]$_{\rm MILES}$
are shown in these plots.
The adopted $lsq$ method takes into account the errors in both variables
by minimizing the sum of distances of all points to the line.
Statistically representative inverse linear transformation expressions 
[Fe/H]$_{\rm MILES}$ = $-A$/$B$ + 1/$B$ [Fe/H]$_{\rm work}$ are obtained
after applying a 95\% confidence level t-test for each fit parameter
($A$$\neq$0? and/or $B$$\neq$1?).
The fitted straight lines found are very close to the 1:1 relationship.
There are tiny systematic differences between the MILES [Fe/H] scale and
the scales from
Fulbright (2000),
Bensby {\it et al.} (2005), and
Luck \& Heiter (2005).
However, these differences are dominated by a few outliers,
as can be seen in the panels (c), (e) and (h),
that are specifically localized either
at the metal-poor regime on the comparison with the
Fulbright (2000)
data,
or over metal-rich stars on the comparisons with the samples of
Bensby {\it et al.} (2005), and
Luck \& Heiter (2005).
If we applied metallicity corrections for these samples,
they would be smaller than the involved uncertainties,
even for the works of
Bensby {\it et al.} (2005), and
Luck \& Heiter (2005).
Therefore no correction has been applied to the [Fe/H] from the consulted works
because no significant systematic deviations from the MILES [Fe/H] scale has been detected.

For the comparison of BM05 and MILES T$_{\rm eff}$ scales,
the $rms$ of 83 K for a linear relationship between the scales and the respective systematic deviation
are comparable to the typical temperature uncertainty in MILES (1$\sigma$ = 100 K).
In the range 4500-7000 K, for instance, the maximum absolute difference reaches around 55 K only.
For the BM05-MILES log $g$ scale comparison, the absolute difference in the interval 3.0-5.0 gets a maximum of 0.13,
which is smaller than the MILES log $g$ uncertainty (1$\sigma$ = 0.20).
The $rms$ of the linear relationship between the gravity scales is smaller than the gravity's uncertainty too (0.145).
The angular coefficients of the linear $lsq$ fits T$_{\rm eff}$(BM05) vs. T$_{\rm eff}$(MILES) and log $g$(BM05) vs. log $g$(MILES)
are different, respectively, 1\% and 11\% only from that of the 1:1 relation.
Therefore the scales of T$_{\rm eff}$ and log $g$ in BM05 and MILES agree quite well between each other,
and no correction has been applied to these stellar parameters too.

\begin{figure*} 
\begin{center} 
\includegraphics[width=73mm]{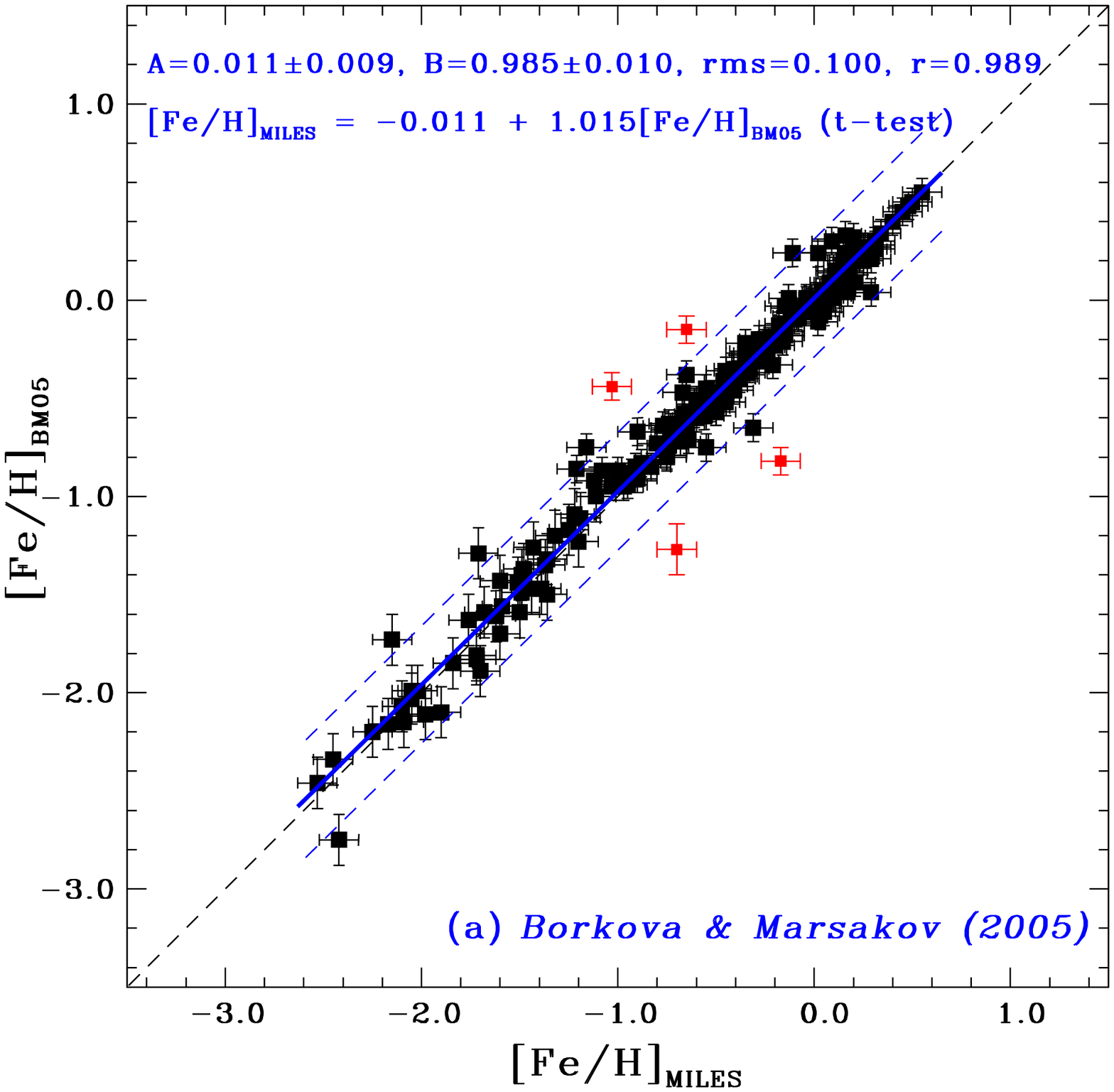} 
\includegraphics[width=73mm]{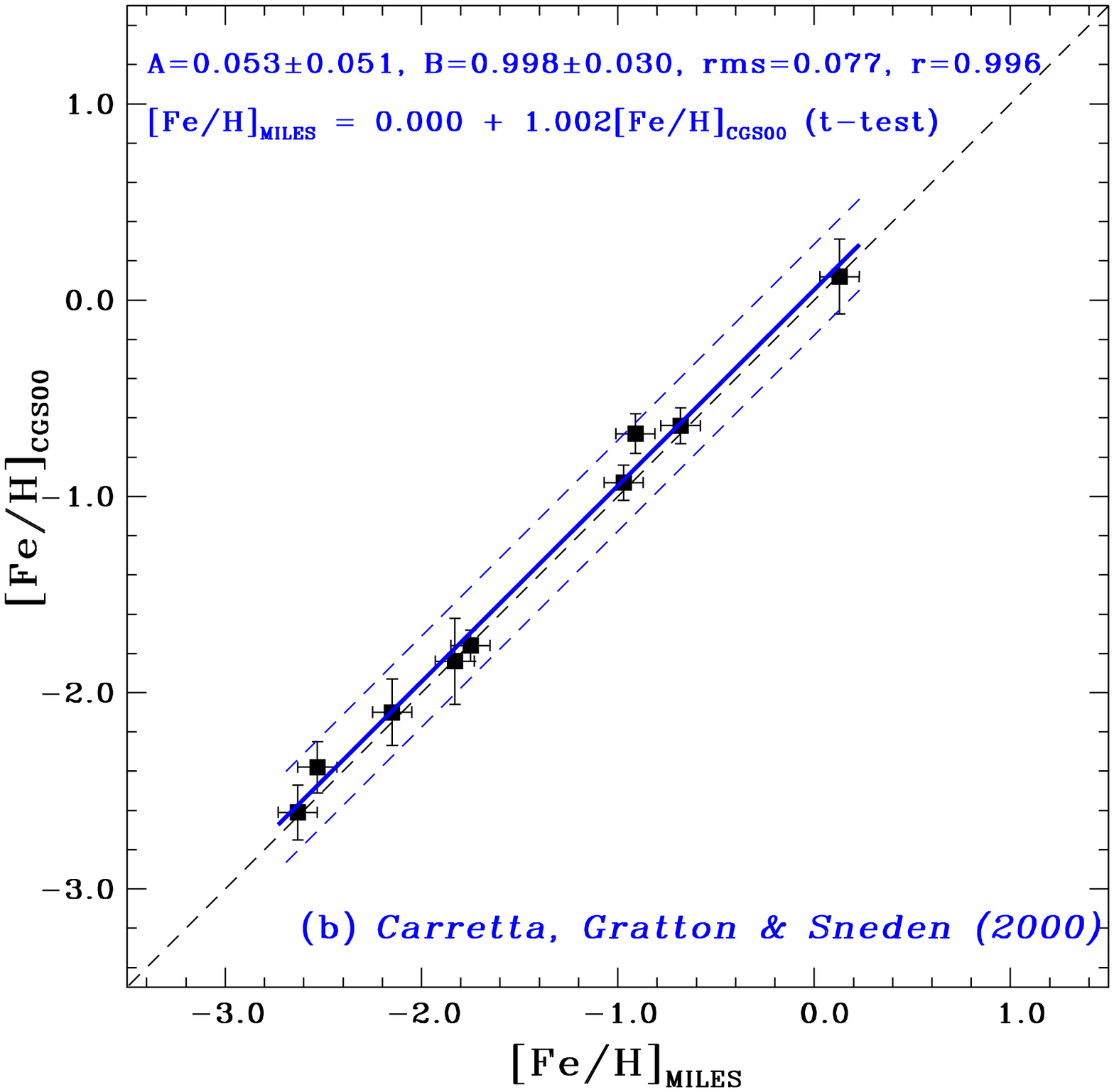} 
\includegraphics[width=73mm]{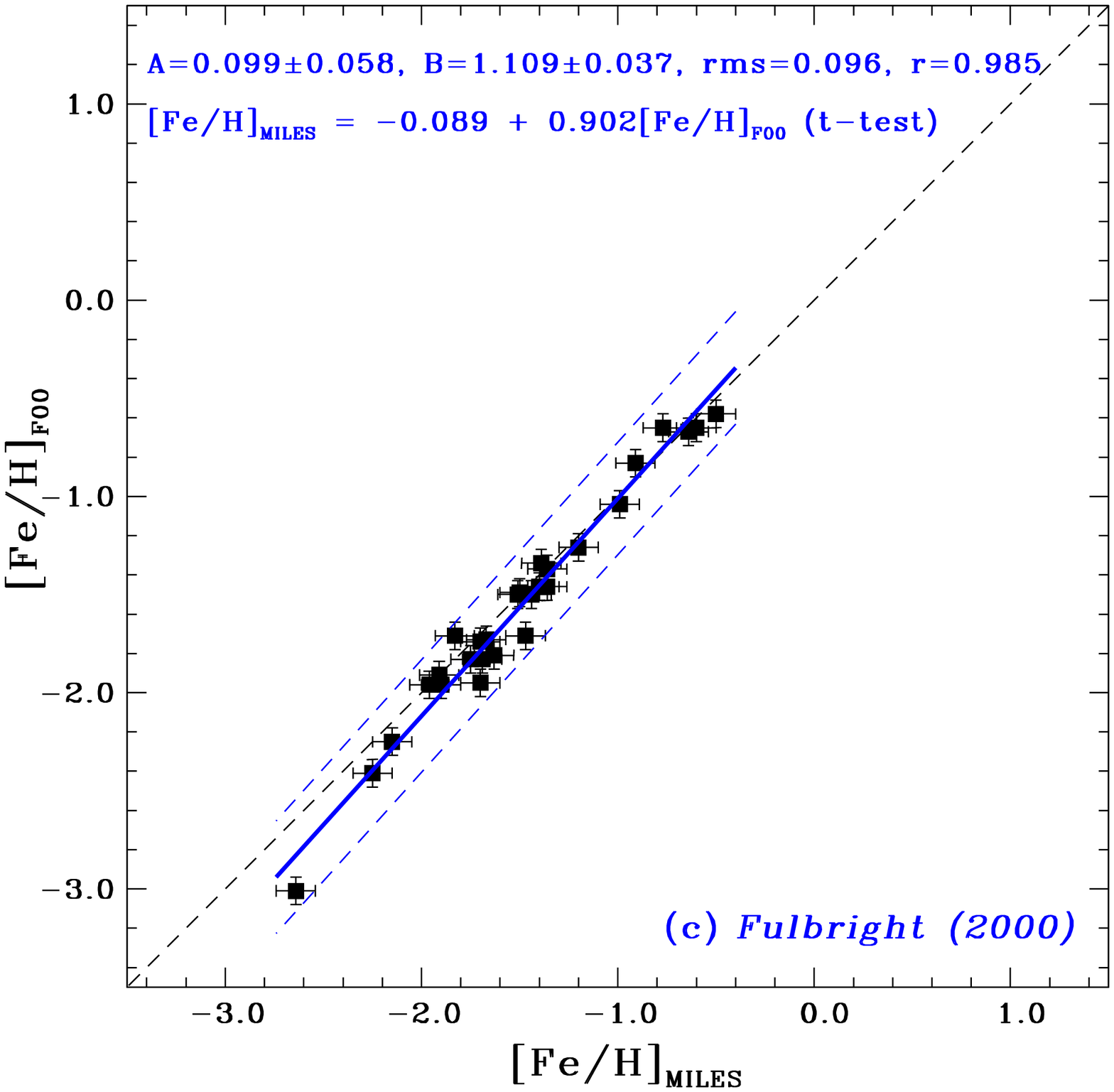} 
\includegraphics[width=73mm]{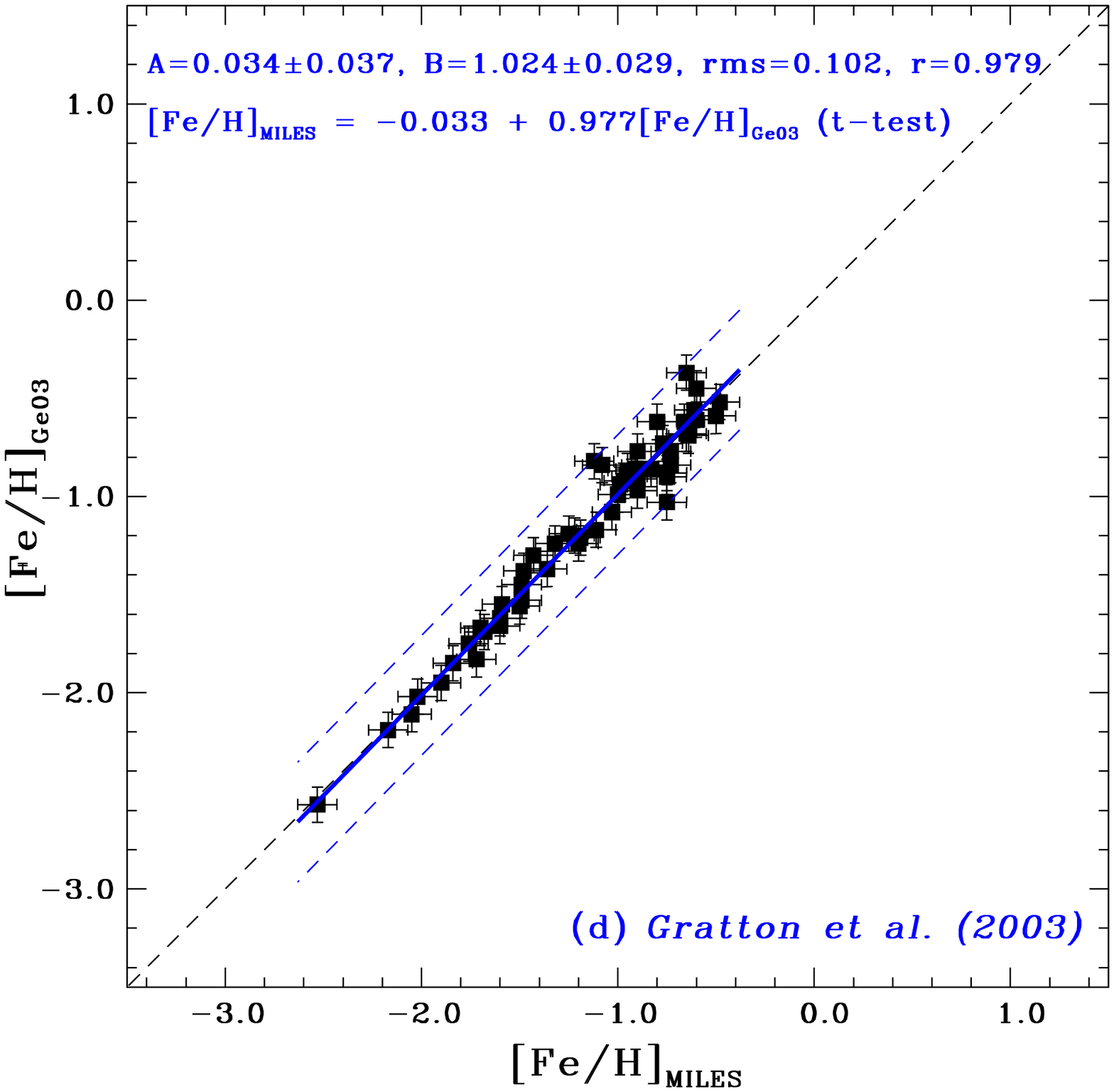} 
\includegraphics[width=73mm]{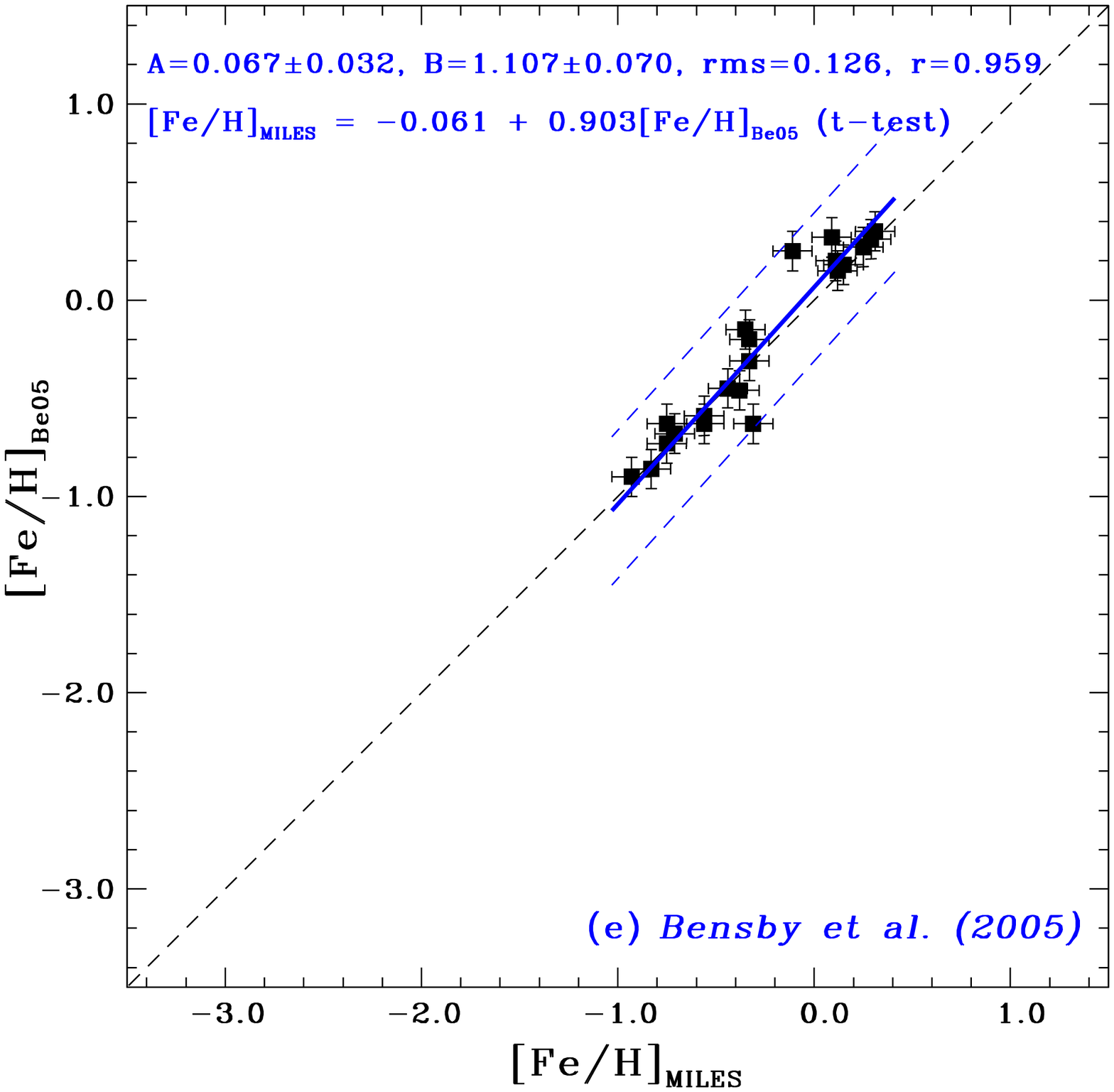} 
\includegraphics[width=73mm]{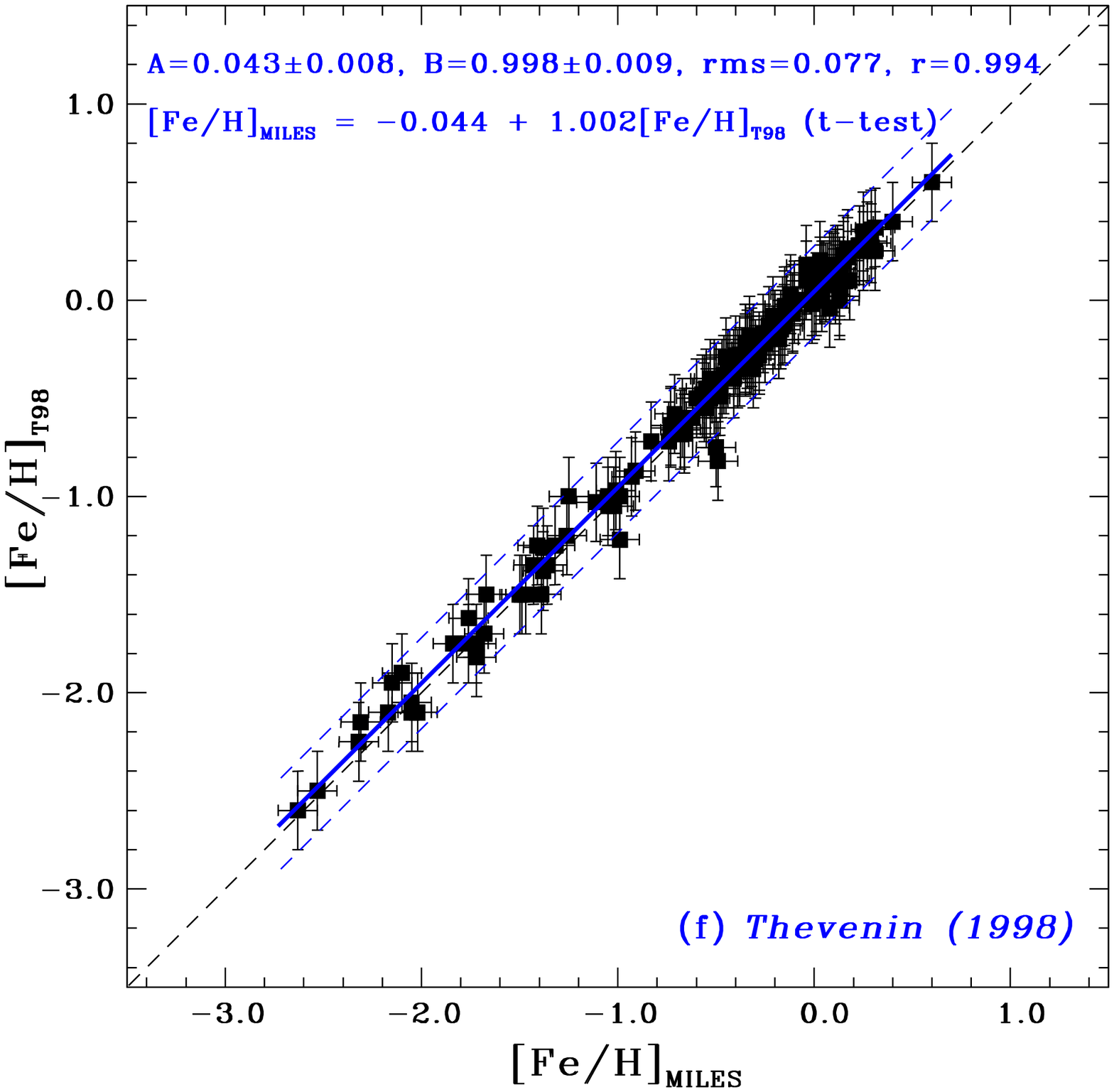} 
\end{center} 
\caption{
[Fe/H]$_{\rm work}$ vs. [Fe/H]$_{\rm MILES}$: 
12 panels (from {\bf a} to {\bf l}) showing comparisons
between the metallicity scales of 
the consulted HR works (designation at the bottom of each panel) with the MILES one.
The linear $lsq$ fittings [Fe/H]$_{\rm work}$ = $A$ + $B$ [Fe/H]$_{\rm MILES}$
with a 3-sigma data clipping are presented
and illustrated by respectively solid blue line plus parallel dashed blue lines with excluded points in red.
The constants $A$ and $B$, $rms$ and correlation coefficient $r$ are shown on top of each plot
as well as the statistically representative inverse expressions
[Fe/H]$_{\rm MILES}$ = $-A$/$B$ + 1/$B$ [Fe/H]$_{\rm work}$, based on 95\% t-tests.
} 
\label{FigB1} 
\end{figure*}

\begin{figure*} 
\begin{center} 
\includegraphics[width=73mm]{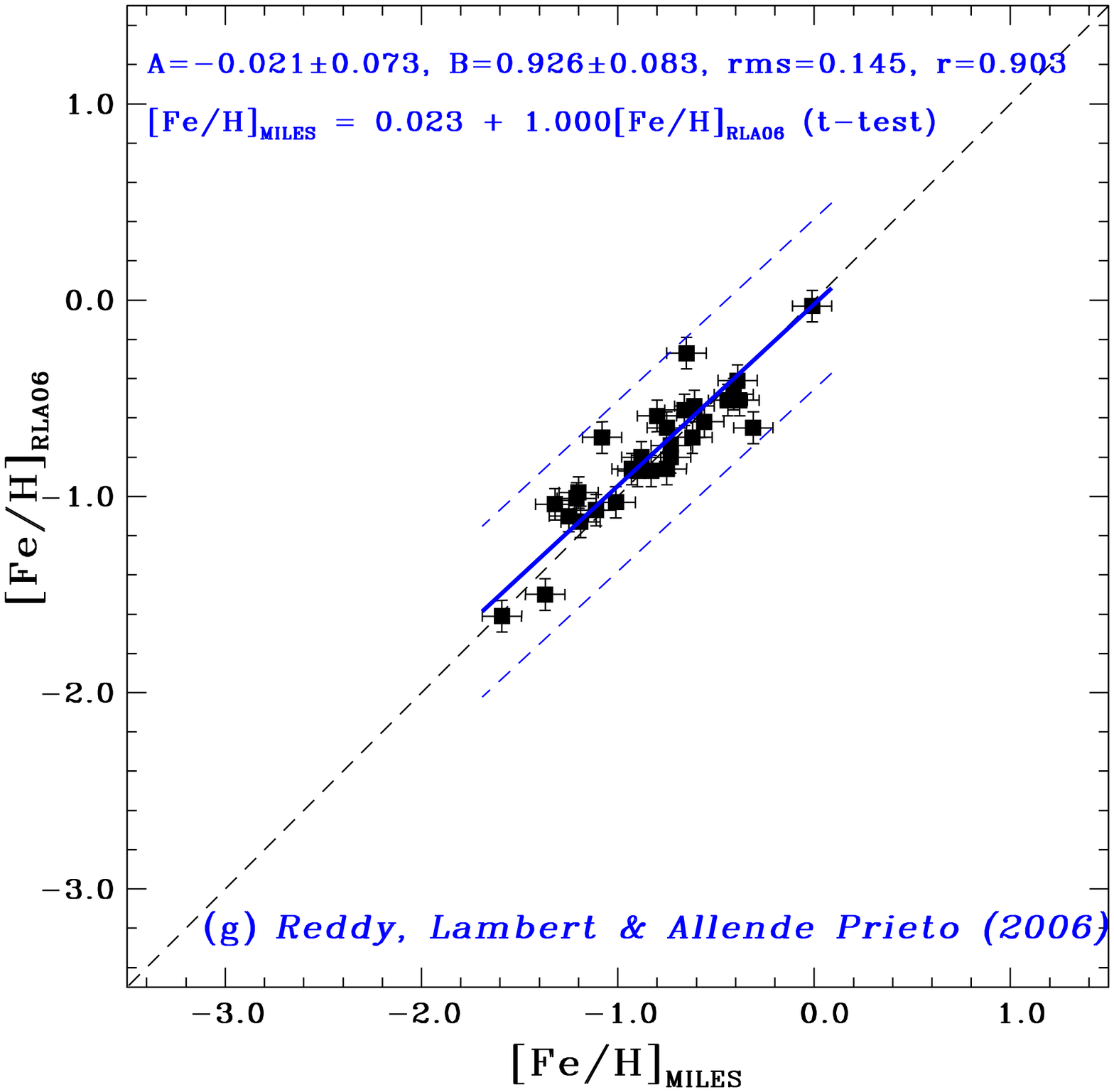} 
\includegraphics[width=73mm]{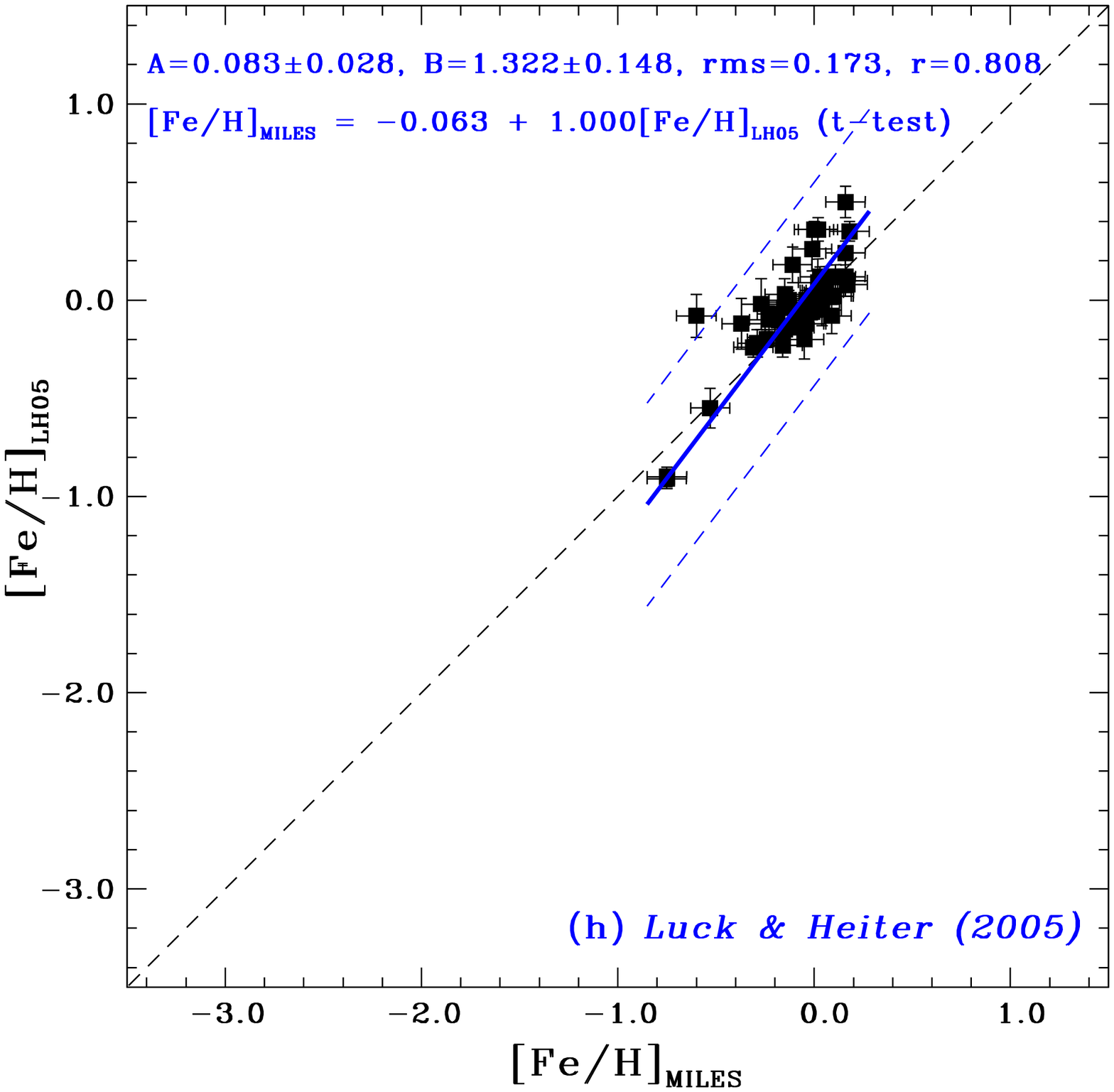} 
\includegraphics[width=73mm]{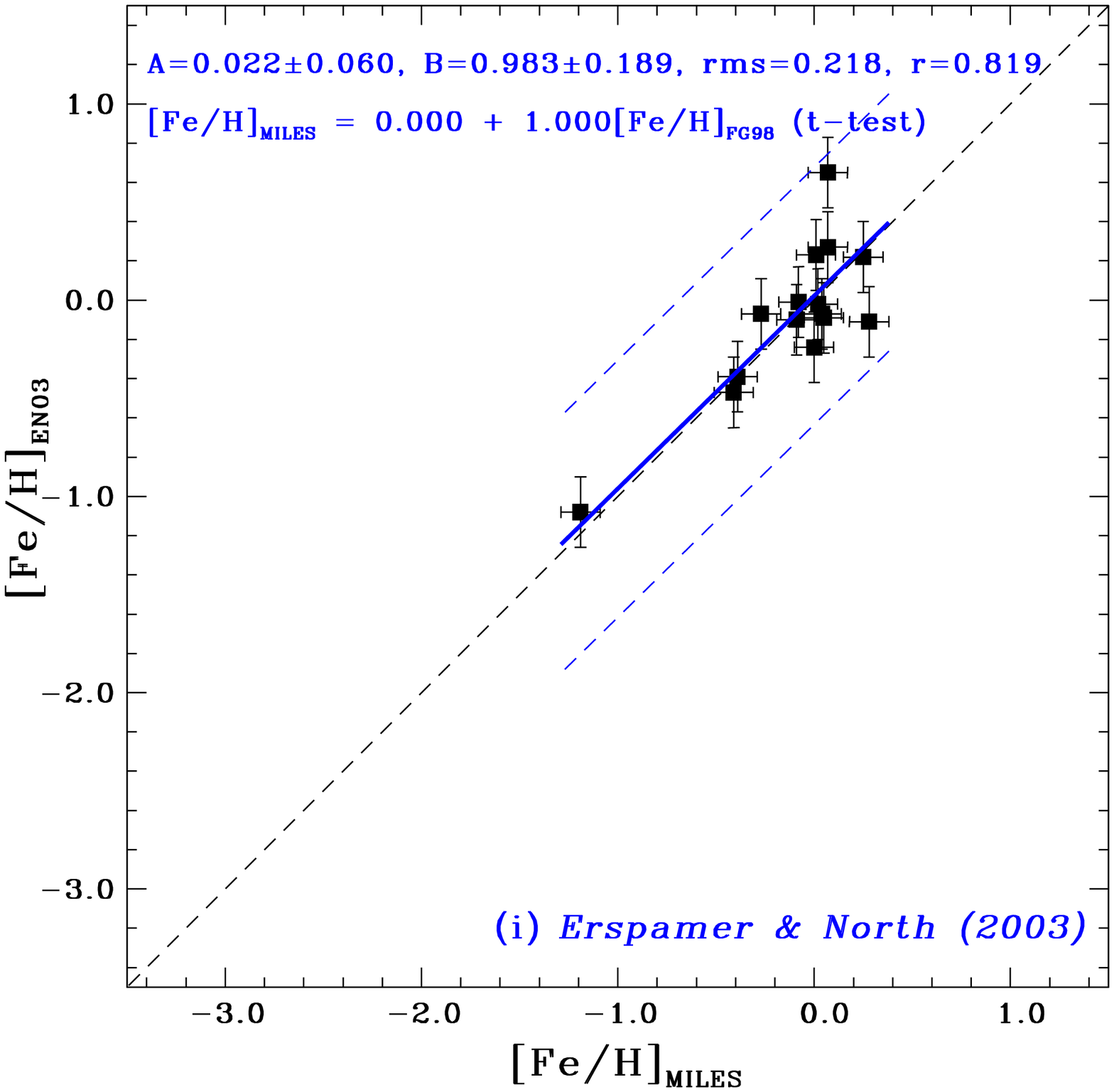} 
\includegraphics[width=73mm]{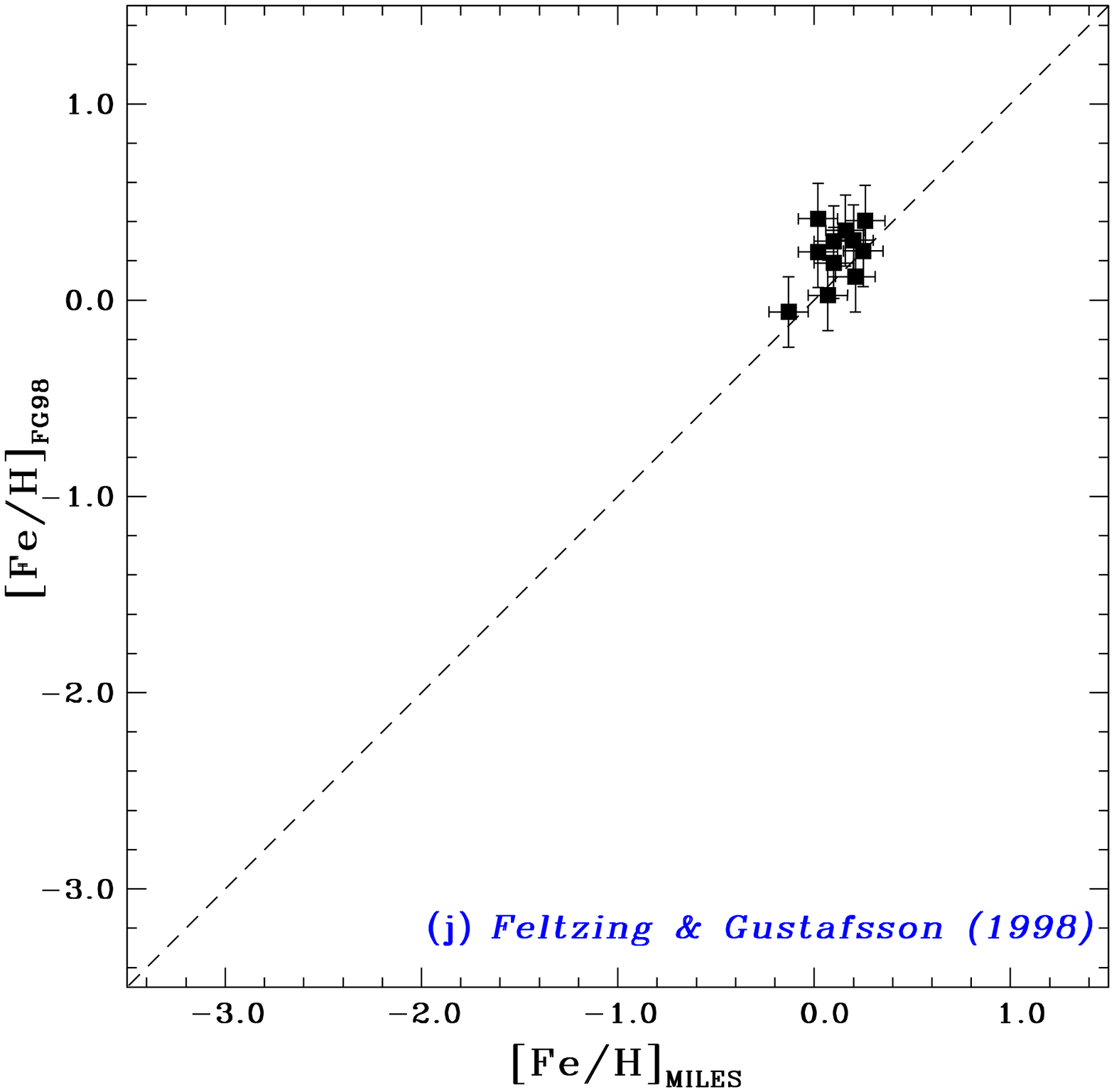} 
\includegraphics[width=73mm]{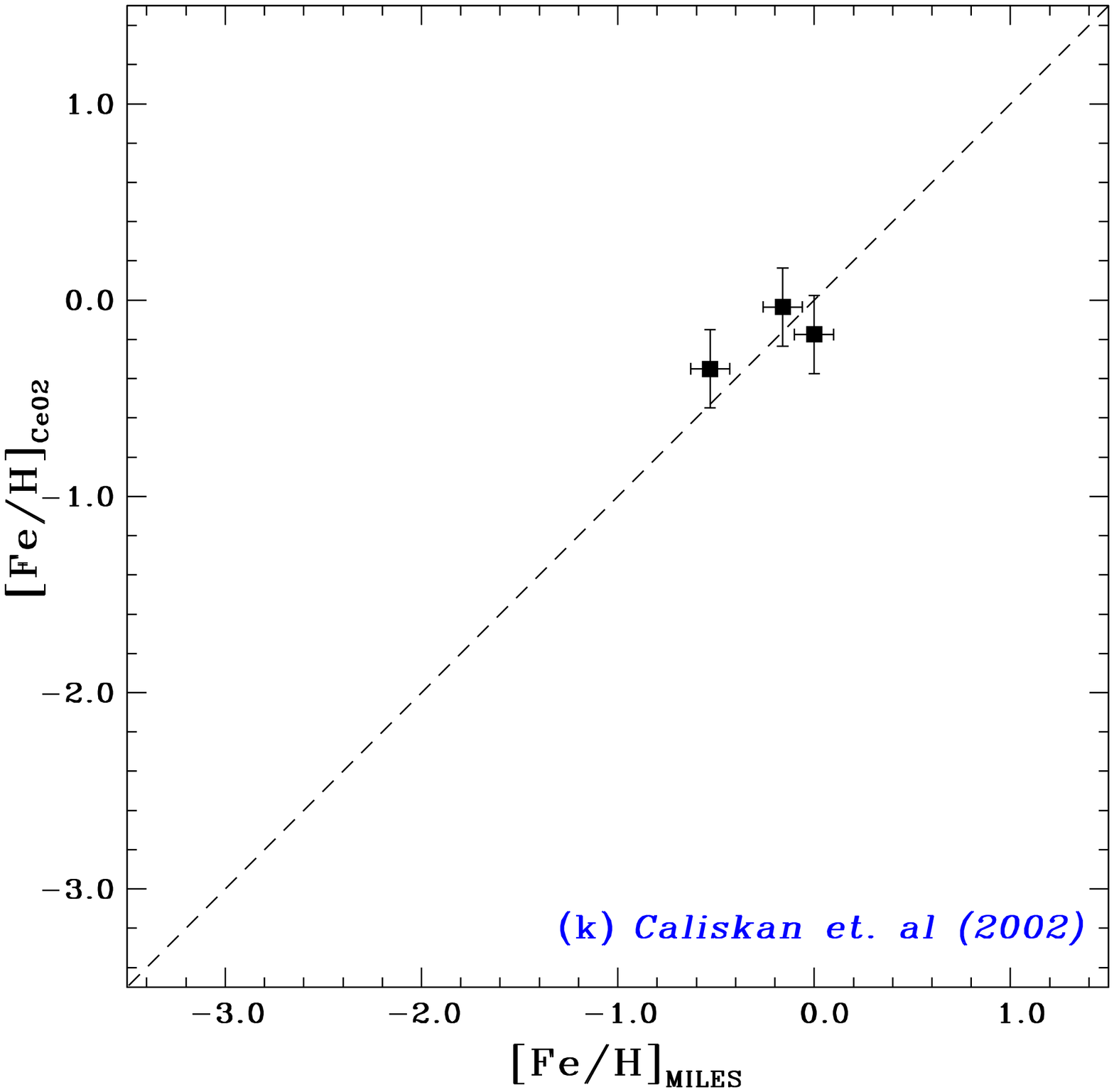} 
\includegraphics[width=73mm]{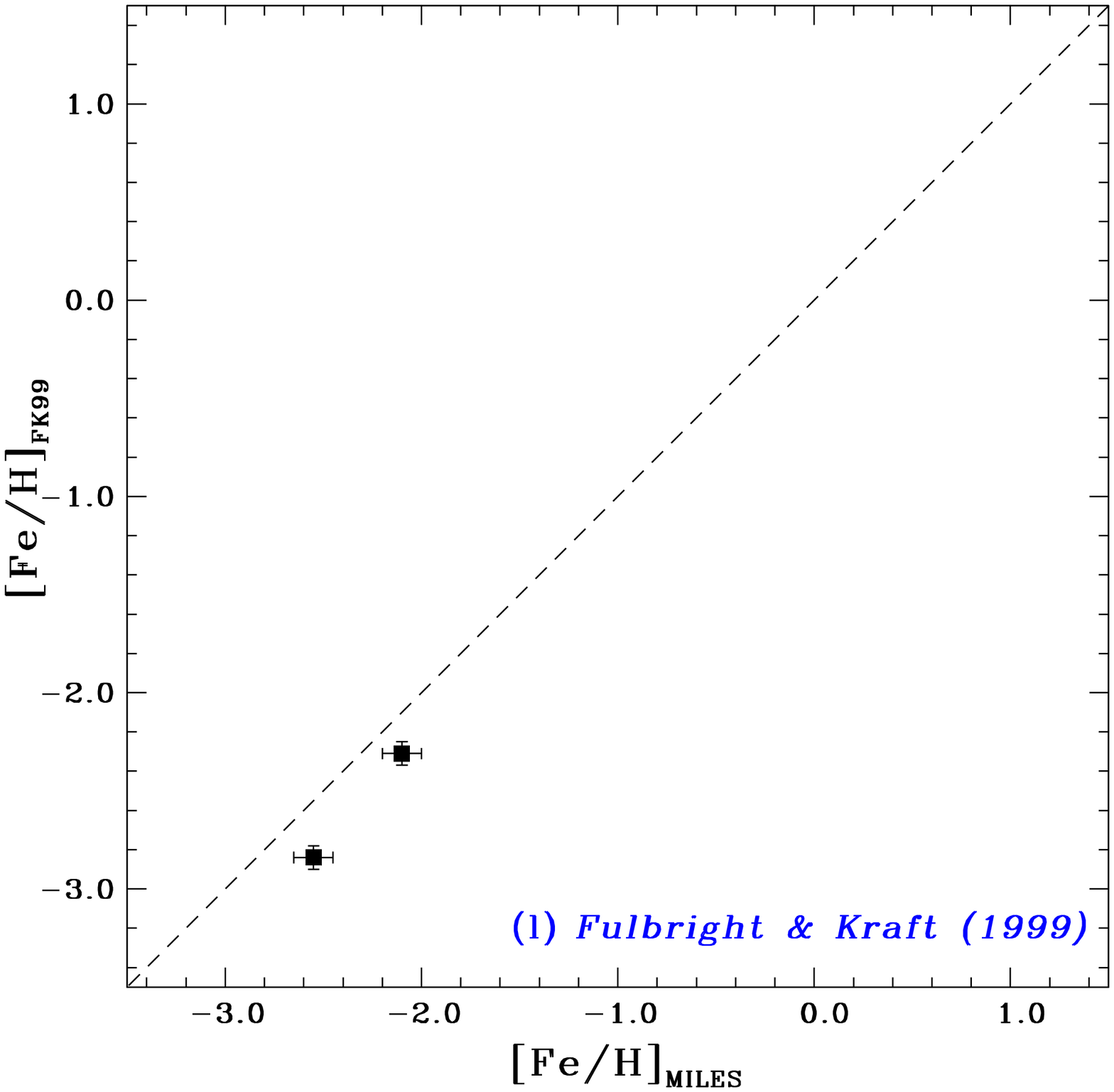} 
\end{center} 
\contcaption{} 
\label{FigB1cont} 
\end{figure*}

\vspace{30mm}

\section{}

In this appendix, comparisons with stars cluster data from high-resolution studies are presented.

There are 89 cluster stars in the MILES database (from 9 open clusters and 8 globular clusters).
Our mid-resolution measurements cover 65 cluster members (73\% of them) including 16 out of 17 clusters of MILES; 
see Table 5 (Sect. 4).
Specifically for three star clusters presented in MILES (Hyades, M71 and NGC7789),
we have obtained [Mg/Fe] for a reasonable number of members
from our spectral synthesis at mid-resolution ($\geq$ 10 stars).
Average values are presented in Table C1 together with the results from other clusters
for which we have done MR measurements for a minimum of two stars,
providing an interesting quality test for our work.
Table  also shows average cluster values of [Mg/Fe] collected from HR studies,
but their abundance ratios are not in the same uniform scale adopted in the current work.
The literature cluster averages [Mg/Fe] are computed from published ratios
to be representative to the star sample analysed in our work for each cluster
in terms of spectral type and luminosity class.

For the open cluster Hyades only ([Fe/H] = $+$0.13 dex),
we could compare our [Mg/Fe] MR measurements directly with the compiled HR data in a same uniform scale.
The average [Mg/Fe] computed from the MR measurements based on the two Mg features
and including 10 members (dwarfs only with 5256 $\leq$ T$_{\rm eff}$ $\leq$ 7634 K)
is $-$0.035 dex with a standard deviation $\sigma$ = 0.169 dex,
whilst the HR data have a average value $-$0.016 (1$\sigma$ = 0.063 dex) from 3 dwarfs only
(all from T98 with T$_{\rm eff}$ = 6486, 6742 and 8850 K).
The average of [Mg/Fe] from both MR and HR data for Hyades is $-$0.030 ($\sigma$ = 0.149 dex).
Schuler, King \& Lih-Sin (2009)
has recently measured, by analysing high-S/N high-resolution spectra,
[Mg/H] = +0.10 ($\sigma$ = 0.02 dex) in 3 main sequence stars (T$_{\rm eff}$ around 5600 K), getting [Mg/Fe] = $-$0.03 dex.
A sample of 55 F-K dwarfs (with 4900 $\leq$ T$_{\rm eff}$ $\leq$ 6450 K) were spectroscopically analysed at HR and high-S/N by
Paulson, Sneden \& Cochran (2003).
They obtained [Fe/H] = +0.13 $\pm$ 0.01 dex and differentially [Mg/Fe] = $-$0.03 $\pm$ 0.04 dex.
The spectroscopic analysis (based on HR and high-S/N spectra) of
Yong, Lambert \& Allende Prieto (2004)
measures for 34 Hyades dwarfs with 4700 $<$ T$_{\rm eff}$ $\leq$ 6200 K
cluster averages of [Fe/H] = $-$0.16 ($\sigma$ = 0.10 dex) and [Mg/Fe] $\approx$ $-$0.09 dex.
All these results corroborate our work.

For the globular cluster M71 ([Fe/H] = $-$0.84 dex),
in which there is a known internal spread of $\alpha$/Fe ratio over the stars
(Ram\'\i rez \& Cohen 2002),
we obtained an average [Mg/Fe] abundance ratio equals to $+$0.485 ($\sigma$ = 0.104 dex)
based on measurements from both Mg features in 19 cluster members,
which are all giants with 4014 $\leq$ T$_{\rm eff}$ $\leq$ 5123 K.
[Mg/Fe] varies in M71 as 
Ram\'\i rez \& Cohen (2002)
measured in 24 giants that were spectroscopically analysed at HR.
This star sample exhibits an average [Mg/Fe] equals to +0.36 ($\sigma$ = 0.09 dex) and a maximum internal spread of 0.18 dex.
Mel\'endez \& Cohen (2009)
recently proved, as concluded in other previous studies,
the existence of two stellar populations in M71; one CN-weak with normal O, Na, Mg, and Al abundances
plus a low isotope abundance ratio of $^{26}$Mg/$^{24}$Mg,
and other CN-strong with enhanced Na and Al accompanied by lower O together with a higher ratio $^{26}$Mg/$^{24}$Mg.
However, they measured a small spread for [Mg/Fe] over 9 giants (0.10 dex at most) belonged to both internal populations
exhibiting an average around +0.20 dex.

For the galactic cluster NGC7789 ([Fe/H] = $-$0.13 dex), the average [Mg/Fe] is $+$0.077 ($\sigma$ = 0.092 dex)
from our Mg5183/Mg5528 measurements in 13 giant stars (with 4020 $\leq$ T$_{\rm eff}$ $\leq$ 4952 K).
A very recent work 
(Pancino {\it et al.} 2010)
has measured for it an average [Fe/H] = +0.04 $\pm$ 0.07 dex with 1$\sigma$ dispersion of 0.10 dex
and [Mg/Fe] = +0.22 $\pm$ 0.07 dex ($\sigma$ = 0.10 dex)
based on spectroscopic measurements of 3 red clump stars with high-quality spectra at HR,
whilst previous studies at lower spectral resolution and through photometry-based techniques 
obtained [Fe/H] $\simeq$ $-$0.2 with [X/Fe] around zero for many elements
(e.g. Friel {\it et al.} 2002,
Pilachowski 1985)

Concerning our star cluster data, the major conclusions are:
(i) the standard deviations of computed average cluster values of [Mg/Fe]
are comparable with the systematic uncertainties of our individual MR measurements
as well as with the dispersion of the cluster average abundance ratios collected from the HR studies,
and
(ii) the [Mg/Fe] averages from the current work are in good agreement with measurements
carried out with high-S/N high-resolution spectra in recent studies.

\begin{table*} 
\caption{
Averages values of [Mg/Fe] for star clusters in the MILES library
computed from two or more star individual measurements at mid-resolution (mr) done in this work,
$\overline{\rm [Mg/Fe]_{mr}}$ and standard deviation $\sigma\overline{\rm [Mg/Fe]_{mr}}$,
and respective average from the high-resolution (HR) studies,
$\overline{\rm [Mg/Fe]_{HR}}$ and standard deviation $\sigma\overline{\rm [Mg/Fe]_{HR}}$.
The number of stars for each average (mr and HR) is informed in the columns six and ten respectively.
The clusters' metallicities adopted in MILES are listed in the third column.
The Mg features adopted in our MR measurements are presented in the seventh column
and the references consulted for the HR data are lied in the last column
(GMR08 for Gebran, Monier \& Richard 2008,
SKL09 for Schuler, King \& Lih-Sin 2009,
YLA04 for Yong, Lambert \& Allende Prieto 2004,
PSC03 for Paulson, Sneden \& Cochran 2003,
T98 for Th\'evenin 1998,
Ke00 for King {\it et al.} 2000,
CM05 for Cohen \& Mel\'endez 2005,
Je05 for Johnson {\it et al.} 2005,
Se04 for Sneden {\it et al.} 2004,
S96 for Shetrone 1996,
MC09 for Mel\'endez \& Cohen 2009,
RC02 for Ram\'\i rez \& Cohen 2002,
GO89 for Gratton \& Ortolani 1989,
and Pe10 for Pancino {\it et al.} 2010,
on the order they appear in the table).
The HR data from T98 (Hyades) comes directly from the MILES [Mg/Fe] catalogue.
Note: the others $\overline{\rm [Mg/Fe]_{HR}}$ are not transformed onto the catalogue's uniform scale.
}
\begin{center} 
\begin{tabular}{@{}llcccrlllrl} 
\hline 
\hline 
Cluster      &  Type    & [Fe/H] &   $\overline{\rm [Mg/Fe]_{mr}}$ & $\sigma\overline{\rm [Mg/Fe]_{mr}}$ & N$_{\rm mr}$ & Mg feature(s)
& $\overline{\rm [Mg/Fe]_{HR}}$ & $\sigma\overline{\rm [Mg/Fe]_{HR}}$ & N$_{\rm HR}$  & Ref \\
\hline 
\hline
             &          & (dex)  &   (dex)                         & (dex)                               &              &              
& (dex)                         & (dex)                               &              &     \\ 
\hline
Coma Ber     & open     &$-$0.05 &   +0.304         &    0.022               &  2       & Mg5528            
&     +0.26      &    0.008 &      2    & GMR08 \\   
Hyades       & open     &  +0.13 & $-$0.035         &    0.169               & 10       & Both
&   $-$0.03      &    0.02  &      3    & SKL09 \\
             &    &     &                  &                        &          &                    
&   $-$0.09      &    0.10  &     34    & YLA04 \\
             &    &     &                  &                        &          &                    
&   $-$0.03      &    0.04  &     55    & PSC03 \\
             &    &     &                  &                        &          &                    
&   $-$0.016     &    0.063 &      3    & T98 \\
Pleiades     & open     & $-$0.03 & $-$0.122         &    0.083               &  2       & Both                
&   $-$0.01      &    0.06  &      2    & Ke00 \\
M3           & globular & $-$1.34 &   +0.299         &    0.183               &  3       & Both       
&     +0.41      &    0.12  &     13    & CM05 \\ 
             &    &     &                  &                        &          &                    
&     +0.17      &    0.15  &     77    & Je05 \\     
             &    &     &                  &                        &          &                    
&     +0.22      &    0.15  &     23    & Se04 \\   
M5           & globular & $-$1.11 &   +0.426         &    0.169               &  2       & Both         
&     +0.16      &    0.07  &      6    & S96  \\   
M71          & globular & $-$0.84 &   +0.485         &    0.104               & 19       & Both   
&     +0.20      &    0.10  &      9    & MC09 \\
             &    &     &                  &                        &          &                    
&     +0.36      &    0.09  &     24    & RC02 \\
             &    &     &                  &                        &          &                    
&     +0.34      &    0.08  &      8    & S96  \\
M79          & globular & $-$1.37 &   +0.493         &    0.009               &  2       & Mg5528             
&     +0.47      &    0.45  &      2    & GO89 \\
M92          & globular & $-$2.16 &   +0.485         &    0.173               &  2       & Mg5528             
&     +0.19      &    0.19  &      6    & S96  \\   
NGC7789      & open     & $-$0.13 &   +0.077         &    0.092               & 13       & Both
&     +0.22      &    0.10  &      3    & Pe10 \\
\hline 
\hline 
\end{tabular} 
\end{center} 
\label{star_cluster_results} 
\end{table*}

\bsp

\label{lastpage} 


\begin{thebibliography}{99}

\bibitem[\protect\citeauthoryear{Adelman et al.}{2006}]{Adelman06}
Adelman S. J., Caliskan H., Gulliver A. F., Teker, A., 2006, \textit{A\&A}, 447, 685 (Ae06)

\bibitem[\protect\citeauthoryear{Adelman et al.}{2001}]{Adelman01}
Adelman S. J., Caliskan H., Kocer D., Kablan H., Y\"uce K., Engin S., 2001, \textit{A\&A}, 371, 1078 (Ae01)

\bibitem[\protect\citeauthoryear{Bensby et al.}{2010}]{Bensby10}
Bensby T., Alves-Brito A., Oey M. S., Yong D., Mel\'endez J., 2010, \textit{A\&A}, 516, L13

\bibitem[\protect\citeauthoryear{Bensby et al.}{2005}]{Bensby05}
Bensby T., Feltzing S., Lundstr\"om I., Ilyin I., 2005, \textit{A\&A}, 433, 185 (Be05)

\bibitem[\protect\citeauthoryear{Bertone et al.}{2008}]{Bertone08}
Bertone E., Buzzoni A., Ch\'avez M., Rodr\'\i guez-Merino L. H., 2008, \textit{A\&A}, 485, 823

\bibitem[\protect\citeauthoryear{Borkova \& Marsakov}{2005}]{BorkovaMarsakov05}
Borkova T. V., Marsakov V.A., 2005, \textit{AZh}, 82, 453 (BM05)

\bibitem[\protect\citeauthoryear{Caliskan et al.}{2002}]{Caliskan02}
Caliskan H., Adelman S. J., Cay M. T., Cay I. H., Gulliver A. F., Tektunali G. H., Kocer D., Teker A.,
2002, \textit{A\&A}, 394, 187 (Ce02)

\bibitem[\protect\citeauthoryear{Cardiel et al.}{1998}]{Cardiel98}
Cardiel N., Gorgas J., Cenarro J., Gonzalez J. J., 1998, \textit{A\&AS}, 127, 597

\bibitem[\protect\citeauthoryear{Carretta et al.}{2000}]{Carretta00}
Carretta E., Gratton R. G., Sneden C., 2000, \textit{A\&A}, 356, 238 (CGS00)

\bibitem[\protect\citeauthoryear{Cenarro et al.}{2009}]{Cenarro09}
Cenarro A. J., Cardiel N., Vazdekis A., Gorgas J., 2009, \textit{MNRAS}, 396, 1895

\bibitem[\protect\citeauthoryear{Cenarro et al.}{2001}]{Cenarro01}
Cenarro A. J., Gorgas J., Cardiel N., Pedraz S., Peletier R. F., Vazdekis A., 2001, \textit{MNRAS}, 326, 981

\bibitem[\protect\citeauthoryear{Cenarro et al.}{2002}]{Cenarro02}
Cenarro A. J., Gorgas J., Cardiel N., Vazdekis, A., Peletier R. F., 2002, \textit{MNRAS}, 329, 863

\bibitem[\protect\citeauthoryear{Cenarro et al.}{2007}]{Cenarro07}
Cenarro A. J., Peletier R. F., S\'anchez-Bl\'azquez P., Selam S. O.,
Toloba E., Cardiel N., Falc\'on-Barroso J., Gorgas J., Jim\'enez-Vicente J., Vazdekis A., 2007, \textit{MNRAS}, 374, 664 

\bibitem[\protect\citeauthoryear{Chavez et al.}{1995}]{Chavez95}
Chavez M., Malagnini M. L., Morossi C., 1995, \textit{ApJ}, 440, 210

\bibitem[\protect\citeauthoryear{Chavez et al.}{1997}]{Chavez97}
Chavez M., Malagnini M. L., Morossi C., 1997, \textit{A\&AS}, 126, 267

\bibitem[\protect\citeauthoryear{Chen et al.}{2000}]{Chen00}
Chen Y. Q., Nissen P. E., Zhao G., Zhang H. W., Benoni T., 2000, \textit{A\&AS}, 141, 491

\bibitem[\protect\citeauthoryear{Coelho et al.}{2005}]{Coelho05}
Coelho P., Barbuy B., Mel\'endez J., Schiavon R. P., Castilho B. V., 2005, \textit{A\&A}, 443, 735

\bibitem[\protect\citeauthoryear{Coelho et al.}{2007}]{Coelho07}
Coelho P., Bruzual G., Charlot S., Weiss A., Barbuy B., Ferguson J. W., 2007, \textit{MNRAS}, 382, 498

\bibitem[\protect\citeauthoryear{Cohen \& Mel\'endez}{2005}]{CohenMelendez05}
Cohen J. G.,\& Mel\'endez J., 2005, \textit{AJ}, 129, 303 (CM05)

\bibitem[\protect\citeauthoryear{Cook et al.}{2007}]{Cook07}
Cook D., Shetrone M., Siegel M., Bosler T., 2007, \textit{AAS}, 211, 9513

\bibitem[\protect\citeauthoryear{Denicolo et al.}{2005}]{Denicolo05}
Denicol\'o G., Terlevich R., Terlevich E., Forbes D. A., Terlevich A., 2005, \textit{MNRAS}, 358, 813

\bibitem[\protect\citeauthoryear{Dotter et al.}{2008}]{Dotter08}
Dotter A., Chaboyer B., Jevremovi\'c D., Kostov V., Baron E., Ferguson J. W., 2008, \textit{ApJS}, 178, 89

\bibitem[\protect\citeauthoryear{Erspamer \& North}{2003}]{ErspamerNorth03}
Erspamer D., North P., 2003, \textit{A\&A}, 398, 112 (EN03)

\bibitem[\protect\citeauthoryear{Feltzing et al.}{1998}]{Feltzing98}
Feltzing S., Gustafsson B., 1998, \textit{A\&AS}, 129, 237 (FG98)

\bibitem[\protect\citeauthoryear{Fiorentin et al.}{2007}]{Fiorentin07}
Fiorentin P., Bailer-Jones C. A. L., Lee Y. S., Beers T. C., Sivarani T., Wilhelm R., Allende Prieto C., Norris J. E.,
2007, \textit{A\&A}, 467, 1373

\bibitem[\protect\citeauthoryear{Fremaux et al.}{2006}]{Fremaux06}
Fr\'emaux J.,  Kupka F.,  Boisson C., Joly M.,  Tsymbal V., 2006, \textit{A\&A}, 449, 109

\bibitem[\protect\citeauthoryear{Friel \& Janes}{1993}]{FrielJanes93}
Friel E. D., Janes K. A., 1993, \textit{A\&A}, 267, 75

\bibitem[\protect\citeauthoryear{Friel et al.}{2002}]{Friel02}
Friel E. D., Janes K. A., Tavarez M., Scott J., Katsanis R., Lotz J., Hong L, Miller N., 2002, \textit{AJ}, 124, 2693

\bibitem[\protect\citeauthoryear{Fulbright}{2000}]{Fulbright00}
Fulbright J. P., 2000, \textit{AJ}, 120, 1841 (F00)

\bibitem[\protect\citeauthoryear{Fulbright \& Kraft}{1999}]{FK99}
Fulbright J. P., Kraft, R. P., 1999, \textit{AJ}, 118, 527 (FK99)

\bibitem[\protect\citeauthoryear{Gebran et al.}{2008}]{Gebran08}
Gebran M., Monier R., Richard O., 2008, \textit{A\&A}, 479, 189 (GMR08)

\bibitem[\protect\citeauthoryear{Gratton et al.}{2003}]{Gratton03}
Gratton R. G., Carretta E., Claudi R., Lucatello S., Barbieri M., 2003, \textit{A\&A}, 404, 187 (Ge03)

\bibitem[\protect\citeauthoryear{Gratton \& Ortolani}{1989}]{GrattonOrtolani89}
Gratton R. G., Ortolani S., 1989, \textit{A\&A}, 211, 41

\bibitem[\protect\citeauthoryear{Grevesse et al.}{2007}]{Grevesse07}
Grevesse N., Asplund M., Sauval A. J., 2007, \textit{Space Science Review}, 130, 205 (GAS07)

\bibitem[\protect\citeauthoryear{Gustafsson et al.}{2008}]{Gustafsson08}
Gustafsson B., Edvardsson B., Eriksson K., J\"orgensen U. G., Nordlund {\AA}., Plez B., 2008, \textit{A\&A} 486, 951

\bibitem[\protect\citeauthoryear{Heiter}{2002}]{Heiter02}
Heiter U., 2002, \textit{A\&A}, 381, 959 (H02)

\bibitem[\protect\citeauthoryear{Ivans et al.}{2003}]{Ivans03}
Ivans I. I., Sneden C., James C. R., Preston G. W., Fulbright J. P., H\"oflich P. A., Carney B. W., Wheeler J. C.,
2003, \textit{ApJ}, 592, 906

\bibitem[\protect\citeauthoryear{Johnson et al.}{2005}]{Johnson05}
Johnson C. I., Kraft R. P., Pilachowski C. A., Sneden C., Ivans I. I., Benman G., 2005, \textit{PASP}, 117, 1308 (Je05)

\bibitem[\protect\citeauthoryear{King et al.}{2000}]{King00}
King J. R., Soderblom D. R., Fischer D. and Jones B. F., 2000, \textit{ApJ}, 533, 944 (Ke00)

\bibitem[\protect\citeauthoryear{Kirby et al.}{2009}]{Kirby09}
Kirby E. N., Guhathakurta P., Bolte M., Sneden C., Geha M. C., 2009, \textit{ApJ}, 705, 328

\bibitem[\protect\citeauthoryear{Korn et al.}{2005}]{Korn05}
Korn A. J., Maraston C., Thomas, D., 2005, \textit{A\&A}, 438, 685 (K05)

\bibitem[\protect\citeauthoryear{Kupka et al.}{1999}]{Kupka99}
Kupka F., Piskunov N. E., Ryabchikova T. A., Stempels H. C., Weiss W. W., 1999, \textit{A\&AS}, 138, 119

\bibitem[\protect\citeauthoryear{Kupka et al.}{2000}]{Kupka00}
Kupka, F., Ryabchikova T. A., Piskunov N. E., Stempels H. C., Weiss W. W., 2000, \textit{Baltic Astronomy}, 9, 590

\bibitem[\protect\citeauthoryear{Kurucz}{1995}]{Kurucz95}
Kurucz R., 1995, An Atomic and Molecular Data Bank for Stellar Spectroscopy, \textit{ASP Conference N$^{o}$ 81}, CD-ROM N$^{o}$ 18

\bibitem[\protect\citeauthoryear{Lee et al.}{2005}]{Lee05}
Lee H-C., Worthey G., 2005, \textit{ApJS}, 160, 176

\bibitem[\protect\citeauthoryear{Lee et al.}{2009}]{Lee09}
Lee H-C., Worthey G., Dotter A., 2009, \textit{ApJ}, 694, 902

\bibitem[\protect\citeauthoryear{Luck \& Heiter}{2005}]{LuckHeiter05}
Luck R. E., Heiter U., 2005, \textit{AJ}, 129, 1063 (LH05)

\bibitem[\protect\citeauthoryear{Marsteller et al.}{2009}]{Marsteller09}
Marsteller B., Beers T. C., Thirupathi S., Rossi S., Placco V., Knapp G. R., Johnson J. A., Lucatello S.,
2009, \textit{AJ}, 138, 533

\bibitem[\protect\citeauthoryear{Martins et al.}{2007}]{Martins07}
Martins L. P., Coelho P., 2007, \textit{MNRAS}, 381, 1329

\bibitem[\protect\citeauthoryear{Martins et al.}{2005}]{Martins05}
Martins L. P.,  Gonz\'alez Delgado R. M., Leitherer C., Cervi\~o M., Hauschildt P., 2005, \textit{MNRAS}, 358, 49

\bibitem[\protect\citeauthoryear{Masseron}{2008}]{Masseron08}
Masseron T., 2008, \textit{PhD Thesis}, Ohio State University (USA) 

\bibitem[\protect\citeauthoryear{Melendez et al.}{2009}]{Melendez09}
Mel\'endez J., Cohen J. G., 2009, \textit{ApJ}, 699, 2017 (MC09)

\bibitem[\protect\citeauthoryear{Mishenina et al.}{2004}]{Mishenina04}
Mishenina T. V., Soubiran C., Kovtyukh V. V., Korotin S. A., 2004, \textit{A\&A}, 418, 551

\bibitem[\protect\citeauthoryear{Munari et al.}{2005}]{Munari05}
Munari U.,  Sordo R.,  Castelli F.,  Zwitter T., 2005, \textit{A\&A}, 442, 1127

\bibitem[\protect\citeauthoryear{Murphy et al.}{2004}]{Murphy04}
Murphy T., Meiksin A., 2004, \textit{MNRAS}, 351, 1430

\bibitem[\protect\citeauthoryear{Neves et al.}{2009}]{Neves09}
Neves V., Santos N. C., Sousa S. G., Correia A. C. M., Israelian G., 2009, \textit{A\&A}, 497, 563

\bibitem[\protect\citeauthoryear{Nissen \& Schuster}{2010}]{Nissen10}
Nissen P. E., Schuster W. J., 2010, \textit{A\&A}, 511, 10

\bibitem[\protect\citeauthoryear{Pagel}{1970}]{Pagel70}
Pagel B. E. J., \textit{Vistas in Astronomy}, 1970, 12-1, 313

\bibitem[\protect\citeauthoryear{Pancino et al.}{2010}]{Pancino10}
Pancino E., Carrera R., Rossetti E., Gallart C., 2010, \textit{A\&A}, 511, 56 (Pe10)

\bibitem[\protect\citeauthoryear{Paulson et al.}{2003}]{Paulson03}
Paulson D. B., Sneden C., Cochran W. D., 2003, \textit{AJ}, 125, 3185 (PSC03)

\bibitem[\protect\citeauthoryear{Pietrinferni et al.}{2004}]{Pietrinferni04}
Pietrinferni A., Cassisi S., Salaris M., Castelli F., 2004, \textit{ApJ}, 612, 168 

\bibitem[\protect\citeauthoryear{Pietrinferni et al.}{2006}]{Pietrinferni06}
Pietrinferni A., Cassisi S., Salaris M., Castelli F., 2006, \textit{ApJ}, 642, 797

\bibitem[\protect\citeauthoryear{Pilachowski}{1985}]{Pilachowski09}
Pilachowski A. A., 1985, \textit{PASP}, 97, 801

\bibitem[\protect\citeauthoryear{Pipino et al.}{2009a}]{Pipino09a}
Pipino A., Chiappini C., Graves G., Matteucci F., 2009a, \textit{MNRAS}, 396, 1151

\bibitem[\protect\citeauthoryear{Pipino et al.}{2009b}]{Pipino09b}
Pipino A., Devriendt J. E. G., Thomas D., Silk J., Kaviraj S., 2009b, \textit{A\&A}, 505, 1075 

\bibitem[\protect\citeauthoryear{Piskunov et al.}{1995}]{Piskunov95}
Piskunov N. E., Kupka F., Ryabchikova T. A., Weiss W. W., Jeffery C. S., 1995, \textit{A\&AS}, 112, 525

\bibitem[\protect\citeauthoryear{Proctor et al.}{2002}]{Proctor02}
Proctor R. N., Sansom A. E., 2002, \textit{MNRAS}, 333, 517

\bibitem[\protect\citeauthoryear{Ram\'\i rez \& Cohen}{2002}]{RamirezCohen02}
Ram\'\i rez S. V., Cohen J. G., 2002, \textit{AJ}, 123, 3277 (RC02)

\bibitem[\protect\citeauthoryear{Reddy et al.}{2006}]{Reddy06}
Reddy B. E., Lambert D. L., Allende Prieto C., 2006, \textit{MNRAS}, 367, 1329 (RLA06)

\bibitem[\protect\citeauthoryear{Rodriguez-Merino et al.}{2005}]{Rodriguez-Merino05}
Rodriguez-Merino L. H. Chavez M., Bertone E., Buzzoni A., 2005, \textit{ApJ}, 626, 411

\bibitem[\protect\citeauthoryear{Ryabchikova et al.}{1997}]{Ryabchikova97}
Ryabchikova T. A., Piskunov N. E., Kupka F., Weiss W. W., 1997, \textit{Baltic Astronomy}, 6, 244

\bibitem[\protect\citeauthoryear{Sanchez-Blazquez et al.}{2006}]{Sanchez-Blazquez06}
S\'anchez-Bl\'azquez P., Peletier R. F., Jim\'enez-Vicente J., Cardiel N., Cenarro A. J.,
Falc\'on-Barroso J., Gorgas J., Selam S., Vazdekis A., 2006, \textit{MNRAS}, 371, 703 

\bibitem[\protect\citeauthoryear{Schiavon et al.}{2007}]{Schiavon07}
Schiavon R. P., 2007, \textit{ApJS}, 171, 146

\bibitem[\protect\citeauthoryear{Schuler et al.}{2009}]{Schuler09}
Schuler S. C., King J. R., Lih-Sin, 2009, \textit{ApJ}, 701, 837 (SKL09)

\bibitem[\protect\citeauthoryear{Serven et al.}{2005}]{Serven05}
Serven J., Worthey G., Briley M. M., 2005, \textit{ApJ}, 627, 754

\bibitem[\protect\citeauthoryear{Shetrone}{1996}]{Shetrone96}
Shetrone M. D., 1996, \textit{AJ}, 112, 1517

\bibitem[\protect\citeauthoryear{Smith et al.}{2009}]{Smith09}
Smith R. J., Lucey J. R., Hudson M. J., 2009, \textit{MNRAS}, 400, 1690

\bibitem[\protect\citeauthoryear{Smith et al.}{2009}]{Smith09}
Smith R. J., Lucey J. R., Hudson M. J., Bridges T. J., 2009, \textit{MNRAS}, 398, 119

\bibitem[\protect\citeauthoryear{Sneden}{2002}]{Sneden02}
Sneden C., 2002, \textit{The MOOG code In: http://verdi.as.utexas.edu/moog.html} 

\bibitem[\protect\citeauthoryear{Sneden et al.}{2004}]{Sneden04}
Sneden C., Kraft R. P., Guhathakurta P., Peterson R. C., Fulbright J. P., 2004, \textit{AJ}, 127, 2162 (Se04)

\bibitem[\protect\citeauthoryear{StephensAnnMerchant}{2002}]{Stephens02}
Stephens A., Ann Merchant B., 2002, \textit{AJ}, 123, 1647

\bibitem[\protect\citeauthoryear{Tantalo et al.}{1998}]{Tantalo98}
Tantalo R., Chiosi C., Bressan A., 1998, \textit{A\&A}, 333, 419

\bibitem[\protect\citeauthoryear{Terndrup et al.}{1995}]{Terndrup95}
Terndrup D. M., Sadler E. M., Rich R. M., 1995, \textit{AJ}, 110, 1774

\bibitem[\protect\citeauthoryear{Thevenin}{1998}]{Thevenin98}
Th\'evenin F., 1998, \textit{yCat}, 3193, 0

\bibitem[\protect\citeauthoryear{Thomas et al.}{2003}]{Thomas03}
Thomas D., Maraston C., Bender R., 2003, \textit{MNRAS}, 339, 897

\bibitem[\protect\citeauthoryear{Tinsley}{1980}]{Tinsley80}
Tinsley B. M., 1980, \textit{Fundamentals of Cosmic Phys.}, 5, 287

\bibitem[\protect\citeauthoryear{Trager et al.}{2000a}]{Trager00a}
Trager S. C., Faber S. M., Worthey G., Gonz\'alez J. J., 2000a, \textit{AJ}, 119, 1645

\bibitem[\protect\citeauthoryear{Trager et al.}{2000b}]{Trager00b}
Trager S. C., Faber S. M., Worthey G., Gonz\'alez J. J., 2000b, \textit{AJ}, 120, 165

\bibitem[\protect\citeauthoryear{Tripicco \& Bell}{1995}]{Tripicco95}
Tripicco M. J., Bell R. A., 1995, \textit{AJ}, 110, 3035

\bibitem[\protect\citeauthoryear{Vazdekis et al.}{1997}]{Vazdekis97}
Vazdekis A., Peletier R. F., Beckman J. E., 1997, \textit{ApJS}, 111, 203

\bibitem[\protect\citeauthoryear{Vazdekis et al.}{2010}]{Vazdekis10}
Vazdekis A., S\'anchez-Bl\'azquez P., Falc\'on-Barroso J., Cenarro A. J.,
Beasley M. A., Cardiel N., Gorgas J., Peletier R. F., 2010, \textit{MNRAS}, 404, 1639

\bibitem[\protect\citeauthoryear{Walcher et al.}{2009}]{Walcher09}
Walcher C. J., Coelho P., Gallazzi A., Charlot S., 2009, \textit{MNRAS}, 398L, 44

\bibitem[\protect\citeauthoryear{Weiss et al.}{1995}]{Weiss95}
Weiss A., Peletier R.F., Matteucci F., 1995, \textit{A\&A}, 296, 73

\bibitem[\protect\citeauthoryear{Yong et al.}{2004}]{Yong04}

Yong D., Lambert D. L., Allende Prieto C., 2004, \textit{ApJ}, 603, 697 (YLA04)


\end{thebibliography}
\end{document}